\documentclass[a4paper,notitlepage]{report}   	%
\renewcommand{\thechapter}{{\textbf{\arabic{chapter}}}}
\renewcommand{\thesection}{\thechapter:$\,$\arabic{section}}

\renewcommand{\appendix}{
	\section*{Appendices}
	\setcounter{section}{0}
	\renewcommand{\thesection}{\thechapter:$\,$\Alph{section}}
}

\usepackage{amsmath}
\usepackage{amssymb}
\usepackage{xr}
\usepackage{refcount}

\usepackage{graphicx}
\usepackage{mathrsfs}
\usepackage{bm}
\usepackage[hyperref]{xcolor} %
\definecolor{darkgreen}{rgb}{0,0.4,0} %
\definecolor{darkred}{rgb}{0.55,0,0} %
\definecolor{navy}{rgb}{0,0,0.55} %
\usepackage[hypertexnames=false,breaklinks=true,colorlinks=true,citecolor=darkgreen,linkcolor=black,urlcolor=navy]{hyperref} %
\usepackage{amsthm}
\usepackage{lscape}
\theoremstyle{definition} %

\newcommand{\F}{\mc{F}}
\newcommand{\G}{\mc{G}}
\newcommand{\Z}{\mc{Z}}

\newcommand{\prm}[1]{\protect{$#1$}}

\newcommand{\sol}[1]{}
\newcommand{\h}{\mathrm{H}}%
\newcommand{\bmh}{{\bm{\mathrm{H}}}}%

\newcommand{\tc}{{\tilde{c}}}

\newcommand{\mrm}[1]{\mathrm{#1}}
\newcommand{\mbf}[1]{\mathbf{#1}}
\newcommand{\mbb}[1]{\mathbb{#1}}
\newcommand{\mc}[1]{\mathcal{#1}}

\newcommand{\mscr}[1]{\mathscr{#1}}
\newcommand{\bgrid}{\left(\begin{array}{rrr}}
\newcommand{\egrid}{\end{array}\right)}
\newcommand{\bgridt}{\left(\begin{array}{rr}}
\newcommand{\egridt}{\end{array}\right)}
\newcommand{\bgridtt}{\left[\begin{array}{rr}}
\newcommand{\egridtt}{\end{array}\right]}
\newcommand{\os}{\!\!\not\!} %
\newcommand{\eref}[1]{(\ref{#1})}
\newcommand{\Eref}[1]{Eq.~(\ref{#1})}

\newcommand{\Erefr}[2]{Eqs.~(\ref{#1}--\ref{#2})}

\newcommand{\erefr}[2]{(\ref{#1}--\ref{#2})}

\newcommand{\Erefs}[2]{Eqs.~(\ref{#1}) and~(\ref{#2})}
\newcommand{\Ereft}[1]{Eqs.~(\ref{#1})}

\newcommand{\cref}[1]{Chapter~\ref{#1}}
\newcommand{\Cref}[1]{Chapter~\ref{#1}}
\newcommand{\crefs}[2]{Chapters~\ref{#1} and~\ref{#2}}
\newcommand{\crefr}[2]{Chapters~\ref{#1}--\ref{#2}}

\newcommand{\fref}[1]{Fig.~\ref{#1}}

\newcommand{\sref}[1]{Sec.~\ref{#1}}

\newcommand{\Sref}[1]{Section~\ref{#1}}
\newcommand{\srefs}[2]{Secs.~\ref{#1} and~\ref{#2}}

\newcommand{\Srefs}[2]{Sections~\ref{#1} and~\ref{#2}}
\newcommand{\srefr}[2]{Secs.~\ref{#1}--\ref{#2}}

\newcommand{\tref}[1]{Table~\ref{#1}}
\newcommand{\bsm}{\bar\sigma^\mu}
\newcommand{\bsmm}{\bar\sigma_\mu}

\newcommand{\bsnn}{\bar\sigma_\nu}

\newcommand{\nn}{\nonumber}
\newcommand{\rmd}{\mathrm{d}}
\newcommand{\rmi}{\mathrm{i}}%
\newcommand{\la}{\langle}
\newcommand{\ra}{\rangle}
\newcommand{\dslash}{{\,\os\partial}}
\newcommand{\Dslash}{{\,\os\! D}}
\newcommand{\bmDslash}{\bm{\Dslash}}

\newcommand{\rcite}[1]{Ref.~\cite{#1}}

\newcommand{\p}[1]{\phantom{#1}}

\newcommand{\Weyl}[2]{\left(\begin{aligned}&#1\\&#2\end{aligned}\right)}

\newcommand{\bg}{\mrm{bg}}
\newcommand{\fg}{\mrm{fg}}
\newcommand{\bgfield}[1]{[#1]_\mrm{bg}}
\newcommand{\fgfield}[1]{[#1]_\fg}
\newcommand{\OO}[1]{\mrm{O}\left(#1\right)}

\newcommand{\eV}{\mrm{eV}}

\newcommand{\MeV}{\mrm{M}\eV}
\newcommand{\GeV}{\mrm{G}\eV}

\newcommand{\SU}[1]{\mrm{SU}(#1)}

\newcommand{\SOp}[1]{\mrm{SO}^+(#1)}
\newcommand{\U}[1]{\mrm{U}(#1)}

\newcommand{\rmU}{\mrm{U}}

\newcommand{\peref}[1]{\protect{(\ref{#1})}}
\newcommand{\pEref}[1]{\protect{Eq.~(\ref{#1})}}

\newcommand{\psref}[1]{\protect{Sec.~\ref{#1}}}

\newcommand{\GL}[2]{\mrm{GL}(#1,\mbb{#2})}

\newcommand{\SL}[2]{\mrm{SL}(#1,\mbb{#2})}

\newcommand{\SLTC}{\mrm{SL}(2,\mbb{C})}

\newcommand{\btheta}{\theta}
\newcommand{\bthetabar}{\bar\theta}
\newcommand{\bthetas}{(\bthetabar,\btheta)}
\renewcommand{\bar}[1]{\overline{#1}}

\newcommand{\ILO}[1]{\mrm{O}(#1)}

\newcommand{\N}{{\mathfrak{n}}}%
\newcommand{\RM}{{{\mbb{R}^{1,3}}}}

\newcommand{\Cw}[1]{{{\mbb{C}^{\wedge #1}}}}
\newcommand{\RT}{{\mbb{R}^{3}}}

\newcommand{\vp}{\varphi}

\newcommand{\Span}[2]{\mrm{Span}\{#1\,|\,#2\}}
\newcommand{\wn}{{\wedge n}}
\newcommand{\wt}{{\wedge 2}}
\newcommand{\Cwt}{{\Cw{2}}}
\newcommand{\Cwn}{{\Cw{\N}}}
\newcommand{\Cwtn}{{\Cw{2\N}}}

\newcommand{\quietGversion}[1]{}
\newcommand{\onlyinsummary}[1]{}

\newcommand{\completed}[1]{}
\newcommand{\chapeight}[1]{}

\newcommand{\chap}[1]{}
\newcommand{\notchap}[1]{#1}

\newcommand{\notstandalone}[1]{}
\usepackage{geometry}
\usepackage[numbers,square,sort&compress]{natbib}

\begin{document}

\title{A Classical Analogue to the Standard Model\\and General Relativity}
\author{Chapter 1\\~\\R. N. C. Pfeifer}

\date{01 January 2024}

\maketitle
\thispagestyle{empty}
\newpage

\newgeometry{left=2.0cm,right=2.0cm,top=2cm,bottom=2.5cm}
\section*{Overview: CASMIR}%

~

The Classical Analogue to the Standard Model In pseudo-Riemannian space-time (CASMIR) \cite{pfeifer2022CASM1,pfeifer2022CASM2,pfeifer2022CASM3,pfeifer2022CASM4,pfeifer2022CASM5} 
is a model designed to assist in the evaluation of certain interactions involving the Standard Model (SM) in a gravitational field, and to reduce to the Standard Model plus general relativity (GR) in appropriate regimes.
By construction it displays high congruence with the Standard Model across almost all experimentally accessible regimes, and reproduces key Standard Model parameters which are not input into the model
to within $0.2\,\sigma$ %
of their observed values.

~

The model has no tuning parameters, and takes as its input only the fine structure constant and the masses of the electron and muon. In terms of these parameters it yields a system of equations which are predictive of (retrodictive of) the values of 
\begin{itemize}
\item the $W$~boson mass as measured by ATLAS~\citep{the-ATLAS-collaboration2023},
\item the $W$~boson mass as measured by CDF~II~\citep{aaltonen2022},
\item the anomalous value of the muon gyromagnetic anomaly $a_\mu$ as measured by Fermilab Muon $g-2$~\citep{aguillard2023},
\item the Particle Data Group (PDG) consensus value of the $Z$~boson mass~\citep{workman2022},
\item the ATLAS high-precision value of the Higgs boson mass~\citep{aad2023},%
\item the PDG consensus value of the tau mass~\citep{workman2022}, and
\item the NIST/CODATA value of Newton's constant~\cite{tiesinga2018},
\end{itemize}
all with tensions of $0.2\,\sigma$ or less
(see \tref{tab:overview}). %

~

\begin{table}[!h]
\caption{Fundamental constants and key experimental results: Observed values, values predicted by CASMIR, and values predicted by the Standard Model (where these exist).\label{tab:overview}}
{\small
\begin{center}\begin{tabular}{|ll|r@{~}r|r|c|l@{~}r|c|} %
\hline
\multicolumn{2}{|c|}{Parameter}&
\multicolumn{2}{c|}{Observed value} &\multicolumn{2}{c|}{CASMIR \cite{pfeifer2022CASM1,pfeifer2022CASM2,pfeifer2022CASM3,pfeifer2022CASM4,pfeifer2022CASM5}}&\multicolumn{3}{c|}{Standard Model}\\\cline{5-9}
&&&& \multicolumn{1}{c|}{Calculated} & Tension $(\sigma)$ &\multicolumn{2}{c|}{Calculated}&Tension $(\sigma)$\\\hline\hline
$m_W$ (ATLAS)&$(\GeV/c^2)$& 80.360(16)& %
\citep{the-ATLAS-collaboration2023}~~& 80.3587(22)~ & $\p{<}\;0.1$ & \multicolumn{2}{c|}{80.356(6)  \citep{workman2022,awramik2004,erler2019}\p{$^\dagger$}} & \p{$^*$}0.2\p{$^*$}\\
$m_W$ (CDF~II)&$(\GeV/c^2)$ & 80.4335(94)& \citep{aaltonen2022}~~& 80.4340(22)~ & $\p{<}\;0.1$ & \multicolumn{2}{c|}{80.356(6) \citep{workman2022,awramik2004,erler2019}\p{$^\dagger$}} & \p{$^*$}6.9\p{$^*$}\\
$a_\mu\times 10^{11}$& & 116592059(22)&
\cite{aguillard2023}~~& 116592071(46)~ &$\p{<}\;0.2$ & 116591810(43)&\citep{aoyama2020}\p{$^\dagger$} & \p{$^*$}5.2$^*$\\
$m_Z$ (LEP)&$(\GeV/c^2)$& 91.1876(21)&  \citep{workman2022}~~& 91.1877(35)~ & $<0.1$ && ---\p{$^\dagger$} & \p{$^*$}---\p{$^*$}\\
$m_Z$ (CDF~II)&$(\GeV/c^2)$& 91.1920(75)& \citep{aaltonen2022}~~& 91.1922(35)~ & $<0.1$ &91.1876(21)&\citep{workman2022}$^\dagger$&0.6\\
$m_\bmh$&$(\GeV/c^2)$& 125.11(11)&   \citep{aad2023}~~& 125.1261(48)~ & $\p{<}\;0.1$ && ---$^\ddagger$ & \p{$^*$}---\p{$^*$}\\ %
$m_\tau$&$(\MeV/c^2)$& 1776.86(12)&  \citep{workman2022}~~& 1776.867413(43)~ & $\p{<}\;0.1$ && ---\p{$^\dagger$} & \p{$^*$}---\p{$^*$}\\
\multicolumn{2}{|l|}{$G_N\times10^{11}$\hfill($\mrm{m}^3\mrm{kg}^{-1}\mrm{s}^{-2}$)} & $6.67430(15)$ &\citep{tiesinga2018}~ & $6.67426(230)~$ & $<0.1$ &&---\p{$^\dagger$}&\p{$^*$}---\p{$^*$}\\\hline
\end{tabular}\end{center}

~

\onlyinsummary{\profomit{Note for lay readers: ``Tension'' is a measure of how well a theoretical result agrees with experiment. A tension below 1 is essentially perfect agreement, while a tension of 5 or higher is a clear indicator of a strong disagreement. A tension between 4 and 5 is highly suggestive of a possible disagreement and is worthy of substantial further investigation.

~}}

$^*$ There have been two important developments since this value was obtained in 2020: A lattice calculation of the Hadron Vacuum Polarisation (HVP) which disagrees with previous data-driven values obtained from $e^+e^-$ experiments~\cite{borsanyi2021}, and a measurement of the $e^+e^-\rightarrow \pi^+\pi^-$ cross-section at CMD-3~\cite{ignatov2023} which also disagrees with previous $e^+e^-$ experiments. Attempts to reconcile these results are ongoing~\cite{colangelo2022}. Pending this reconciliation, it is likely that a contemporary re-evaluation of the Standard Model prediction for $a_\mu$ would have greater uncertainty and thus smaller tension with experiment.

~

$^\dagger$ The Standard Model predicts no discrepany between the values of the $Z$~boson mass in LEP and CDF~II, thus the observed value at LEP may be taken as a prediction for the value at CDF~II.

~

$^\ddagger$ The mass of the Higgs in the Standard Model is a measured rather than a derived quantity. Theoretical considerations provided some broad constraints on the Higgs mass prior to its detection in 2012, but these are not a calculated value in the sense considered here.}
\end{table}

\restoregeometry

~

{\small This is paper one in a series of seven papers describing the $\Cw{18}$ Classical Analogue to the Standard Model In pseudo-Riemannian space-time (CASMIR).} %
{%
The latest version of this document may always be found at \href{https://www.academia.edu/65931513}{https://www.academia.edu/65931513}, and may incorporate additions and amendments to the arXiv versions.

~

Versions of this work may be found at: %
\begin{center}
\begin{tabular}{c|c}
Chapter & \multicolumn{1}{c}{Available from} %
\\~&\\\cline{1-2}\cline{1-2}
1   & arXiv:2106.08130 \cite{pfeifer2022CASM1} \\ %
2  & arXiv:2108.07719 \cite{pfeifer2022CASM2} \\ %
3 & arXiv:0805.3819\p{1} \cite{pfeifer2022CASM3} \\ %
4-7,9-11  & arXiv:2008.05893 \cite{pfeifer2022CASM4} \\
8 & arXiv:2212.01255 \cite{pfeifer2022CASM5} \\
& (For Ch.~8 see also {Int. J. Mod. Phys. A},\\
& \href{https://doi.org/10.1142/S0217751X23400043}{doi:10.1142/S0217751X23400043})
\end{tabular}
\end{center}

~

Significant portions of this work were carried out at:
\begin{itemize}
\item{School of Mathematics and Physics, The University of Queensland, St.\ Lucia, QLD~4072, Australia}
\item{Department of Physics, The University of Western Australia, 35 Stirling Highway, Perth, WA~6009, Australia}
\item{Perimeter Institute for Theoretical Physics, 31 Caroline St. N, Waterloo, ON, Canada}
\item{Dept.~of Physics \& Astronomy, Macquarie University, NSW~2109, Australia}
\item{School of Physics, The University of Sydney, Sydney, NSW~2006, Australia}
\end{itemize}
}

\section*{Updates}

\subsubsection*{31 December 2023}
\begin{itemize}
\item Reintroduced some missing magnitude signs in \sref{sec:sxBRspecifics}.
\item Clarified table headings appearing in \tref{tab:sxBRshifts}. Discussed the meanings of the calculated values. Corrected some values in \tref{tab:sxBRshifts} (the conclusions drawn remain unchanged). Added comment on $p\bar{p}$ collisions.
\item Added publication note for \cref{ch:CDF2}.
\item Added rudimentary \cref{ch:Higgs} on the physics of the Higgs boson, including a comment on production cross-sections. Referenced this from \sref{sec:addconstrs}.
\item Fixed some minor typos in \sref{sec:scalbosmass}. %
\item Clarified point~\ref{item:scalbosonloopspara_kind2term2} in \sref{sec:scalbosonloopspara}.
\item Found missing factor of~3 in the $\ILO{{N_0}^{-1}}$ and $\ILO{\alpha {N_0}^{-1}}$ corrections to the Higgs mass [\srefs{sec:scalbosonmass}{sec:scalbosmass}; in particular \Erefs{eq:Hwithk2}{eq:WHratio}]. Higgs mass reduced by $0.1\%$.
\item Clarified explanation of factor of~$\frac{1}{9}$ in \sref{sec:mHgluonbosonloops}.
\item Brief discussion of the strength of the strong interaction added to \sref{sec:strongint}.
\item \rcite{aad2023} is now the
state-of-the-art published measurement of the Higgs mass.
\end{itemize}

\subsubsection*{05 November 2023}
\begin{itemize}
\item Revised and extended \cref{ch:CDF2}. Added discussion of $W$~boson and $Z$~boson production cross-sections (\sref{sec:sxBR}).
\end{itemize}

\subsubsection*{04 September 2023}
\begin{itemize}
\item Identified $2\mc{L}_\Omega$ as the shortest scale over which particles consistently take part in mass interactions, in contrast with $\mc{L}_0$ which is the largest scale over which mass vertices correlate (\sref{sec:Wmass5v}).
\item Substantially updated understanding of weak massive boson formation, with corrections to the values of $m_{Z^{\tc}}$ and $m_{Z^{\dot cc}}$ in \sref{sec:colouredbosonmasses} and to the application of CASMIR to CDF~II in \cref{ch:CDF2}. Chromatic $Z$ boson effects are predicted to be present in two-lepton CDF~II $Z$~boson events. Predictions are consistent with observation.
\item Incorporated updated results from Muon~$g-2$ experiment into \cref{ch:CDF2}. Acknowledged updated analysis of Higgs mass from ATLAS.
\item Tidied up the discussion in \srefs{sec:pairdecay}{sec:higherorder} around masslessness of particles in \fref{fig:persistentfigures}(iv).
\end{itemize}

\subsubsection*{01 August 2023}
\begin{itemize}
\item Clarified comments about physical results being independent of convention choice in \sref{sec:QCcorr}.
\item Added further discussion of some aspects of the gravitational calculation as \sref{sec:observations}.
\end{itemize}

\subsubsection*{26 June 2023}
\begin{itemize}
\item Corrected typos in \Eref{eq:Zcadjustment}, \Eref{eq:Zccbaremass}, and the expression for ${m_{Z^{\tc}}^2}$ in the Overview.
\end{itemize}

\subsubsection*{19 June 2023}
\begin{itemize}
\item Merged Discussion and Conclusion in \cref{ch:CDF2}
\end{itemize}

\subsubsection*{11 June 2023}
\begin{itemize}
\item Full treatment of chromatic weak vector bosons and their impact on all particle mass calculations. This major update supercedes all previous versions.
\end{itemize}

\tableofcontents

\chapter{Normalisable quasiparticles on a manifold with anticommuting co-ordinates\label{ch:simplest}}

\begin{abstract}
Quasiparticles and analog models are ubiquitous in the study of physical systems. Little has been written about quasiparticles on manifolds with anticommuting co-ordinates, yet they are capable of emulating a surprising range of physical phenomena. This chapter introduces a classical model of free fields on a manifold with anticommuting co-ordinates, identifies the region of superspace which the model inhabits, and shows that the model emulates the behaviour of a five-species interacting quantum field theory on $\RM$. The Lagrangian of this model arises entirely from the anticommutation property of the manifold co-ordinates. Later chapters extend this toy model to incorporate greater numbers of species, to provide the construction underpinning the Classical Analogue to the Standard Model In pseudo-Riemannian space--time (CASMIR).
\end{abstract}

\section{Introduction\label{sec:intro}}

Quasiparticles are ubiquitous throughout physics. 
Hadrons are collections of quarks which collectively admit a quasiparticle description in the low-energy regime. Atomic nuclei, the same. Indeed, if seeking %
to be contentious, one could argue that the entirety of condensed matter physics is no more than the study of collective excitations in all their multiplicity---scaling dimensions and critical exponents describing the collective properties of these excitations, phase transitions the boundaries between different excitation regimes, and so forth. %
Quasiparticles exist in (literally) infinite variety, from commonplace entities such as phonons \cite{srivastava1990}, solitons \cite{drazin1989}, and Cooper pairs \cite{cooper1956,bardeen1957} through to more exotic topological phenomena such as skyrmions \cite{skyrme1961,skyrme1962,tokura2020}, hopfions  \cite{gobel2020}, hedgehogs \cite{tokura2020,gobel2020}, and anyons (abelian and nonabelian, each including families infinite in their own right) \cite{kitaev2006,bonderson2007}.

Intersecting this topic, especially when viewed through the lens of quasiparticle phenomenology, is the study of analogue models \cite{maynard2001,dragoman2004,lewenstein2007}. Indeed, any system supporting quasiparticles may be considered as an analogue system for the species described, whether or not such a species otherwise appears in nature. The field of analogue models is %
far broader than this, however, and also includes such notable results as the mapping between the one-dimensional quantum Ising chain and the two-dimensional classical Ising model \cite{suzuki1976}, allowing application of Onsager's exact solution \cite{onsager1944} to the quantum system, and more recently, the study of gravitational analogues \cite{visser2002,liberati2009,barcelo2011} in systems as varied as fluid flows \cite{unruh1981}, Bose--Einstein condensates \cite{garay2000,garay2001,lahav2010}, optics \cite{gordon1923,leonhardt1999,leonhardt2000}, superfluid helium \cite{jacobson1998}, electromagnetic systems \cite{reznik2000,schutzhold2005}, and even water waves \cite{schutzhold2002}.

Given this ubiquity, %
it might seem curious that so little %
has been written about quasiparticle models on manifolds with anticommuting co-ordinates. %
Perhaps this is because Taylor series on such manifolds %
truncate after a finite number of terms, %
so
these manifolds appear to be %
unlikely candidates for hosting normalisable excitations.

This chapter considers a manifold $M$ constructed from a vector space with four real dimensions, supplememted by a product operation under which its %
basis elements anticommute. %
A homomorphism is constructed from this manifold to $\RM$. Choosing an appropriate set of classical fields on $M$ then
gives rise to quasiparticle excitations on $\RM$ which emulate a simple interacting quantum field theory.
This demonstrates that manifolds with anticommuting co-ordinates can %
exhibit %
normalisable quasiparticle excitations, and also provides a familiar framework for the study of these excitations on $M$. Given the presence of anticommuting co-ordinate fields, it is perhaps inevitable that manifold $M$ is shown to be a submanifold of a superspace \cite{dewitt1986,buchbinder1998}.
Curiously, the Lagrangian of the emulated quantum field theory (with interactions) is seen to arise entirely from the anticommutation relationships of co-ordinates on $M$.

Choices of conventions are specified in \sref{sec:I:conventions}. \Srefs{sec:vecspace}{sec:anticommmanif} describe the anticommuting vector space, its embedding in the space of supernumbers $\Lambda_\infty$, the corresponding manifold, and its embedding in the %
superspace $\mbb{R}^{4|4}$. The mapping between $M$ and $\RM$ is established in \sref{sec:I:mapping}, and the field content of the microscopic model is introduced in \sref{sec:quantumdust}. \Srefs{sec:indivfields}{sec:prodfields} identify the physically relevant degrees of freedom in the pullback of these fields to $\RM$. \Sref{sec:QLintro} then begins construction of the analogue model, specifying a family of field configurations on $M$ which correspond to pseudovacuum states on $\RM$, while \sref{sec:quasi} %
introduces small perturbations to these configurations and shows that they correspond to quasiparticles on $\RM$. \Sref{sec:entropic} then relaxes some of the requirements imposed during the initial construction of the analogue model. \Sref{sec:semiclassical} makes the surprising observation that these quasiparticles behave as a \emph{semi}classical system, and provided certain criteria are met corresponding to the domain of validity of the analogue model, the quasiparticle fields %
emulate a simple quantum field theory.
\Sref{sec:physapp} speculates on %
a number of situations where the model presented here and its close relatives may provide useful physical insight.
Finally, \sref{sec:lookforward} 
anticipates the directions of subsequent chapters.

\section{Conventions\label{sec:I:conventions}}

\subsection{Notation}

This chapter uses two-component (Weyl) spinor notation%
. Spinor indices are represented by greek letters from the beginning of the alphabet, with and without dots, and are frequently left implicit, %
\begin{equation}
\theta\theta=\theta^{\alpha}\theta_{\alpha}\qquad\bar\theta\bar\theta=\bar\theta_{\dot\alpha}\bar\theta^{\dot\alpha}.\label{eq:indexsuppression}
\end{equation}
Spinor indices are raised and lowered using the totally antisymmetric tensors $\varepsilon^{\alpha\beta}$, $\varepsilon_{\alpha\beta}$, $\varepsilon^{\dot\alpha\dot\beta}$, and $\varepsilon_{\dot\alpha\dot\beta}$.
Sigma matrices follow the convention 
\begin{align}
\sigma_\mu&\equiv\{\mbb{I}_2,\bm\sigma\}\\
\bar\sigma_\mu&\equiv\{\mbb{I}_2,-\bm\sigma\}
\end{align}
where $\bm\sigma$ are the Pauli matrices
\begin{equation}
\begin{pmatrix}0&1\\1&0\end{pmatrix}\qquad
\begin{pmatrix}0&-\rmi\\\rmi&\phantom{-}0\end{pmatrix}\qquad
\begin{pmatrix}1&\phantom{-}0\\0&-1\end{pmatrix}. 
\end{equation}
The barred and unbarred sigma matrices are related by
\begin{equation}
\bar\sigma^{\dot\alpha\alpha}_\mu=\varepsilon^{\dot\alpha\dot\beta}\varepsilon^{\alpha\beta}\sigma_{\mu\beta\dot\beta}.
\end{equation}
The signature of the space--time metric is $-,+,+,+$ and the value of $c$ is 1.

Where it does not cause ambiguity, indices (both spinor and vector) may be left implicit when writing functions of the components of a vector or tensor, e.g.
\begin{align}
\F(x^\mu)&\longrightarrow \F(x)\\
\G(%
\theta_\alpha)&\longrightarrow \G(%
\theta).
\end{align}

Regarding sign conventions for higher-order derivative operators on $\RM$, the 
Laplacian and the vector boson derivative operator are defined as
\begin{equation}
\square:=\partial^\mu\partial_\mu\quad\textrm{and}\quad\triangle^{\mu\nu}:= g^{\mu\nu}\square-\partial^\mu\partial^\nu
\end{equation}
respectively.

To avoid confusion with enumerated roman counting indices e.g.~$i$, $j$, etc., when referring to a model in the series $\Cwtn$ the parameter of the series is written as $\N$, not $n$.

\subsection{Vectors and forms}

Following conventions of differential geometry,
tangent vectors of a manifold carry a lowered index (so their components have raised indices), while covectors and differential forms have raised indices (so their components have lowered indices) \cite{frankel2004}. Thus, for example, the components of a co-ordinate vector $x$ on $\RM$ carry a raised index ($x^\mu$) and contract with the tangent vectors $\vec{\partial}_\mu$, while a 1-form $\mbf{f}$ has components $f_\mu$ which contract with the cotangent (1-form) basis $\rmd x^\mu$. [The exception is for undotted two-component complex spinors, where consistency between index suppression~\eref{eq:indexsuppression} and matrix multiplication implies that the components $\theta_\alpha$, which have a low index, nevertheless behave as a column vector.]

A ``$k$-vector'' is an object whose components carry $k$ vector-style indices, e.g.~$V^{\mu\nu}$ contracting with $\vec{\partial}_\mu\vec{\partial}_\nu$. A 1-vector with $n$ components (e.g.~$v^\mu~|~\mu\in\{1,\ldots,n\})$ is an ``$n$-component vector'', not an $n$-vector. Where $k$ is not stated, a ``vector'' is a 1-vector.

A form acts on a vector from the left. Thus, for example, given complementary orthogonal bases $\vec{\partial}_\mu$ and $\rmd x^\mu$, a 1-form $\mbf{f}$ and a vector $\mbf{v}$ satisfy
\begin{equation}
\mbf{f}=f_\mu\rmd x^\mu\qquad\mbf{v}= v^\mu\vec{\partial}_\mu\qquad
\mbf{f}(\mbf{v})=f_\mu v^\mu.
\end{equation}
If notation requires that a vector act on a form from the left, this is represented using $i$-notation,
\begin{equation}
i_{\mbf{v}}(\mbf{f}) = (-1)^\phi\mbf{f}(\mbf{v})\label{eq:inotation}
\end{equation}
where $\phi$ is 0 if the vector and form commute and 1 otherwise.

Tangent vectors to a manifold $M$ inhabit a tangent space $TM$, and 1-forms inhabit the cotangent space $T^*M$ dual to $TM$. When $TM$ is identified with some other space, say $\mbb{R}^4$, then $T^*M$ may be denoted similarly, e.g.~$\mbb{R}^{4*}$.

Unless otherwise specified, a reference to ``the space of $p$-forms'' refers to the space of all forms across all values of $p$, and likewise ``the space of $k$-vectors'' includes 1-vectors, 2-vectors, etc. across all values of $k$.

\subsection{Grassmann algebras}

The elements of the infinite-dimensional Grassmann algebra $\Lambda_\infty$ over a field $\mbb{C}$ are termed the supernumbers \cite{buchbinder1998}, and %
any element $z\in\Lambda_\infty$ may be decomposed into a sum of even ($z_c$) and odd ($z_a$) parts,
\begin{equation}
z=z_c+z_a.
\end{equation}
Even parts commute with all parts of all supernumbers, while odd parts anticommute with other odd parts. An element $z$ having vanishing $z_c$ or $z_a$ is a pure supernumber, termed a $c$-number if even, or an $a$-number if odd. 
The spaces of even and odd pure supernumbers are denoted $\mbb{C}_c$ and $\mbb{C}_a$ respectively, and satisfy
\begin{equation}
\mbb{C}_c\cdot\mbb{C}_c=\mbb{C}_c\quad~
\mbb{C}_a\cdot\mbb{C}_a\!\subset\!\mbb{C}_c~\quad
\mbb{C}_a\cdot\mbb{C}_c=\mbb{C}_c\cdot\mbb{C}_a=\mbb{C}_a.
\end{equation}
Supernumbers admit an operation of complex (hermitian) conjugation, and the product of two $a$-numbers is an imaginary $c$-number.

The space of supernumbers may be generated by the collection
\begin{equation}
\bm\{1,\{\zeta_i|i\in\mbb{Z}^+\};\cdot\,;+;\mbb{C}\bm\},\label{eq:gensupernums}
\end{equation}
comprising a product operator, a sum operator, a set of anticommuting basis elements $\zeta_i$, 
\begin{equation}
\zeta_i\zeta_j=-\zeta_j\zeta_i,
\end{equation}
and a field. The notation $\zeta_i$ is reserved for {generators} of the superalgebra, and thus while $z\in\Lambda_\infty$ represents an arbitrary supernumber, $\zeta_i\in\Lambda_\infty$ always specifically refers to one of the anticommuting generators in \Eref{eq:gensupernums}.

It may be useful to refer to the \emph{tier} of a supernumber. A supernumber which is a product of $n$ anticommuting generators $\zeta_i$ and a complex number is a \emph{tier-$n$} supernumber. A sum of tier-$n$ supernumbers is also a tier-$n$ supernumber. %
For example,
\begin{equation}
C^{i_1\ldots i_k}\zeta_{i_1}\ldots\zeta_{i_k}
\end{equation}
is a general tier-$k$ supernumber. Supernumbers which are a sum of supernumbers from different tiers do not have a well-defined tier.

Returning to \Eref{eq:gensupernums}, now let $\Lambda^F_n$ denote the subalgebra obtained on restricting the number of anticommuting basis elements to $n$, constructed over a field $F$. That is, $\Lambda^F_n$ is generated by
\begin{equation}
\bm\{1,\{\zeta_i|i\in\mbb{Z}_n\};\cdot\,;+;F\bm\}.
\end{equation}
If field $F\subseteq\Lambda_\infty$, then this algebra satisfies $\Lambda^F_n\subseteq\Lambda_\infty$.

\section{The Model\label{sec:themodel}}

\subsection{Anticommuting vector spaces\label{sec:vecspace}}

\subsubsection{Construction}

An anticommuting vector space is a vector space whose basis vectors are elements of an anticommuting algebra, supplemented with an outer product operation which is the product operation on that algebra.
To describe %
such a space, begin with $\mbb{R}^n$, the prototypical vector space with $n$ basis vectors, and let these be represented by the elementary column vectors $\hat{\mbf{e}}_i$, $i\in\mbb{Z}^n$. If a space of 1-forms is constructed dual to these vectors, this %
is denoted $\mbb{R}^{n*}$, 
and is represented by the elementary row vectors $\hat{\mbf{e}}^i$,
\begin{equation}
\hat{\mbf{e}}^i\hat{\mbf{e}}_j=\delta^i_j.\label{eq:prodrowcolumn}
\end{equation}
Anticommutation may be introduced by replacing each basis vector $\hat{\mbf{e}}_i$ of $\mbb{R}^n$ by an anticommuting basis element $\zeta_i\in\Lambda_\infty$, %
\begin{equation}
\hat{\mbf{e}}_i\rightarrow \zeta_i.
\end{equation}
The resulting vector space, here denoted $\mbb{R}^{\wedge \N}$, is isomorphic to $\mbb{R}^n$ but has elements which are real tier-1 $a$-numbers in $\Lambda^\mbb{R}_n$ instead of column vectors. 
The space of 1-forms dual to $\mbb{R}^{\wedge \N}$ is denoted $\mbb{R}^{\wedge n*}$, and
may be represented by the %
real tier-1 $a$-numbers in a second copy of $\Lambda^\mbb{R}_n$, denoted $\Lambda^{\mbb{R}*}_n$. The elements of a basis of 1-forms are written $\zeta^i$, with raised indices, and their action on vectors from the left is defined by
\begin{equation}
\zeta^i(\zeta_j) = \zeta^i\zeta_j = \delta^i_j, \label{eq:zetaaction}
\end{equation}
analogous to \Eref{eq:prodrowcolumn}.
The exchange statistics of 1-vectors and 1-forms become relevant when an outer product is introduced, permitting construction of $k$-vectors and $p$-forms. %
This product operation is inherited from the supernumbers.
Anticommutation of the forms $\{\zeta^i\}$ automatically ensures antisymmetry of $p$-forms under index exchange, eliminating the need to specifically define a wedge product, and thus if $\mbf{F}$ is a 2-form 
\begin{align}
\mbf{F}&=F_{ij}\zeta^i\zeta^j
\end{align}
then $F_{ij}$ is antisymmetric.
Likewise, if $\mbf{v}$ and $\mbf{w}$ are 1-vectors, then $\mbf{vw}$ is a 2-vector
\begin{align}
\mbf{v}&=v^i\zeta_i\qquad\mbf{w}=w^i\zeta_i\label{eq:Rw4vectors}\\
\mbf{vw}&=v^iw^j\zeta_i\zeta_j=-\mbf{wv}\label{eq:contractPK}
\end{align}
where only the antisymmetric portion of $v^iw^j$ makes a nonvanishing contribution to the 2-vector $\mbf{vw}$. If a 2-vector $\mbf{A}$ is defined by $\mbf{A}=\mbf{vw}$, then
\begin{equation}
\begin{split}
\mbf{A}&=A^{ij}\zeta_i\zeta_j\qquad A^{ij}=\frac{1}{2}(v^iw^j-v^jw^i).%
\end{split}
\end{equation}
Antisymmetry under index exchange makes it simple to evaluate the action of a $p$-form on a $k$-vector from left, as it suffices to act each basis form on the basis vector to its immediate right:
\begin{align}
\mbf{F}(\mbf{A})&=F_{ij}A^{kl}~\zeta^i\zeta^j\zeta_k\zeta_l\label{eq:F(A)}
=F_{ij}A^{kl}~\zeta^i\,\delta^j_k\,\zeta_l\\
&=F_{ij}A^{jl}~\zeta^i\zeta_l=F_{ij}A^{jl}\delta^i_l
\nn\\
&=F_{ij}A^{ji}.\nn
\end{align}
Further, exploiting commutation of $\{\zeta^i\}$ and $\{\zeta_i\}$ with the reals %
yields the equation
\begin{align}
\zeta^i(\zeta^j\zeta_k)\zeta_l = &\zeta^j(\zeta^i\zeta_l)\zeta_k = (\zeta^i\zeta_l)(\zeta^j\zeta_k)
\end{align}
which exposes the additional anticommutation relationship
\begin{equation}
\{\zeta^i,\zeta_j\}=
0,\label{eq:extrazetacomm}
\end{equation}
though the utility of this relationship is limited %
to reorderings which do not change which basis forms $\zeta^i$ act on which basis vectors $\zeta_j$.

This construction also generalises to complex vector spaces, with $\mbb{C}^{\wedge \N}$ again denoting a vector space constructed on $n$ anticommuting basis vectors, this time satisfying
\begin{equation}
\begin{split}
\mbf{v}&= v^i\zeta_i,\qquad\forall~i,~v^i\in\mbb{C} %
\end{split}
\end{equation}
for all $\mbf{v}$ in $\mbb{C}^{\wedge \N}$.

\subsubsection{Symmetries of \protect{$\mbb{C}^n$} and \protect{$\mbb{C}^{\wedge \N}$}\label{sec:symCwn}}

Consider the vector space $\mbb{C}^n$. Its symmetries are well-known, but will provide a starting point for exploring the symmetries of $\Cwn$ so they are reviewed here. First, consider that the space $\mbb{C}^n$ is invariant under the action of the (complex) translation group on $\mbb{C}^n$, 
\begin{equation}
\begin{split}
t&:\mbb{C}^n\longrightarrow\mbb{C}^n\\
t&(\mbb{C}^n)=\mbb{C}^n
\end{split}
\qquad\forall\quad t\in T^n_{\mbb{C}^n},
\end{equation}
where a translation operator $T^n_V$ carries a subscript $V$ to show that it acts on vector space $V$.
By construction the elements of the translation group are in 1:1 correspondence with points in $\mbb{C}^n$, and an element $t$ which maps the origin $\mbf{0}$ to point $\mbf{t}$ may be associated with the vector $\mbf{t}$. Let this association define a mapping
\begin{equation}
\begin{split}
&\mbb{C}(\cdot):T^n_{\mbb{C}^n}\longrightarrow\mbb{C}^n\qquad
\mbb{C}(T^n_{\mbb{C}^n})=\mbb{C}^n\\
\end{split}
\end{equation}
such that
\begin{equation}
\begin{split}
\forall~t\in T^n_{\mbb{C}^{n}}:\qquad \begin{split}&t(\mbf{0}) = \mbf{t}\qquad \{\mbf{0},\mbf{t}\}\in\mbb{C}^{n}\\
&\mbb{C}(t)=t(\mbf{0})=\mbf{t}=t^i\hat{\mbf{e}}_i.\end{split}
\end{split}
\end{equation}
The action of $T^n_{\mbb{C}^n}$ on $\mbb{C}^n$ then maps to vector addition,
\begin{equation}
\begin{split}
&\forall\quad t\in T^n_{\mbb{C}^n},\quad\mbf{v}\in \mbb{C}^n\\
&t(\mbf{v}) = \mbb{C}(t)+\mbf{v}=\mbf{t}+\mbf{v}.
\end{split}
\end{equation}

In addition to invariance under translation, vector space $\mbb{C}^n$ is also invariant under the action of $\GL{n}{C}$,
\begin{equation}
\begin{split}
\begin{split}
g&:\mbb{C}^n\longrightarrow\mbb{C}^n\\
g&(\mbb{C}^n)=\mbb{C}^n
\end{split}
\qquad\forall\quad g\in\GL{n}{C}.
\end{split}
\end{equation}
If vector space $\mbb{C}^n$ is represented by the $n$-element complex column vectors, then the fundamental representation of $\GL{n}{C}$ is the space of nonsingular $n\times n$ complex matrices, acting on $\mbb{C}^n$ by matrix multiplication from the left:
\begin{align}
\begin{split}
&\{\mbf{v},\mbf{w}\}\in\mbb{C}^n\qquad\mbf{g}\in\GL{n}{C}\\
&\mbf{v}=v^i\hat{\mbf{e}}_i\qquad\quad~~\mbf{g}=g^i_{\p{i}j}\hat{\mbf{e}}_i\hat{\mbf{e}}^j\\
&\mbf{w}=\mbf{g}(\mbf{v}):=\mbf{g}\mbf{v}=g^i_{\p{i}j}v^j\hat{\mbf{e}}_i\\
\Rightarrow ~&\mbf{w}=w^i\hat{\mbf{e}}_i,\quad w^i=g^i_{\p{i}j}v^j.
\end{split}
\end{align}

These symmetries may also be identified on $\mbb{C}^{\wedge \N}$ and their representations constructed explicitly. For the translation group, elements are once again associated with the vector onto which they map the origin. That is, if the mapping from translation group elements to vectors on $\mbb{C}^{\wedge \N}$ is denoted $\mbb{C}^\wedge$, then
\begin{equation}
\begin{split}
\forall~t\in T^n_{\mbb{C}^{\wn}}:\qquad \begin{split}&t(\mbf{0}) = \mbf{t}\qquad \{\mbf{0},\mbf{t}\}\in\mbb{C}^{\wn}\\
&\mbb{C}^\wedge(t)=t(\mbf{0})=\mbf{t}=t^i\zeta_i.\end{split}
\end{split}
\end{equation}
The action of the translation group then corresponds to addition on the supernumbers,
\begin{equation}
\begin{split}
&\forall\quad t\in T^n_{\mbb{C}^\wn},\quad\mbf{v}\in \mbb{C}^\wn\\
&t(\mbf{v}) = \mbb{C}^\wedge(t)+\mbf{v}=\mbf{t}+\mbf{v}.
\end{split}
\end{equation}
To describe the action of $\GL{n}{C}$ on $\mbb{C}^{\wedge \N}$, define 
\begin{equation}
\mbf{X}:=\hat{\mbf{e}}_i\zeta^i\qquad\mbf{X}^{-1}=\zeta_i\hat{\mbf{e}}^i
\end{equation}
where $\mbf{X}$ is a mapping from a vector $\mbf{v}_\zeta\in\mbb{C}^{\wedge \N}$ %
to a vector $\mbf{v}_\mbf{e}\in\mbb{C}^n$, %
\begin{equation}
\begin{split}
\mbf{v}_\zeta&=v^i\zeta_i\\
\mbf{v}_\mbf{e}&=\mbf{X}(\mbf{v}_\zeta)=\hat{\mbf{e}}_i\zeta^i (v^j\zeta_j)= v^i\hat{\mbf{e}}_i.
\end{split}
\end{equation}
This allows the action of $\GL{n}{C}$ on the anticommuting vector space to be written
\begin{align}
\begin{split}
&\{\mbf{v},\mbf{w}\}\in\mbb{C}^{\wedge \N}\qquad\mbf{g}\in\GL{n}{C}\\
&\mbf{v}=v^i\zeta_i\qquad\quad~~~~\mbf{g}=g^i_{\p{i}j}\hat{\mbf{e}}_i\hat{\mbf{e}}^j\\
&\mbf{w}=\mbf{g}(\mbf{v}):=(\mbf{X}^{-1}\mbf{gX})\mbf{v}=g^i_{\p{i}j}v^j\zeta_i\\
\Rightarrow~&w^i=g^i_{\p{i}j}v^j.
\end{split}\label{eq:GLnRaction}
\end{align}
The set of elements $\{\mbf{X}^{-1}\mbf{g}\mbf{X}\}$ also comprise a fundamental representation of $\GL{n}{C}$. %

Vectors in $\mbb{C}^\wn$ have so far been represented as tier-1 supernumbers. 
It is now convenient to introduce a column vector notation for vectors in $\mbb{C}^\wn$, while preserving the anticommuting property of the supernumber representation. Given a vector $\mbf{v}\in\mbb{C}^\wn$, $\mbf{v}=v^i\zeta_i$, define
\begin{equation}
\begin{split}
\vec{\mbf{v}} &:= \sum_i \zeta_i\hat{\mbf{e}}_i\zeta^i\mbf{v}.
\end{split}
\end{equation}
Object $\vec{\mbf{v}}$ is an anticommuting column vector, with the entry in row $i$ being $v^i\zeta_i$ (no sum over $i$). Its components are written $\vec{\mbf{v}}^i$. If $\mbf{g}$ is an element of the fundamental $n\times n$ complex matrix representation of $\GL{n}{C}$, its action on the anticommuting column vector representation of $\mbb{C}^\wn$ is given by
\begin{align}
\begin{split}
&\{\vec{\mbf{v}},\vec{\mbf{w}}\}\in\mbb{C}^\wn\qquad\mbf{g}\in\GL{n}{C}\qquad
\mbf{g}=g^i_{\p{i}j}\hat{\mbf{e}}_i\hat{\mbf{e}}^j\\
&\vec{\mbf{g}}^i_{\p{i}j}:= g^i_{\p{i}j}\zeta_i\zeta^j\quad\textrm{(no sums over $i$ or $j$)}
\\
&\vec{\mbf{w}}=\mbf{g}(\vec{\mbf{v}}):=\vec{\mbf{g}}\vec{\mbf{v}}\qquad
\vec{\mbf{w}}^i=\vec{\mbf{g}}^i_{\p{i}j}\vec{\mbf{v}}^j.~~\end{split}\label{eq:indexprod}
\end{align}
The objects $\{\vec{\mbf{g}}^i_{\p{i}j}\}$ consequently also form a fundamental matrix representation of $\GL{n}{C}$, which acts on the anticommuting column vector representation of $\mbb{C}^\wn$ by matrix multiplication.
Similarly, the action of $T^n_{\mbb{C}^\wn}$ on this representation of $\mbb{C}^\wn$ corresponds to addition of supernumber-valued vectors.

An advantage of using an index-based notation is that it can be used to eliminate dependency on left-to-right ordering. Where indices are paired, e.g.~$j$ in the final line of \Eref{eq:indexprod}, this indicates that the vector and covector (1-form) basis elements associated with these indices are contracted pairwise to yield a factor of~1 as per \Eref{eq:zetaaction}. Basis elements associated with unpaired indices are not contracted. %
The use of indices to denote pairwise contractions permits factors in tensor products %
to be freely reordered %
without ambiguity, and evaluation can always be achieved by restoring an ordering of the basis elements such that the forms all act to the right on the appropriate vectors, or are acted on through the use of $i$-notation~\eref{eq:inotation} if preferred. In supernumber notation, residual (uncontracted) vector basis elements can be collected to the left of residual forms.

Now specialise to $n=2$. Let $\vec{\bm{\theta}}$ be a supernumber-valued column vector representation of a vector in $\mbb{C}^\wn$, and for consistency with implicit spinor index contraction~\eref{eq:indexsuppression} let its components $\vec{\bm{\theta}}^{\,i}$ be rewritten
\begin{equation}
\vec{\bm{\theta}}^{\,i}\rightarrow\vec{\theta}_\alpha\qquad\alpha=i,
\end{equation}
where the vector is unbolded and the index is written low and greek. 
Similarly, let the numeric components $\theta^i$ in $\vec{\bm{\theta}}^{\,i}$ be rewritten
\begin{equation}
\begin{split}
&\vec{\bm{\theta}}^{\,i}=\theta^i\zeta_i\quad\longrightarrow\quad\vec{\theta}_\alpha=\theta_\alpha\zeta_\alpha\qquad\alpha=i\\
&\textrm{(no sum over $i$)}.
\end{split}
\end{equation}
The arrow over $\vec{\theta}_\alpha$ is retained for now, to distinguish between the supernumber-valued vector $(\vec{\theta}_\alpha)$ and its complex-valued numeric components $({\theta}_\alpha)$. Note that the positions of the indices on the anticommuting basis elements remain unchanged. 
The elements of $\GL{2}{C}$ are rewritten similarly, with the upper index being lowered and the lower index being raised,
\begin{equation}
\begin{split}
&\vec{\mbf{g}}^i_{\p{i}j}\longrightarrow \vec{g}_\alpha^{\p{\alpha}\beta}=g_\alpha^{\p{\alpha}\beta}\zeta_\alpha\zeta^\beta,\\
&g_\alpha^{\p{\alpha}\beta}=g^i_{\p{i}j}\qquad\alpha=i,~\beta=j,%
\end{split}
\end{equation}
and the resulting representation of $\GL{2}{C}$ acts by matrix multiplication as expected,
\begin{equation}
[\vec{g}(\vec{\theta})]_\alpha = \vec{g}_\alpha^{\p{\alpha}\beta}\vec{\theta}_\beta.
\end{equation}

Next, %
decompose $\GL{2}{C}$ as
\begin{equation}
\GL{2}{C}\cong\SL{2}{C}\oplus\mbb{R}^+\oplus\U{1}.\label{eq:I:GL2Cdecomp}
\end{equation}
For the fundamental matrix representation of $\GL{2}{C}$, the $\U{1}$ subgroup corresponds to the phase of the matrix determinant and $\mbb{R}^+$ to its magnitude. Collectively, $\U{1}\oplus\mbb{R}^+$ is the space of nonzero complex numbers. %
It is straightforward to construct the associated contragradient, conjugate, and contragradient conjugate representations of $\SL{2}{C}$ as per \rcite{buchbinder1998}, and this is outlined below. Extension to $\GL{2}{C}$ then follows immediately as shown in \sref{sec:symCwtRm}. %

As per \rcite{buchbinder1998}, the contragradient representation is constructed by introducing the number-valued totally antisymmetric tensor $\varepsilon^{\alpha\beta}$ and its inverse $\varepsilon_{\alpha\beta}$, $\varepsilon^{12}=\varepsilon_{21}=1$, both of which are invariant under the action of $\SL{2}{C}$. %
For $\mbb{C}^{\wt}$ these tensors become %
\begin{equation}
\begin{split}
&\vec{\varepsilon}^{\,\alpha\beta}=\varepsilon^{\alpha\beta}\zeta^\alpha\zeta^\beta\\
&\vec{\varepsilon}_{\alpha\beta}=\varepsilon_{\alpha\beta}\zeta_\alpha\zeta_\beta. %
\end{split}
\end{equation}
The index on $\vec{\theta}_\alpha$ may then be raised through
\begin{equation}
\vec{\theta}^{\,\alpha}:=\vec{\varepsilon}^{\,\alpha\beta}\vec{\theta}_\beta.
\end{equation}
If elements of the complex-valued fundamental matrix representation $\SL{2}{C}\subset\GL{2}{C}$ are denoted $N_\alpha^{\p{\alpha}\beta}$ as in \rcite{buchbinder1998}, then elements of the supernumber-valued matrix representation of $\SL{2}{C}$ are given by
\begin{equation}
\begin{split}
\vec{N}_\alpha^{\p{\alpha}\beta}&:={N}_\alpha^{\p{\alpha}\beta}\zeta_\alpha\zeta^\beta.
\end{split}
\end{equation}
By unimodularity, $(N^{-1})_\alpha^{\p{\alpha}\beta}=(N^T)_\alpha^{\p{\alpha}\beta}$ and thus equivalent representations of $\SL{2}{C}$
act on $\vec{\theta}_\alpha$ from the left and on $\vec{\theta}^{\,\alpha}$ from the right,
\begin{align}
\vec{\theta}^{\,\prime}_\alpha&=\vec{N}_\alpha^{\p{\alpha}\beta}\vec{\theta}_\beta\label{eq:actN1}\\
\vec{\theta}^{\,\prime\alpha}&=\vec{\theta}^{\,\beta}(\vec{N}^{-1})_\beta^{\p{\beta}\alpha},
\end{align} 
where $\vec{N}^{-1}$ is defined by
\begin{equation}
\vec{N}_\alpha^{\p{\alpha}\beta}(\vec{N}^{-1})_\beta^{\p{\beta}\gamma}=\delta_\alpha^\gamma\,\zeta_\alpha\zeta^\gamma.
\end{equation}
To obtain the conjugate and contragradient conjugate representations, first identify
\begin{align}
\begin{split}
\vec{\bar\theta}&:=(\vec{\theta})^\dagger\\
\vec{\bar\theta}_{\dot\alpha}&=(\vec{\theta}_\alpha)^\dagger\quad |\quad\dot\alpha=\alpha.
\end{split}
\end{align}
The conjugation operator $\dagger$ maps vectors into their duals, e.g.~row vectors into column vectors, and thus it also interchanges elements of $\Lambda^{\mbb{C}}_2$ and $\Lambda^{\mbb{C}*}_2$. Consequently $\vec{\bar\theta}_{\dot\alpha}$ is an element of a row vector which takes a value in $\Lambda^{\mbb{C}*}_2$, %
\begin{align}
\begin{split}
\vec{\theta}_\alpha&=\theta_\alpha\zeta_\alpha\quad\Rightarrow\quad
\vec{\bar{\theta}}_{\dot\alpha}=(\theta_\alpha)^*\zeta^{\dot\alpha}\quad|\quad\alpha=\dot\alpha.
\end{split}
\end{align}
Next, introduce
\begin{equation}
\begin{split}
&\vec{\varepsilon}^{\,\dot\alpha\dot\beta}=\varepsilon^{\dot\alpha\dot\beta}\zeta_{\dot\alpha}\zeta_{\dot\beta}\quad\textrm{(no sums over $\dot\alpha$,~$\dot\beta$)} %
\end{split}
\end{equation}
and write
\begin{equation}
\vec{\bar{\theta}}^{\,\dot\alpha}:=\vec{\varepsilon}^{\,\dot\alpha\dot\beta}\vec{\bar{\theta}}_{\dot\beta}.
\end{equation}
The corresponding representations of $\SL{2}{C}$ are obtained by taking the conjugate of $\vec{N}$,
\begin{equation}
\vec{N}^{\dagger\dot\alpha}_{\p{*\dot\alpha}\dot\beta}:=(\vec{N}_\beta^{\p{\beta}\alpha})^\dagger\quad|\quad\alpha=\dot\alpha,~\beta=\dot\beta.
\end{equation}
These act as
\begin{align}
\vec{\theta}^{\,\prime}_{\dot\alpha}&=\vec{\theta}_{\dot\beta}\vec{N}^{\dagger\dot\beta}_{\p{\dagger\dot\beta}\dot\alpha}\\
\vec{\theta}^{\,\prime\dot\alpha}&=(\vec{N}^{\dagger-1})^{\dot\alpha}_{\p{\dot\alpha}\dot\beta}\vec{\theta}^{\,\dot\beta}.
\end{align} 

A further simplification of notation may now be obtained by recognising that on objects bearing arrows:
\begin{enumerate}
\item A lowered undotted greek index e.g.~$_\alpha$ is always associated with the corresponding generator on $\Lambda^{\mbb{C}}_2$, $\zeta_\alpha$.
\item A raised undotted greek index e.g.~$^\alpha$ is always associated with the corresponding generator on $\Lambda^{\mbb{C}*}_2$, $\zeta^\alpha$.
\item A lowered dotted greek index e.g.~$_{\dot\alpha}$ is always associated with the corresponding generator on $\Lambda^{\mbb{C}*}_2$, $\zeta^{\dot\alpha}$.
\item A raised dotted greek index e.g.~$^{\dot\alpha}$ is always associated with the corresponding generator on $\Lambda^{\mbb{C}}_2$, $\zeta_{\dot\alpha}$.
\end{enumerate}
This allows immediate reconstruction of objects such as $\vec{\varepsilon}^{\,\alpha\beta}$ and $\vec{\varepsilon}^{\,\dot\alpha\dot\beta}$ from their components ($\varepsilon^{\alpha\beta}$ and $\varepsilon^{\dot\alpha\dot\beta}$ respectively). It is therefore convenient to drop the arrows, 
\begin{equation}
\textrm{e.g.}\quad\vec{\varepsilon}^{\,\alpha\beta}\rightarrow {\varepsilon}^{\alpha\beta}, 
\end{equation}
and recognise that in the context of a model on $\mbb{C}^{\wt}$ an object carrying greek indices also carries the associated generators of $\Lambda^{\mbb{C}}_2$ and $\Lambda^{\mbb{C}*}_2$.

\subsection{Anticommuting and commuting manifolds\label{sec:anticommmanif}}

\subsubsection{Anticommuting manifold \protect{$\Cwt$}}

The anticommuting objects $\theta_\alpha$, $\theta^\alpha$, $\bar\theta_{\dot\alpha}$, and $\bar\theta^{\dot\alpha}$ are all acted on by representations of $\SL{2}{C}$, and may thus be thought of as anticommuting two-component Weyl spinors. All of these sets of objects are interconvertible through combinations of index raising, lowering, and Hermitian conjugation, and thus to obtain a set of co-ordinates covering all $\mbb{C}^{\wt}$ it suffices to take $\{\theta_\alpha\}$. The tensor $\varepsilon^{\alpha\beta}$ then acts as a metric, allowing definition of the (complex) interval
\begin{equation}
\rmd s^2 = \varepsilon^{\alpha\beta}\rmd\theta_\beta\rmd\theta_\alpha,
\end{equation}
and together the vector space $\mbb{C}^\wt$ and metric $\varepsilon^{\alpha\beta}$ define a manifold, also denoted $\mbb{C}^\wt$.

\subsubsection{Mapping to \protect{$\mbb{R}^{1,3}$}\label{sec:I:mapping}}

Now introduce the sigma matrices, $\sigma_{\mu\alpha\dot\alpha}$. On vector space $\mbb{C}^\wt$ these implicitly incorporate elements of the anticommuting bases as described above, for example
\begin{equation}
\sigma_0=\left(\begin{array}{cc}1\cdot\zeta_1\zeta^1&0\cdot\zeta_1\zeta^2\\0\cdot\zeta_2\zeta^1&1\cdot\zeta_2\zeta^2\end{array}\right)=\left(\begin{array}{cc}\zeta_1\zeta^1&0\\0&\zeta_2\zeta^2\end{array}\right).
\end{equation}
Recall that $\SL{2}{C}$ is a double cover of $\SOp{1,3}$, the proper orthochronous Lorentz group, so any element $N\in\SL{2}{C}$ may be mapped to an element $\Lambda(N)\in\SOp{1,3}$. It may then be observed that $\sigma_\mu$ is invariant under the action of $N$
\begin{equation}
\sigma_\mu\rightarrow N\sigma_\nu N^\dagger[\Lambda(N)^{-1}]^{\nu}_{\p{\nu}\mu}.\label{eq:sigmainvar}
\end{equation}

The sigma matrices may be used to construct a mapping from co-ordinates on vector space $\mbb{C}^{\wt}$ to co-ordinates on a vector space isomorphic to $\mbb{R}^4$. The sigma matrices encode the Minkowski metric \cite[pp.~11-12]{buchbinder1998}, and thus it is expected that such a mapping will be from manifold $\mbb{C}^\wt$ to $\RM$,
\begin{equation}
\F:\Cwt\longrightarrow\RM.\label{eq:Fmanifoldmap}
\end{equation}
As an initial candidate for $\F$, consider the mapping
\begin{equation}
\begin{split}
\F&(\theta)=-\frac{1}{2}\bar\theta\bsm\theta=:x^\mu.\label{eq:defxmuprovisional}
\end{split}
\end{equation}
where the factor of $-\frac{1}{2}$ is introduced for later convenience.
Using \Eref{eq:sigmainvar}, an $\SL{2}{C}$ transformation on $\theta$ induces the associated $\SOp{1,3}$ transformation on $x^\mu$,
\begin{equation}
\theta_\alpha\rightarrow (N\theta)_\alpha\quad\Rightarrow\quad x^\mu\rightarrow x^\nu[\Lambda(N)]_\nu^{\p{\nu}\mu}.
\end{equation}
Mapping $\F$ therefore satisfies \Eref{eq:Fmanifoldmap}. Similarly, a translation on $\Cwt$ is readily seen to induce a translation (possibly null) on $\RM$.
Less conveniently, however, mapping $\F$ is insensitive to transformations which multiply $\theta$ by a phase. To capture this phase, define the periodic co-ordinate
\begin{equation}
\alpha=\frac{\theta\theta}{|\theta\theta|},\label{eq:alpha}
\end{equation}
and map $\alpha$ to a linear co-ordinate $\tau$ using
\begin{equation}
\tau=\pm\ln{\left[\pm\frac{\rmi}{\pi}\ln{(\alpha)}\right]}\label{eq:tmapping}
\end{equation}
where $\rmi\ln{(\alpha)}$ is chosen to lie in the range $(-\pi,\pi]$. As $\alpha$ ranges from $\exp{(-\rmi\pi)}$ to $\exp{(\rmi\pi)}$, $\tau$ ranges from 0 to $-\infty$ then from $+\infty$ to 0.
This degree of freedom is isomorphic to $\mbb{R}$, and is projected out by mapping $\F$, so 4-vector $x^\mu$ retains at most three real degrees of freedom. Also note that the mapping from $\theta$ to $\tau$ is a double cover, as $\alpha$ is invariant under $\theta\rightarrow -\theta$.

On $\Cwt$, considering co-ordinate basis $\theta$ as a four-dimensional real vector space, the vectors
\begin{equation}
\Weyl{1}{0}\qquad\frac{1}{\sqrt{2}}\Weyl{1}{1}\qquad\frac{1}{\sqrt{2}}\Weyl{1}{\rmi}\label{eq:Cwtvecs}
\end{equation}
over a real field are linearly independent and can construct any vector in $\Cwt$ up to an overall phase. Under mapping $\F$, these vectors map to
\begin{equation}
-\frac{1}{2}\left(\begin{array}{c}1\\0\\0\\1\end{array}\right)\qquad-\frac{1}{2}\left(\begin{array}{c}1\\1\\0\\0\end{array}\right)
\qquad-\frac{1}{2}\left(\begin{array}{c}1\\0\\1\\0\end{array}\right)\label{eq:RMvecs}
\end{equation}
forming a linearly independent co-ordinate basis for the past light cone in the co-ordinate sector defined by
\begin{equation}
\begin{split}
x^1&\leq0\qquad x^2\leq0\qquad x^3\leq0, 
\end{split}
\end{equation}
henceforth the ``negative-$\mbf{x}$ sector''.

Eight such co-ordinate patches, related by 90$^\circ$ rotations, suffice to ensure that all of $\RM$ lies within the negative-$\mbf{x}$ sector of some co-ordinate patch. As the 90-degree rotations are contained within $\SOp{1,3}$, and thus within $\SL{2}{C}$, corresponding rotations may also be identified on $\Cwt$. %
Extend the definition of mapping $\F$ using these eight co-ordinate patches, such that at all points it takes form~\eref{eq:defxmuprovisional} in the co-ordinate patch in which $x^1$, $x^2$, and $x^3$ are all negative. Mapping $\F$ is now a many-to-one mapping from $\Cwt$ to the past light cone of the origin of $\RM$.

To escape the restriction to the past light cone, first adopt a preferred rest frame. Restrict attention to the submanifold $\mbb{R}^{1^-,3}$ defined by $x^0\leq0$ in that frame, and replace the previous definition of $\F$ in the negative-$\mbf{x}$ sector with
\begin{align}
\begin{split}
\F(\theta)&=-\frac{1}{2}(\bar\theta\bsm\theta+T^\mu) =:x^\mu\\
T^\mu&:=\delta^{\mu0}\left[\sqrt{(\bar\theta\bar\sigma^0\theta)^2+4s^2}-\bar\theta\bar\sigma^0\theta\right]
\\
s^2&:=\frac{1}{4}\left[\tau^2-(\bar\theta\bar\sigma^0\theta)^2\right], 
\end{split}\label{eq:defxmulong}
\end{align}
where %
where $\tau$ is defined in \Eref{eq:tmapping}.
Co-ordinate rotations by 90$^\circ$ extend this mapping to the entirety of $\mbb{R}^{1^-,3}$.
The value of $\tau$ goes to zero on the surface $x^0=0$, while the value of $s$ corresponds to the interval between a point and the origin of $\RM$.
The entirety of $\mbb{R}^{1^-,3}$ is then covered by the $\mbb{R}^3$ parameterisation of the past light cone, plus half of the range of parameter $\tau$ (corresponding to $\tau>0$). Although it is useful to make $s^2$ explicit as per \Eref{eq:defxmulong}, $x^\mu$ in the negative-$\mbf{x}$ sector may also be written more concisely as
\begin{equation}
x^\mu:=-\frac{1}{2}[\bar\theta\bsm\theta+\delta^{\mu0}(\tau-\bar\theta\bar\sigma^0\theta)].\label{eq:defxmu}
\end{equation}

Next, reintroduce the other half of $\RM$. This is a time-reflected copy of $\mbb{R}^{1^-,3}$, denoted $\mbb{R}^{1^+,3}$, and defined by $x^0\geq0$. When $\tau<0$, map a co-ordinate $\theta$ to $\mbb{R}^{1^+,3}$ instead of $\mbb{R}^{1^-,3}$ by first performing mapping $\F$ then taking $x^0\rightarrow -x^0$.

Let \Eref{eq:defxmu} in the negative-$\mbf{x}$ sector, its 90$^\circ$ rotations within $\mbb{R}^{1^-,3}$, and their time reflections onto $\mbb{R}^{1^+,3}$ %
collectively comprise the full definition of mapping ${\F}$, which is onto by construction. 
The coefficients of the vectors of \Erefs{eq:Cwtvecs}{eq:RMvecs} are in 1:1 correspondence in each octant, and as observed below \Eref{eq:tmapping}, co-ordinates $\theta$ have a 2:1 mapping onto $\alpha$. Thus ${\F}:\Cwt\rightarrow\RM$ is a double cover.

It is also clear that the form of mapping $\F$ given in \Eref{eq:defxmu} is not invariant under boost. However, having initially adopted a preferred inertial frame in which to write mapping $\F$ in a given sector, it may readily be reexpressed in a different rest frame by recognising that under $\theta\rightarrow N\theta$, $T^\mu$ in \Eref{eq:defxmulong} transforms as a 4-vector,
\begin{equation}
T^\mu\rightarrow T^\nu\Lambda_\nu^{\p{\nu}\mu}.
\end{equation}

\subsubsection{Symmetries of \protect{$\Cwt$} and \protect{$\RM$}\label{sec:symCwtRm}}

So far, only the mapping of $\SL{2}{C}$ from $\Cwt$ to $\RM$ has been explicitly addressed. Extension to $\GL{2}{C}$ is relatively straightforward. 
Invariance of $\Cwt$ under proper rescaling 
\begin{equation}
\theta\longrightarrow f\theta\qquad f\in\mbb{R}^+
\end{equation}
maps to invariance of $\RM$ under proper rescaling
\begin{equation}
\begin{split}
&x^\mu\longrightarrow f^2x^\mu\qquad f^2\in\mbb{R}^+,
\end{split}
\end{equation}
while invariance of $\Cwt$ under the action of the $\U{1}$ subgroup maps to invariance of $\RM$ under an unusual transformation which cycles surfaces of equal interval. %
On transitioning from positive to negative $s^2$ these surfaces appear to transition from a connected manifold to a disjoint bipartite manifold, but this is an artefact of the co-ordinate system introduced in \Eref{eq:tmapping} and both groups of surfaces are closed and connected %
on the 1,3-generalisation of the Riemann sphere.

Elements of the translation group on $\Cwt$ are identified with position vectors on $\Cwt$. Under mapping $\F$ these map directly to position vectors on $\RM$, and thence to translations on $\RM$. Since $\F$ is 2:1, $T^2_\Cwt$ is a double cover of $T^4_\RM$.

Drawing all of this together, the symmetries of $\Cwt$ and $\RM$ may be written
\begin{align}
\Cwt&:\quad\SL{2}{C}\oplus T^2_\Cwt\oplus\mbb{R}^+\oplus\U{1}\label{eq:globalsym1}\\
\RM&:\quad\SOp{1,3}\oplus T^4_\RM\oplus\mbb{R}^+\oplus\mbb{R}\label{eq:globalsym2}
\end{align}
where $\SL{2}{C}\oplus T^2_\Cwt$ on $\Cwt$ is the universal cover of the Poincar\'e group $\SOp{1,3}\oplus T^4_\RM$ on $\RM$. Invariance of $\Cwt$ under proper rescalings maps to invariance of $\RM$ under proper rescalings, and invariance of $\Cwt$ under multiplication of co-ordinates by a phase maps to invariance of $\RM$ under the %
transformation relating surfaces of equal interval%
, parameterised by %
$s^2$ in \Eref{eq:defxmulong} where $s^2$ ranges over all of $\mbb{R}$.

\subsubsection{Representations of \protect{$\Cwn$} vector spaces\label{sec:repCwt}}

The space of anticommuting vectors $\Cwn$ is defined as $\Span{\zeta_i\in\Lambda^\mbb{C}_n}{\mbb{C}}\subset\Lambda^\mbb{C}_n\subset\Lambda_\infty$, where $\Span{\ldots\!}{F}$ specifies that the span is over a field $F$. This locates vector space $\Cwn$ as a subspace of the supernumbers. 
Further, on introducing an outer product operation, the space of all $k$-vectors is also a subspace of $\Lambda^\mbb{C}_n$: 
For $k_0$ a specific value of $k$, the space of $k_0$-vectors corresponds to the space of tier-$k_0$ supernumbers in $\Lambda^\mbb{C}_n$.

While vector space $\Cwn$ can be realised as a subspace of $\Lambda_\infty$, the space of supernumbers, manifold $\Cwn$ may be recognised as a subspace of $\mbb{R}^{4|4}$. In this context the real co-ordinate is restricted to a single point on the Minkowski submanifold e.g.~$\mbb{R}^{4|4}|_{x=(0,0,0,0)}$, and the %
$\Cwn$ manifold %
is a subspace of the supersymmetric extension of that point.

It may also be illuminating to provide a less abstract construction for $\Cwn$. %
To this end, consider an $\mbb{C}^n$ manifold whose co-ordinates are complex vectors, let $T\mbb{C}^n$ be the space of vectors tangent to $\mbb{C}^n$, and let $T^*\mbb{C}^n$ be the space of covectors, or 1-forms. The basis of $T^*\mbb{C}^n$ may be equipped with an anticommuting (wedge) product to allow the construction of higher-order differential forms, and the algebra (over a real field) of all differential forms over $\mbb{C}^n$ is then a subalgebra of $\Lambda^\mbb{C}_n$. The 1-forms $\rmd x^i$ anticommute under the action of the wedge product. %
Noting that in this model it is never necessary to add $k$-vectors with different values of $k$, or $p$-forms with different values of $p$, it follows that an equivalency may be identified,
\begin{align}
\Cwn
\leftarrow\,&
\bm\{1,
\{\zeta_i\in\Lambda_\infty|i\in\mbb{Z}_n\};\cdot\,;+;\mbb{C}\bm\}\\
&%
\equiv\bm\{1,
\{\delta_{ij}\rmd x^j|\rmd x^j\in T^*\mbb{C}^n,i\in\mbb{Z}_n\};\wedge;+;\mbb{C}\bm\}.\nn
\end{align}
That is, $\Cwn$ is the space of 1-forms cotangent to $\mbb{C}^n$, viewed as a vector space.

\subsubsection{Pullback from \protect{$M\subset\Cwt$} to \protect{$\RM$}\label{sec:defpullback}}

It now useful to define a restriction of mapping ${\F}$ to a submanifold $M\subset\Cwt$, where $M$ is chosen such that ${\F}:M\rightarrow\RM$ is both 1:1 and onto. Use $\G$ to denote the restriction of ${\F}$ to manifold $M$. %
By construction, $\G$ is an isomorphism. %
It therefore has an associated pullback
\begin{equation}
\begin{split}
\G^{-1*}[f_{\dot\alpha}^{\p{\dot\alpha}\alpha}(\theta,\bar\theta)\,\rmd\theta_{\alpha}\rmd\bar\theta^{\dot\alpha}] &= f_\mu(x)\,\rmd x^\mu\\
f_\mu(x)&:=f^{\dot\alpha\alpha}(\theta,\bar\theta)\sigma_{\mu\alpha\dot\alpha}
\label{eq:pullGinv} %
\end{split}
\end{equation}
where $\rmd x^\mu$ are the 1-forms corresponding to infinitesimal translations on $\RM$, and satisfy
\begin{align}
\label{eq:R13-1form}\rmd x^\mu&=-\frac{1}{2}[\rmd\bar\theta\bsm\rmd\theta+\delta^{\mu 0}\rmd(\tau-\bar\theta\bar\sigma^0\theta)]\\
&=\rmd x^\mu_\mrm{I}+\rmd x^\mu_{\theta\theta}\nn\\
\rmd x^\mu_\mrm{I}&:=-\frac{1}{2}\rmd\bar\theta\bsm\rmd\theta~~\quad\rmd x^\mu_{\theta\theta}:=-\frac{\delta^{\mu 0}}{2}\rmd(\tau-\bar\theta\bar\sigma^0\theta).
\end{align}
On performing an infinitesimal translation on $\Cwt$, 
\begin{equation}
\theta\rightarrow\theta+\rmd\theta, 
\end{equation}
the term $\rmd x^\mu_\mrm{I}$ is a shift in co-ordinates $x^\mu$ along a surface of constant interval with respect to the origin. This shift also induces a change in $\tau$, since $\tau$ is not in general constant on these surfaces. The remainder of \Eref{eq:R13-1form}, $\rmd x^\mu_{\theta\theta}$ then further breaks down into two parts. The first is the overall change in $\tau$, and the second offsets the portion of the change in $\tau$ which is attributable to translation along a surface of constant interval. Combining these terms, the whole of $\rmd x^\mu_{\theta\theta}$ is thus purely a function of co-ordinate $\alpha$, the phase of $\theta\theta$.

\subsection{Real scalar fields}

\subsubsection{Real scalar fields on $\Cwt$\label{sec:quantumdust}} %

Now let there exist a set of $N$ real unitless classical scalar fields on $\Cwt$, each of which may be written as the action of a 1-form on the co-ordinate vector, i.e.{}
\begin{equation}
\varphi_q=\bm{\vp}_q(\theta)\quad|\quad q\in\{1,\ldots,N\}. \label{eq:scalarfields}
\end{equation}
Although $\bm{\vp}_q$ is a complex 1-form, by construction the field $\vp_q$ is real. %
In terms of numerical components, %
$\vp_q$ is therefore conventionally written as a function of both $\theta^\alpha$ and $\bar\theta_{\dot\alpha}$, i.e.~$\vp_q\bthetas$. %

Exploiting the anticommuting nature of co-ordinates on $\Cwt$, Taylor expansion of any such field $\varphi_q\bthetas$ may be performed exactly to yield
\begin{align}
\varphi_q\bthetas&=c_q^{(0,0)}+\sum_{i=1}^{2}\left(\,\prod_{k=1}^i\theta^{\alpha_k}\!\right) [c_q^{(i,0)}]_{\alpha_1\ldots \alpha_i}\nn\\
&+\sum_{j=1}^{2}\left(\,\prod_{l=1}^j\bar\theta_{\dot\alpha_l}\!\right) [c_q^{(0,j)}]^{\dot\alpha_1\ldots\dot\alpha_j}\label{eq:superfield}
\\
&+\sum_{i=1}^{2}\sum_{j=1}^{2}\left(\,\prod_{k=1}^i\bar\theta_{\dot\alpha_k}\!\right)\left(\,\prod_{l=1}^j\theta^{\alpha_l}\!\right) [c_q^{(i,j)}]_{\alpha_1\ldots \alpha_j}^{\p{\alpha_1\ldots \alpha_j}\dot\alpha_1\ldots\dot\alpha_i}\nn
\end{align}
where coefficient $c_q^{(i,j)}$ carries $i$ holomorphic and $j$ antiholomorphic indices which are contracted with the co-ordinate spinors, %
e.g.{}
\begin{equation}
\ldots + \bar\theta_{\dot\alpha}\theta^{\alpha}\theta^{\beta} [c_q^{(2,1)}]^{\p{\alpha\beta}\dot\alpha}_{\alpha \beta}+\ldots\,.
\end{equation}
Counting the incidence of spinor co-ordinates $\bthetas$, it is easy to see that under mapping to $\RM$ each excitation may depend at most quadratically on $x^\mu$. 
The pullback of a form $\bm{\vp}_q$ %
to $\RM$, and hence the mapping of a field $\varphi_q\bthetas$ to $\RM$, is therefore not normalised on any spacelike submanifold. %

For a single real scalar field $\varphi_q$ the configuration space is seen from \Eref{eq:superfield} to be $\mbb{R}^9$. The configuration space of the entire system is then the tensor product of the configuration spaces of the $N$ real scalar fields, giving a configuration space isomorphic to $\mbb{R}^{9N}$. However, as an assumption of the model, let %
the fields themselves be %
physically distinct. Thus, although
a field configuration may be specified in any co-ordinate system on $\mbb{R}^{9N}$, there exists a privileged factorisation
\begin{equation}
\mbb{R}^{9N}\rightarrow (\mbb{R}^9)^{\otimes N}
\end{equation}
where each $\mbb{R}^9$ subspace corresponds to the parameter space of a distinct physical field. In other words, active linear mixing of the fields does \emph{not} yield an equivalent state on $\Cwt$.
One consequence of this assumption is that for a given field configuration, the quantity
\begin{equation}
\prod_{q=1}^N\vp_q(x)\label{eq:phiprodpreempt}
\end{equation}
is well-defined at each point on $\RM$. %

Regarding this physically favoured set of fields $\{\vp_q\}$, in \sref{sec:QL} %
the \emph{centre} of a field on $\RM$ is defined as the point at which its derivatives on $\RM$ vanish %
(provided this point is unique). It is reasonable but not obligatory to consider configurations in which all fields $\{\vp_q\}$ have centres, and the centres of the fields %
form a regular lattice on some submanifold of $\RM$, perhaps corresponding to the entirety of $\RM$. %
The grid of field centres is then %
conceptually analogous to pixels of the analogue model.
The existence of a favoured set of fields $\{\vp_q\}$ enables the number of field centres on a given submanifold of $\RM$ to be a well-defined quantity.

\subsubsection{Pullbacks of individual fields onto \protect{$\RM$}\label{sec:indivfields}}

As per \sref{sec:quantumdust}, the fundamental fields of the model are scalar fields on $\Cwt$%
, expressed as 1-forms acting on the co-ordinate vector. Manifold $\Cwt$ is timeless, and specifying the power series expansion of each field $\varphi_q$ fixes the values of the fields over all $\Cwt$. In addition to the fundamental, unitless scalar fields \{$\varphi_q\bthetas\}$ one may also describe gradient fields of the general form
\begin{equation}
\left(\prod_{i}\bar\partial_{\dot{\alpha}_i}\right)\left(\prod_{j}\partial_{\alpha_j}\right)\varphi_q\bthetas\label{eq:derivativefields}
\end{equation}
where there are between one and four Grassmann derivative operators present (made up of between zero and two chiral derivatives $\partial_{\alpha_i}$ and between zero and two conjugate chiral derivatives $\bar\partial_{\dot{\alpha}_i}$). These fields are not independent of the unitless scalar fields, being completely fixed by full knowledge of $\{\varphi_q\bthetas\}$.

On mapping from $\Cwt$ to $\RM$, specification of the fields $\{\varphi_q\bthetas\}$ likewise fully constrains the values of $\{\varphi_q(x)\}$ and all derivatives across all of $\RM$. However, when working on manifolds with a time axis it is more usual to introduce a Lagrangian and confine attention to a Cauchy surface $C$. 
To fully describe a set of fields across all of $\RM$, it then suffices to specify on $C$ both the family of fields under study and their derivatives. In conjunction with an appropriate Lagrangian on $\RM$, this once again fully constrains those fields across all of $\RM$. 
The construction of such a
set of %
fields, derivative operators, and Lagrangian %
begins here, and is completed in \sref{sec:quasi}. %

Recall that co-ordinates $x^\mu$ are defined through \Eref{eq:defxmu}. Requiring
\begin{equation}
\partial_\mu x^\nu = \delta_\mu^\nu\label{eq:reqdelta}
\end{equation}
implies that the elements of derivative operator $\partial_\mu$ on $\RM$ (enumerated by $\mu$)
may be identified with the expressions on $\Cwt$
\begin{equation}
\partial_\mu:%
=\bar\partial_{\dot\alpha}\bar\sigma^{\dot\alpha\alpha}
_{\mu}\partial_{\alpha}
.\label{eq:mudiv}
\end{equation}
This operator is not chiral, but acts independently on the holomorphic and antiholomorphic spinors in $x^\mu$. It is well-defined across all of $\Cwt$ and thus all of $\RM$, is real (hermitian), and satisfies \Eref{eq:reqdelta}. %
An infinitesimal translation on $\Cwt$ maps to an infinitesimal translation on $\RM$, and thus \Eref{eq:mudiv} 
is %
a satisfactory derivative operator on $\RM$. 
However, it is also possible to construct another two linearly independent derviative operators on $\RM$. These operators are chiral, being sensitive to changes in $\U{1}$ co-ordinate $\alpha$, they are scalar, and they form a conjugate pair. They are defined as
\begin{equation}
\partial_\mrm{U}:=\partial^\alpha\partial_\alpha, \qquad\bar{\partial}_\mrm{U}:=\bar{\partial}_{\dot\alpha}\bar{\partial}^{\dot\alpha}\label{eq:Udivs}
\end{equation}
and will be used in addition to $\partial_\mu$.

Next, note that repeated application of $\partial_\mu$ to $\varphi_q$ on $\RM$ yields the values of the nonchiral derivatives $\bar\partial_{\dot\alpha}\partial_\alpha\varphi_q$ and $\bar\partial_{\dot\alpha}\bar\partial_{\dot\beta}\partial_\alpha\partial_\beta\varphi_q$ on $\Cwt$ but provides no access to the chiral derivatives of $\varphi_q$. However, if new fields are defined on $\RM$ according to
\begin{align}
\left[\psi^\alpha_{q}(x)\right]_\RM&:=\left[\partial^\alpha\varphi_q(\bar\theta,\theta)\right]_\Cwt\label{eq:newfields1}\\
\left[\bar\psi_{q\dot\alpha}(x)\right]_\RM&:=\left[\bar\partial_{\dot\alpha}\varphi_q(\bar\theta,\theta)\right]_\Cwt\\
\left[\h^{\alpha\beta}_{q}(x)\right]_\RM&:=\left[\partial^\alpha\partial^\beta\varphi_q(\bar\theta,\theta)\right]_\Cwt\\
\left[\bar\h_{q\dot\alpha\dot\beta}(x)\right]_\RM&:=\left[\bar\partial_{\dot\alpha}\bar\partial_{\dot\beta}\varphi_q(\bar\theta,\theta)\right]_\Cwt\label{eq:newfields4}
\end{align}
and the notation is introduced
\begin{equation}
\varphi_{q\mu}:=-\rmi\partial_\mu\varphi_q,\label{eq:defvecphi}
\end{equation}
then the following fields on $\RM$ %
correspond to $\varphi_q$ and its full family of derivative fields~\eref{eq:derivativefields}:
\begin{align}
\begin{array}{lll}
\varphi_q\\
\varphi_{q\mu}\qquad&\partial_\mu\varphi_{q\nu}\\
\psi_q^{\alpha}\qquad&\partial_\mu\psi_q^{\alpha}\\
\bar\psi_{q\dot\alpha}\qquad&\partial_\mu\bar\psi_{q\dot\alpha}\\
\h_q^{\alpha\beta}\qquad&\bar\h_{q\dot\alpha\dot\beta}. %
\end{array}\label{eq:allderivs}
\end{align}

To construct a Lagrangian, note that on $\Cwt$ all derivatives of higher order than those shown in \Eref{eq:allderivs} must vanish. This follows from anticommutation of Grassmann derivative operators acting on any field $\Phi$
\begin{equation}
\partial^\alpha\partial^\beta \Phi=-\partial^\beta\partial^\alpha \Phi\label{eq:Phiacomm}
\end{equation}
where $\Phi$ is arbitrary, so may itself carry indices or be a derivative, e.g.~$\Phi\equiv\partial^\gamma\varphi_q$.
Sufficiently high-order derivatives necessarily contain repeated spinor indices on either or both of the holomorphic and antiholomorphic sectors ($\partial^\alpha$ and $\bar\partial_{\dot\alpha}$), and thus when $\Phi$ contains a spinor derivative $\partial^\gamma$, \Eref{eq:Phiacomm}
evaluates to zero.

In keeping with power series expansion~\eref{eq:superfield}, %
a field $\varphi_q$ therefore satisfies the constraints
\begin{align}
\partial_\mu\partial_\nu\partial_\rho\varphi_q&=0\label{eq:RGconstr1}\\
\partial_\mu\partial_\nu\partial^\alpha\varphi_q&=0\\
\partial_\mu\partial_\nu\bar\partial_{\dot\alpha}\varphi_q&=0\label{eq:RGconstr3}\\
\partial_\mu\partial^\alpha\partial^\beta\varphi_q&=0\label{eq:RGconstr4}\\
\partial_\mu\bar\partial_{\dot\alpha}\bar\partial_{\dot\beta}\varphi_q&=0\label{eq:RGconstr5}
\end{align}
for which the corresponding single-species constraints on $\RM$ are
\begin{align}
\rmi\partial_\mu\partial_\nu(\varphi_{q\rho})&=0\label{eq:RMconstr1}\\
\partial_\mu(\partial_\nu\psi^\alpha_{q})&=0\label{eq:RMconstr2}\\
\partial_\mu(\partial_\nu\bar\psi_{q\dot\alpha})&=0\label{eq:RMconstr3}\\
\partial_\mu(\h^{\alpha\beta}_{q})&=0\label{eq:RMconstr4}\\
\partial_\mu(\bar\h_{q\dot\alpha\dot\beta})&=0.\label{eq:RMconstr5}
\end{align}
These
contain as a subset the constraints
\begin{align}
(g^{\mu\sigma}g^{\nu\rho}-g^{\mu\nu}g^{\rho\sigma})&\partial_\mu\partial_\nu(\varphi_{q\rho})=\triangle^{\sigma\rho}\vp_{q\rho}=0
\label{eq:RMconstr1a}\\
\dslash^{\dot\alpha\alpha}(\partial_\nu\psi_{q\alpha})&=0\label{eq:RMconstr2a}\\
\dslash_{\dot\alpha\alpha}(\partial_\mu\bar\psi_{q}^{\dot\alpha})&=0\label{eq:RMconstr3a}\\
\square\h_{q}=0\qquad&\h_{q}=\varepsilon^{\beta\alpha}\h_{q\alpha\beta}\label{eq:RMconstr4a}\\
\square\h^*_{q}=0\qquad&\h^*_{q}=(\h_q)^*=\varepsilon_{\dot\beta\dot\alpha}\bar\h_q^{\dot\alpha\dot\beta},\label{eq:RMconstr5a}
\end{align}
which %
are readily seen to be the equations of motion arising from a Lagrangian of the form
\begin{align}
\begin{split}
\mscr{L}_q=%
&\p{+}\frac{K_{q}^{(1)}}{2}\varphi_{q\mu}\triangle^{\mu\nu}\varphi_{q\nu}\\
&+\rmi K_{q}^{(2)}(%
\partial_\mu\bar\psi_q)\,\dslash\,(%
\partial^\mu\psi_q)\\
&+K_{q}^{(3)}\h_q^*\square\h_q
\end{split}
\label{eq:Lq0}
\end{align}
where the $K_{q}^{(i)}$ are %
numerical constants. 

However, this family of constraints may be rewritten in a more useful form. %
Noting that the fields %
$\h^{\alpha\beta}_q$, and $\bar\h_{q\dot\alpha\dot\beta}$ are themselves defined in terms of derivative operators, that
\begin{equation}
\partial_\mu\partial_\nu\equiv-\frac{1}{2}\eta_{\mu\nu}\bar\partial\bar\partial\partial\partial\equiv-\frac{1}{2}\eta_{\mu\nu}\bar\partial_\rmU\partial_\rmU,\label{eq:makedU}
\end{equation}
and that $\partial_\rmU$ is linearly independent of $\partial_\mu$ and thus
\begin{equation}
\partial_\rmU\partial_\mu\vp_q=\partial_\mu\partial_\rmU\vp_q=\frac{1}{\vp_q}(\partial_\rmU\vp_q)(\partial_\mu\vp_q), 
\end{equation}
any pair of derivative operators $\partial_\mu\partial_\nu$ in term~1 of \Eref{eq:Lq0} may be replaced according to the rule
\begin{equation}
\partial_\mu\partial_\nu\rightarrow-\frac{\eta_{\mu\nu}}{2\vp_q^{\p{q}2}}\h^*_q\h_q\label{eq:HHrewrite}
\end{equation}
to yield alternative constraints on $\RM$ additional to those given in \Erefr{eq:RMconstr1a}{eq:RMconstr5a}. The extra terms in these constraints are implicit in the original antisymmetry constraint on $\Cwt$ \eref{eq:Phiacomm}, but are subsequently missing from the $\RM$ counterparts [in particular \Eref{eq:RGconstr1}, which gives rise to term~1 of \Eref{eq:Lq0}], as the pairs of nonchiral derivatives in \Erefr{eq:RGconstr1}{eq:RGconstr3} and \erefr{eq:RMconstr4a}{eq:RMconstr5a} are insensitive to the chiral co-ordinate $\alpha$. 

Further, recognise that fields $\psi^\alpha_q$ and $\bar\psi_{q\dot\alpha}$ are at most linear in $x^\mu$, and therefore either they are constant, or there exists a closed boundary $B$ on the Riemann sphere %
on which they go to zero. %
In either case %
the second term of Lagrangian~\eref{eq:Lq0} necessarily admits integration by parts, permitting the derivative operators $\partial^\mu$ and $\partial_\mu$ to be rearranged as $\bar\partial_\rmU$ acting on $\psi^\alpha_q$ and $\partial_\rmU$ acting on $\bar\psi_{q\dot\alpha}$. Once again, linear independence of $\bar\partial_\rmU$ and $\psi^\alpha_q$ (and likewise for conjugates $\partial_\rmU$ and $\bar\psi_{q\dot\alpha}$) permits these scalar derivatives $\partial_\rmU$ and $\bar\partial_\rmU$ also to be rewritten as factors of $\h_q$ and $\h_q^*$. Finally, in the third term of \Eref{eq:Lq0}, $\partial_\mu\h_q$ always vanishes by \Eref{eq:RGconstr4} so the dynamics of $\h_q$ on $\RM$ are trivial. In contrast, however, $\bar\partial_\rmU\h_q$ need not vanish. If it does, then $\h_q$ is totally trivial. If it does not, then it is at most linear in $\alpha$. The parameter space of $\alpha$ is already a 1-sphere so once again there exists a boundary on which $\h_q$ goes to zero. Integration by parts over the corresponding boundary on $\Cwt$ then permits $\square$ to be rewritten in terms of $\h_q\h_q^*$ in accordance with \Eref{eq:HHrewrite}. 

It turns out that this %
set of substitutions may be collectively written in a fairly simple form, %
\begin{align}
\partial_\mu&\rightarrow D_{q\mu}\\
D_{q\mu}&:=\partial_\mu-\frac{\rmi}{2\vp_q}(\Upsilon_\mu\h_q)-\frac{\rmi}{2\vp_q}(\bar\Upsilon_\mu\h_q^*)\\ %
\bar\Upsilon_\mu\Upsilon_\nu&:=\eta_{\mu\nu},\label{eq:defUpsilon1}
\end{align}
plus a superselection criterion requiring  that any term in any expression obtained from this Lagrangian must contain equal numbers of the scalar field $\h_q$ and its conjugate $\h^*_q$.
 On making these substitutions in Lagrangian~\eref{eq:Lq0} there appears at most one pair $\bar\Upsilon_\mu\Upsilon_\nu$. In later contexts there may appear multiple such pairs, therefore note that such expressions are evaluated as the pairwise sum over all combinations of the symbols, representing the different ways of pairing up the derivative operators, and hence the sigma matrices %
present in their construction. For example,
\begin{equation}
\bar\Upsilon_\mu\Upsilon_\nu\bar\Upsilon_\rho\Upsilon_\sigma=\eta_{\mu\nu}\eta_{\rho\sigma}+\eta_{\mu\sigma}\eta_{\rho\nu}.\label{eq:Upsilonexpansion}
\end{equation}
If desiring to avoid $\Upsilon$-notation, the same effect may be achieved %
less concisely but more conventionally using sigma matrices: %
\begin{align}
\partial_\mu&\rightarrow D_{q\mu}^{\bm{\dot\alpha\alpha}}\\
D_{q\mu}^{\bm{\dot\alpha\alpha}}&:=\partial_\mu\bar\sigma_0^{\bm{\dot\alpha\alpha}}-\frac{\rmi}{2\vp_q}(\bar\sigma_\mu^{\bm{\dot\alpha\alpha}}\h_q)-\frac{\rmi}{2\vp_q}(\bar\sigma_\mu^{\bm{\dot\alpha\alpha}}\h_q^*)\nn\\
D^\mu_{q\bm{\alpha\dot\alpha}}&:=-\partial^\mu\sigma^0_{\bm{\alpha\dot\alpha}}+\frac{\rmi}{2\vp_q}(\sigma^\mu_{\bm{\alpha\dot\alpha}}\h_q)+\frac{\rmi}{2\vp_q}(\sigma^\mu_{\bm{\alpha\dot\alpha}}\h_q^*)\nn\\
\bar\psi_q\bsm\psi_q&\rightarrow\bar\sigma_0^{\bm{\dot\alpha\alpha}}\bar\psi_q\bsm\psi_q~~\textrm{or}~~-\sigma^0_{\bm{\alpha\dot\alpha}}\bar\psi_q\bsm\psi_q.\nn
\end{align}
Again it is required that complex scalar bosons appear in conjugate pairs, and that pairwise tracing is performed across all sigma matrices with bold indices. Sigma matrices with unfixed vector indices ($\bar\sigma^{\bm{\dot\alpha\alpha}}_{\mu}$, $\sigma_{\bm{\alpha\dot\alpha}}^{\nu}$) are required to be traced pairwise with other sigma matrices with unfixed vector indices, and any expression is summed over all possible such pairings. Only one arbitrarily-selected pairing need be considered for sigma matrices with indices fixed at zero ($\bar\sigma_0^{\bm{\dot\alpha\alpha}}$ and $\sigma^0_{\bm{\alpha\dot\alpha}}$) as these %
are just included to ensure consistency of indices and scaling factors on all terms of $D^{\bm{\dot\alpha\alpha}}_{q\mu}$.

Using $\Upsilon$-notation, and therefore avoiding sigma matrices with bolded spinor indices, the Lagrangian becomes %
\begin{align}
\mscr{L}'_q=&\p{+}\frac{K_{q}^{(1)}}{2}\varphi_{q\mu}\underline{\triangle}_{q}^{\mu\nu}\varphi_{q\nu}\nn\\
&+\rmi K_{q}^{(2)}(D_q^{\mu}\bar\psi_{q})\,\Dslash_q(D_{q\mu}\psi_{q})\label{eq:Lq1}\\
&+K_{q}^{(3)}\h_q^*\underline{\square}_q\h_q\nn
\\
\underline{\triangle}_{q}^{\mu\nu}&:=\eta^{\mu\nu}\underline{\square}_q-D_q^{\mu} D_{q}^\nu\\
\underline{\square}_q&:=D_q^{\mu} D_{q\mu}\label{eq:usquareq}
\end{align}
and is subject to a superselection criterion that any term in %
this Lagrangian must contain equal numbers of the scalar field $\h_q$ and its conjugate $\h^*_q$. 

This Lagrangian now contains implicit derivatives with respect to the chiral co-ordinate $\alpha$, in the form of the complex scalar field $\h_q$, but no explicit scalar derivatives. And indeed, no non-redundant terms involving explicit scalar derivatives can be constructed. If a scalar derivative acts on $\vp_q$, this defines the scalar field $\h_q$ or its conjugate, e.g.~$\partial_\rmU\vp_q\rightarrow\h_q$. If the scalar derivative acts again, this term vanishes, e.g.{}
\begin{equation}
\partial_\rmU\h_q=0,
\end{equation}
and if it acts on any other species, including the conjugate $\h^*_q$, then once the Lagrangian is constructed it may always be brought to act on a copy of $\vp_q$, either as already described above for terms~1 and~2, or in the case of $\h_q$ or $\h^*_q$, directly by linear independence:
\begin{equation}
\partial_\rmU\h^*_q=\bar\partial_\rmU\h_q=\frac{1}{\vp_q}\h_q\h^*_q.\label{eq:multipleH}
\end{equation}
When constraint~\eref{eq:Phiacomm} is written as a Lagrangian expressed in terms of the species of \Eref{eq:allderivs}, the space--time dynamics of species $\vp_{q\mu}$, $\psi_q^\alpha$, and $\bar\psi_{q\dot\alpha}$ are therefore expressed in terms of the hermitian derivative operator $\partial_\mu$ only,
and the derivative of $\vp_q$ with respect to the chiral co-ordinate $\alpha$---i.e.~$\h_q$---accompanies the space--time derivative $\partial_\mu$ in a manner which somewhat resembles a gauge field, but takes values as a complex scalar.

The constraints which are obtained from Lagrangian~$\mscr{L}'_q$~\eref{eq:Lq1}, namely
\begin{align}
\underline\triangle^{\mu\nu}_q\varphi_{q\nu}=0
\label{eq:RMconstr1b}\\
\Dslash_q^{\dot\alpha\alpha}(D_{q\mu}\psi_{q\alpha})&=0\label{eq:RMconstr2b}\\
\Dslash_{q\alpha\dot\alpha}(D_{q\mu}\bar\psi_{q}^{\dot\alpha})&=0\label{eq:RMconstr3b}\\
\underline\square_q\h_{q}=0\qquad\qquad&\underline\square_q\h^*_{q}=0,\label{eq:RMconstr5b}
\end{align}
are weaker than \Erefr{eq:RGconstr1}{eq:RGconstr5}.
Where additional constraints exist beyond \Erefr{eq:RMconstr1b}{eq:RMconstr5b} (e.g.~$\partial_\mu\h_q=0$), these are manifest as restrictions on the field configurations which can be constructed on any submanifold of $\RM$ while still remaining consistent with some valid configuration of fields on $\Cwt$.

Examining the structure of $\mscr{L}'_q$, it is seen to resemble a Lagrangian on $\RM$ for five types of interacting quasiparticle excitation, namely the real vector field $\varphi_{q\mu}$ %
with units $L^{-1}$, %
the conjugate pair of spin-$\frac{1}{2}$ fields $\psi^\alpha_q$ and $\bar\psi_{q\dot\alpha}$
with %
units $L^{-\frac{1}{2}}$, and the complex scalar field $\h_q$ %
with units $L^{-1}$ and its conjugate $\h^*_q$.
At this stage, this resemblance is slightly misleading both because of the limitations on powers in $x^\mu$ and because couplings between the different species are weighted by a factor of $\vp_q^{-1}(x)$ which is not homogeneous across $\RM$. However, 
this structure nevertheless foreshadows the subsequent construction of an effective field theory on $\RM$.

\subsubsection{Pullbacks of product fields onto \protect{$\RM$}\label{sec:prodfields}}

While a Lagrangian~$\mscr{L}'_q$~\eref{eq:Lq1} may resemble a Lagrangian capable of supporting the wavefunctions of propagating and interacting particles, this resemblance is only superficial as any field appearing in $\mscr{L}'_q$ is at most quadratic in space-time co-ordinate $x^\mu$ on $\RM$, and the coupling strength $\vp_q^{-1}$ is dependent on space-time co-ordinate $x^\mu$ in a manner not explicitly articulated in $\mscr{L}'_q$. These limitations may be overcome on identifying quantities on $\RM$ which admit a quasiparticle description of arbitrary order in $x^\mu$.

First, introduce the all-fields Lagrangian
\begin{equation}
\mscr{L}_\Sigma=\sum_q\mscr{L}'_q\label{eq:LSum}
\end{equation}
and restrict attention to terms involving only the bosonic fields $\{\vp_{q\mu}\}$. Introduce the product field %
\begin{align}
\varphi(x) &:= \prod_q\varphi_q(x)\label{eq:defvp},%
\end{align}
pre-empted in \Eref{eq:phiprodpreempt}, which is a well-defined field over $\RM$ even though it cannot be constructed as a single field over $\Cwt$. Further define
\begin{equation}
\varphi_\mu(x) := -\rmi\partial_\mu\varphi(x),
\label{eq:defvarphimu}
\end{equation}
and consider the Lagrangian on $\RM$
\begin{equation}
\mscr{L}_\vp=\varphi_{\mu}\triangle^{\mu\nu}\varphi_{\nu}\label{eq:Lvector}
\end{equation}
which yields the equations of motion
\begin{equation}
\triangle^{\mu\nu}\varphi_{\nu}=0.\label{eq:EOM}
\end{equation}
The derivative operator in \Eref{eq:defvarphimu} may be expanded as per \Eref{eq:mudiv}, with the holomorphic and antiholomorphic spinor derivatives acting on different components of the product field $\vp(x)$. However, its action then defines a quantity $\vp_\mu(x)$ on $\RM$, and the subsequent derivatives in \Eref{eq:EOM} are space--time derivatives on $\RM$ describing the behaviour of this field. While $\vp(x)$ in \Eref{eq:EOM} may still be expanded in terms of its component fields, and these space--time derivatives may distribute across these components, they do so as a whole (i.e.~as $\partial_\mu$) and not independently as their spinor parts ($\partial^\alpha$ and $\bar\partial_{\dot\alpha}$).

Expansion of \Eref{eq:EOM} %
includes terms incorporating factors having the form of \Eref{eq:RMconstr1a}, which vanish, and also terms 
incorporating products of two or more different underlying vector fields, such as
\begin{equation}
(\partial^\mu\varphi_q^\nu)\varphi_{q'\nu}\quad\mrm{or}\quad\varphi^\mu_q\varphi^\nu_{q'}\varphi_{q''\nu}\label{eq:mixterms1}
\end{equation}
or incorporating a pair of spinor fields from the expansion of $\vp_\nu$, e.g.{}
\begin{equation}
(\partial^\mu\varphi_q^\nu)(\bar\partial\vp_{q'}\bsnn\partial\vp_{q''})\quad\mrm{or}\quad\varphi^\mu_q\varphi^\nu_{q'}(\bar\partial\vp_{q''}\bsnn\partial\vp_{q'''}).\label{eq:mixterms1a}
\end{equation}
However, provided the fields $\{\varphi_q\}$ everywhere satisfy four additional constraints (discussed further in \sref{sec:QLintro}), namely
\begin{align}
\!\forall~\mu,\nu,x~&:~\sum_{q'\not=q} \varphi_{q'\nu}(x)\,\varphi_{q\mu}(x)=0\label{eq:addconstr1}\\
\!\forall~\mu,\nu,\rho,x~&:~\sum_{q'\not=q} [\partial_\rho\varphi_{q'\nu}(x)]\,\vp_{q\mu}(x)=0,\label{eq:addconstr2v}
\\\!\forall~\mu,\nu,x~&:~\sum_{q''} \varphi_{q''\nu}(x)\,[\bar\partial\vp_{q'}\bsmm\partial\vp_q(x)]_{q'\not=q}=0\label{eq:addconstr1s}
\\\!\forall~\mu,\nu,\rho,x~&:~\sum_{q''} [\partial_\rho\varphi_{q''\nu}(x)]\,[\bar\partial\vp_{q'}\bsmm\partial\vp_q(x)]_{q'\not=q}=0,\label{eq:addconstr2s}
\end{align}
then satisfaction of %
\Eref{eq:RMconstr1a} also implies satisfaction of~\Eref{eq:EOM}. 

As noted above, again recall that $\partial_\mu$ is a derivative on $\RM$, with respect to the co-ordinates of $\RM$, i.e.~$x^\mu$, and so by construction the spinor derivatives $\bar\partial_{\dot\alpha}$ and $\partial_\alpha$ which appear in the equivalent expression on $\Cwt$ \eref{eq:mudiv} do not act on separate fundamental fields $\vp_q,~\vp_{q'}$ when evaluating \Eref{eq:EOM}. %

On $\RM$, the product field $\varphi$ is a real polynomial of degree $2N$ in $x^\mu$. Any polynomial of degree $2N$ with real coefficients may be expressed as a product of $N$ quadratics, and thus in the limit $N\rightarrow\infty$, if the coefficients of the fields $\{\varphi_q\bthetas\}$ are unconstrained, the field $\varphi$ may be set to any real polynomial of arbitrarily high degree in $x^\mu$. %
That is, there always exists a choice of fields $\{\varphi_q\bthetas\}$ on $\Cwt$ which realises any choice of field $\vp_\mu(x)$ on %
$\RM$. With the form of $\vp_\mu(x)$ unconstrained,
Lagrangian~\eref{eq:Lvector} is truly %
the Lagrangian for a free bosonic vector field $\vp_{\mu}$ on $\RM$. In conditions under which \Erefr{eq:addconstr1}{eq:addconstr2s} are satisfied, this Lagrangian follows directly from the vector boson portion of \Eref{eq:LSum}.

Now reintroduce the scalar bosons $\h_q$ and $\h_q^*$. 
Following the definition of $\vp_\mu$ it is also useful to introduce a complex scalar field of arbitrary order in $x$, %
\begin{equation}
\h(x) := \partial_\rmU\vp(x),\qquad\h^*(x):=\bar{\partial}_\rmU\vp(x)\label{eq:defHproduct}
\end{equation}
and consider Lagrangian
\begin{equation}
\mscr{L}_\h=\h^*\square\h
\end{equation}
where the derivatives in $\square$ are once again derivatives on $\RM$.
Once again, the equations of motion arising from this Lagrangian will contain terms in which all derivatives act on a single fundamental field, which are consistent with Lagrangian~\eref{eq:LSum} and vanish by \Erefr{eq:RMconstr4a}{eq:RMconstr5a}, and terms involving more than one fundamental field, e.g.
\begin{equation}
(\partial^\mu\vp_{q'\mu})\h_q\qquad\vp^\mu_{q'}\partial_\mu\h_q\qquad\vp^\mu_{q''}\vp_{q'\mu}\h_q.\label{eq:mixterms2}
\end{equation}
If these terms (henceforth ``cross-terms'') vanish (see \sref{sec:QLintro}), then $\h$ again behaves comparably to $\h_q$ in \Eref{eq:Lq1}. Once again the complex scalar boson terms may be incorporated into the covariant derivative, %
\begin{align}
\partial_\mu&\rightarrow D_{\mu}\\
D_{\mu}&:=\partial_\mu-\frac{\rmi}{2\vp}\Upsilon_\mu\h-\frac{\rmi}{2\vp}\bar\Upsilon_\mu\h^*,
\end{align}
this time with coupling coefficient $\vp^{-1}$. Let $\la\vp^{-1}(x)\ra_\mc{L}$ be the mean value of $\vp^{-1}$ evaluated over some region of space--time characterised by a length $\mc{L}$ (and a time $c^{-1}\mc{L}$) and centred on $x$. A set of fields $\{\vp_q\}$ is deemed \emph{sufficiently homogeneous} across some manifold $M\subset\RM$ iff $\la\vp^{-1}(x)\ra_\mc{L}$ varies across this manifold only by fluctuations small compared to some chosen threshold. If this threshold is small enough to render the fluctuations irrelevant to phenomena of interest, then $\la\vp^{-1}\ra_\mc{L}$ may be taken to define an effective coupling constant
\begin{equation}
f:={\la\vp\ra_\mc{L}}^{-1}.\label{eq:deff}
\end{equation}

To proceed similarly for the spinors, define
\begin{equation}
\psi^\alpha(x):=\partial^\alpha\vp(x)\qquad\bar\psi_{\dot\alpha}(x):=\bar\partial_{\dot\alpha}\vp(x). \label{eq:defspinors}
\end{equation}
Following the arguments above, and again assuming vanishing of cross terms, the Lagrangian must contain the spinor propagation term
\begin{align}
\begin{split}
\mscr{L}_{\bar\psi,\psi}&=\rmi (D^\mu\bar\psi)\,\Dslash(D_{\mu}\psi)
\end{split}
\label{eq:Dslash}
\end{align}
where the partial derivatives in $D_\mu$ are again operators on $\RM$.
The product spinors are in fact capable of participating in a broader range of interactions than those generated by \Eref{eq:Dslash}, but the structure of these interactions is not yet apparent, pending further exploration of the conditions under which %
the cross-terms %
vanish (\srefr{sec:fgspinorscalar}{sec:masses}). Therefore suppress those interactions for now, and let \Eref{eq:Dslash} be assumed to describe the propagation of a free fermion of dimension $L^{-1/2}$ in the presence of the complex scalar field. %

The terms $\mscr{L}_\h$ and $\mscr{L}_{\bar\psi,\psi}$ complete the construction of a Lagrangian analogous to $\mscr{L}'_q$~\eref{eq:Lq1}, in which all fields interact only with conjugate pairs of scalar bosons, noting that
\begin{itemize}
\item instead of being constructed on a specific set of fields $\vp_{q\mu}$, $\psi^\alpha_q$, $\bar\psi_{q\dot\alpha}$, $\h_q$, and $\h^*_q$, with $\vp_q^{-1}$ acting as a coordinate-dependent coupling strength, it is constructed on the fields $\vp_{\mu}$, $\psi^\alpha$, $\bar\psi_{\dot\alpha}$, $\h$, and $\h^*$ with an approximately coordinate-independent coupling strength $f=\la\vp^{-1}\ra_\mc{L}$, and
\item it is only consistent with $\mscr{L}_\Sigma$~\eref{eq:LSum} in regions where the gradients of the fundamental fields $\{\vp_q\}$ are uncorrelated (or systematically correlated and anticorrelated) such that cross-terms in the equations of motion %
vanish on summation. %
\end{itemize}
This Lagrangian is
\begin{align}
\begin{split}
\mscr{L}_\Pi=&\p{+}\frac{K^{(1)}}{2}\varphi_{\mu}\underline{\triangle}^{\mu\nu}\varphi_{\nu}\\
&+\rmi K^{(2)}(D^\mu\bar\psi)\,\Dslash(D_{\mu}\psi)\\
&+K^{(3)}\h^*\underline{\square}\h
\end{split}\label{eq:LPi}\\
\underline{\triangle}^{\mu\nu}&:=\eta^{\mu\nu}\underline{\square}-D^\mu D^\nu\\
\underline{\square}&:=D^\mu D_{\mu}\\
D_{\mu}&:=\partial_\mu-\frac{\rmi}{2\vp}\Upsilon_\mu\h-\frac{\rmi}{2\vp}\bar\Upsilon_\mu\h^*.
\end{align}
Let the participating fields ($\vp_\mu$, $\psi^\alpha$, $\bar\psi_{\dot\alpha}$, $\h$, and $\h^*$) be called \emph{product fields} due to their construction from the scalar product field $\vp$~\eref{eq:defvp}. Under Lagrangian $\mscr{L}_\Pi$ their equations of motion are
\begin{align}
\underline\triangle^{\mu\nu}\vp_\nu=0
\label{eq:RMconstr1c}\\
\Dslash^{\dot\alpha\alpha}(D_{\mu}\psi_{\alpha})&=0\label{eq:RMconstr2c}\\
\Dslash_{\alpha\dot\alpha}(D_{\mu}\bar\psi^{\dot\alpha})&=0\label{eq:RMconstr3c}\\
\underline\square\h=0\qquad\underline\square&\h^*=0.\label{eq:RMconstr5c}
\end{align}

Where to next? %
Although \Eref{eq:LPi} resembles the antisymmetry-induced Lagrangians for individual fields~\eref{eq:Lq1}, it is only equivalent to the corresponding Lagrangian~$\mscr{L}_\Sigma$~\eref{eq:LSum} in a context in which cross-terms such as \Erefs{eq:mixterms1}{eq:mixterms2} vanish. 
Therefore the next step is to introduce a field configuration on which cross-terms vanish, and which plays the role of a pseudovacuum (\sref{sec:QLintro}). Then, introduce quasiparticle excitations about this pseudovacuum state (\sref{sec:quasi}) while maintaining the requirement that cross-terms must continue to vanish. 

In the process of maintaining this requirement, some further terms are added to the Lagrangian. Additionally, Eqs.~\erefr{eq:mudiv}{eq:Udivs}, \eref{eq:defvarphimu}, \eref{eq:defHproduct}, and \eref{eq:defspinors} are noted to imply that any vector or scalar boson may be rewritten as a pair of spinors. In conjunction these modifications yield the additional particle interactions mentioned in passing above, which are given in full in \sref{sec:fgspinorscalar}. The incorporation of these terms, plus mass terms in \sref{sec:masses}, then completes the effective Lagrangian of the quasiparticle excitations.

\subsection{Pseudovacuum\label{sec:QLintro}} %

\subsubsection{Construction\label{sec:QL}}

As the purpose of this %
chapter is to 
identify a regime in which excitations on $\Cwt$ behave as
an analogue model of a quantum field theory on manifold $\RM$ (or, arguably, some reasonably large submanifold thereof),
a %
pseudovacuum background %
may be purposely chosen with the intent of %
realising this outcome, being imposed as part of the setup of the $\Cwt$ system into the analogue state. This Section describes the chosen pseudovacuum on $\Cwt$ in terms of its macroscopic properties on $\RM$. The imposition of the pseudovacuum as part of the initial setup of the model is actually quite natural, as will be seen when this is revisited in \sref{sec:entropic}.

Let the pseudovacuum configuration be a state extending over sufficient of $\Cwt$ such that its pullback covers all of %
$\RM$, and let it be chosen such that its pullback is a thermal state on $\RM$: In some rest frame on $\RM$, termed the isotropy frame, %
let this state be macroscopically homogeneous, %
isotropic, and %
stable over time (making it a maximum of any entropy function). It cannot be invariant under boost if the temperature (energy) of the thermal state is greater than zero. Macroscopic observables evaluated on the pseudovacuum in the isotropy frame, such as the mean of an expectation value over some sufficiently large 4-volume of pseudovacuum, are, by definition and construction, %
invariant under rotations and spacetime translations. %

To describe the microscopic construction of a pseudovacuum %
of energy $\mc{E}_0$ (temperature $k_B^{-1}\mc{E}_0$), first
define the \emph{centre} of a field $\varphi_q(x)$ as the point on $\RM$ at which the vector derivative $\partial_\mu\varphi_q(x)$ %
vanishes.\footnote{For any fields \protect{$\varphi_q$} for which the gradient vanishes everywhere, the centre may be chosen arbitrarily as these fields are homogeneous across all space--time %
and thus have vanishing Lagrangians on \protect{$\RM$}.} 
Let the number of fields $N$ on $\RM$ tend to infinity
while ensuring that distribution of the field centres is macroscopically homogeneous in the isotropy frame,
with the number of field centres in a region of $\RM$ characterised by length $\mc{L}$ being finite. %
(For convenience, this region may be taken as an arbitrarily-oriented hypercube with side length $\mc{L}$.) 
For $\mc{L}\gg\mc{L}_0$, where 
\begin{equation}
\mc{L}_0=hc\,\mc{E}_0^{-1},\label{eq:Hprinc} %
\end{equation}
the number of field centres within such a region approaches $N_0\mc{L}^4{\mc{L}_0}^{-4}$, where $N_0$ is a constant corresponding to the mean number of field centres within a hypercube of side length $\mc{L}_0$. %

Next, mandate that no systematic correlation be imposed between the position of the centre of a field and the sign of that field or any derivative thereof, either at the centre or in any far-field limit. 
When evaluating cross-terms such as \Erefr{eq:addconstr1}{eq:addconstr2s}, which compute correlations between a given field $\vp_{q\mu}$ or composite $\bar\partial\vp_{q'}\bsmm\partial\vp_q$ and all other fields in the model, these requirements ensure that far-field contributions to these expressions (i.e.~correlations with fields $\vp_{q'\mu}$ or $\bar\partial\vp_{q'}\bsmm\partial\vp_q$ having centres far from $x$) receive an infinite number of contributions of random and hence uncorrelated sign and magnitude, and their sums therefore %
vanish.

Near-field terms may persist, and may have definite sign, but again for any given instance of \Erefr{eq:addconstr1}{eq:addconstr2s} (corresponding to a specific field $\varphi_{q\mu}$ or $\bar\partial\vp_{q'}\bsmm\partial\vp_q$) this sign is equally likely to be positive or negative. 
Therefore %
let the centre of $\varphi_q$ be denoted $C(q)$, and recognise that %
over a %
region $R$ of volume $\mc{L}^4$, %
the average values of \Erefr{eq:addconstr1}{eq:addconstr2s} (now evaluated over multiple fields $\vp_{q\mu}$ and $\bar\partial\vp_{q'}\bsmm\partial\vp_q$) are %
\begin{align}
\frac{{\mc{L}_0}^4}{{\mc{L}}^{4}N_0}\sum_{q\,|\,C(q)\in R}\,\,\sum_{q'\not=q}&\varphi_{q'\nu}(x)\,\varphi_{q\mu}(x)\label{eq:nearfield1}\\
\frac{{\mc{L}_0}^4}{{\mc{L}}^{4}N_0}\sum_{q\,|\,C(q)\in R}\,\,\sum_{q'\not=q}&[\partial_\rho\varphi_{q'\nu}(x)]\,\varphi_{q\mu}(x)\label{eq:nearfield2v}
\\\frac{{\mc{L}_0}^4}{{\mc{L}}^{4}N_0}\sum_{\substack{q\,|\,C(q)\in R\\q'\,|\,C(q')\in R\\q'\not=q}}\,\,\sum_{q''}&\varphi_{q''\nu}(x)\,[\bar\partial\vp_{q'}\bsmm\partial\vp_q(x)]\label{eq:nearfield1s}\\
\frac{{\mc{L}_0}^4}{{\mc{L}}^{4}N_0}\sum_{\substack{q\,|\,C(q)\in R\\q'\,|\,C(q')\in R\\q'\not=q}}\,\,\sum_{q''}&[\partial_\rho\varphi_{q''\nu}(x)]\,[\bar\partial\vp_{q'}\bsmm\partial\vp_q(x)]\label{eq:nearfield2s}
\end{align}
respectively. Since far-field contributions are already known to vanish, it suffices to consider $\mc{L}\leq\mc{L}_0$.
By random variation of the sign of the near-field terms, independent of their magnitude, and by independence of the spinor and scalar derivatives, expressions~\erefr{eq:nearfield1}{eq:nearfield2s} then also tend to zero provided $N_0$ is sufficiently large, and provided $\mc{L}$ is not small compared with $\mc{L}_0$. For sufficiently large $N_0$, this then suffices to also eliminate the short-range contributions to the cross-terms \emph{on average} over length scales $\mc{L}\geq\mc{L}_0$.
Identical arguments apply to cross-terms arising from spinor and complex scalar boson terms in \Erefr{eq:RMconstr1c}{eq:RMconstr5c}, and thus these equations of motion hold on average over scales $\mc{L}\gg\mc{L}_0$ for the pseudovacuum state. Similarly, the threshold for interaction strength $f$ \eref{eq:deff} to approach a constant is likewise $\mc{L}\gg\mc{L}_0$.

Now consider the values
\begin{equation}
\la\vp^\mu(x)\ra_{\mc{L}}
\quad\mrm{and}\quad\la\vp^\mu(x)\vp_\mu(y)\ra_{\mc{L}}
\end{equation}
where $\la\,\cdot\,\ra_{\mc{L}}%
$ denotes the mean value obtained as the field co-ordinates range over 4-volumes characterised by a length $\mc{L}\gg\mc{L}_0$ and centred on $x$ or $y$ as appropriate.
Large-scale homogeneity implies that for a sufficiently large volume on $\RT$ and/or duration on the time axis, the value of $\la\vp^\mu(x)\vp_\mu(y)\ra_{\mc{L}}%
$ will tend towards some function of $x$ and $y$ with units of $L^{-2}$, whereas arbitrariness of sign and gradient of the fields making up $\vp^\mu$ implies %
\begin{equation}
\la\vp^\mu(x)\ra_{\mc{L}}\rightarrow0\quad\forall\quad\mu.\label{eq:<>0}
\end{equation}

Stability of the pseudovacuum implies maximisation of entropy. This in turn requires minimisation of numbers of degrees of freedom, implying equilibrium between field and spatial modes, 
and hence 
\begin{equation}
\la\vp^\mu(x)\vp_\mu(x)\ra_{\mc{L}}
= -\left(hc\right)^{-2}{\mc{E}_0}^2\label{eq:E0}
\end{equation}
in the isotropy frame. 

Since field $\vp_\mu(x)$ takes units of inverse length, it is convenient to adopt units %
such that $h=c=1$. %
It then follows that for a thermal pseudovacuum in the isotropy frame,
\begin{equation}
\la\vp^\mu(x)\vp_\mu(y)\ra_{\mc{L}}
= -\bm{f}(x-y)\,{\mc{E}_0}^2\label{eq:Exy}
\end{equation}
where $\bm{f}(x-y)$ is a Gaussian distribution which reaches a maximum at $x=y$ and satisfies %
\begin{equation}
\sol{c^{-3}}\!\!\int\!\rmd^4x%
\,{\mc{E}_0}^4\,\bm{f}(x-y)=1\quad\forall\quad y.\label{eq:xynorm}
\end{equation}
In %
calculations in which the form of this distribution is unimportant it is %
often convenient to approximate 
\begin{equation}
\begin{split}
\bm{f}(x)= %
\delta^4_{\mc{L}_0}(x)&:=\prod_{\mu=0}^3\delta_{\mc{L}_0}(x^\mu)\\
\delta_L(a)&:=\left\{
\begin{array}{ll}1&\mrm{if}~|a|\leq L/2\\0&\mrm{if}~|a|>L/2,\end{array}\right.
\end{split}\label{eq:window}
\end{equation} 
corresponding to perfect correlation of all fields within a fixed, arbitrarily-oriented %
hypercube only, having side length $\mc{L}_0$ in the isotropy frame, and centred on the peak of the Gaussian which is being approximated. %
The transition between near and far field regimes in this approximation is abrupt, taking place at the boundary of the hypercube, and thus %
correlators are %
computed 
entirely from the %
fields whose centres lie within the hypercube. %
This approximation %
is consequently only valid when %
the number of field centres $N_0$ within such a hypercube is
sufficiently %
large, or if the evaluation of a quantity is averaged over a sufficiently large number of such hypercubes. %

Equivalent correlators may be obtained for spinor and complex scalar fields by decomposing and substituting
\begin{equation}
\begin{split}
\vp^\mu(x)\vp_\mu(y)&\rightarrow
(\bar\partial\bsm\partial)\vp(x)~(\bar\partial\bsmm\partial)\vp(y)\\
&\rightarrow f^2~\bar\partial\vp(x)\bsm\partial\vp(x)~\bar\partial\vp(y)\bsmm\partial\vp(y)\\
\vp^\mu(x)\vp_\mu(x)&\rightarrow-2~\bar\partial\bar\partial\vp(x)~\partial\partial\vp(x)
\end{split}
\end{equation}
and rearranging (integrating by parts as required) to yield 
\begin{align}
\la\bar\psi(x)\bar\psi(y)\psi(y)\psi(x)\ra_\mc{L}&=2f^2\bm{f}(x-y)\mc{E}_0^2\label{eq:<psipsipsipsi>}\\
\la\h^*(x)\h(y)\ra_\mc{L}&=2\bm{f}(x-y)\mc{E}_0^2.\label{eq:<hh>}
\end{align}
Note that integration by parts depends on the existence of boundaries on $\RM$ where the product fields vanish---this is discussed in \sref{sec:partsQL}. %
The $(x-y)$ dependency in \Eref{eq:<hh>} follows from maximisation of entropy---sigma matrix identities determine the leading factor, and the normalised co-ordinate dependency must have the same form as that in \Eref{eq:Exy}.

\subsubsection{Integration by parts with a pseudovacuum background\label{sec:partsQL}}

In field theories without a pseudovacuum background of the sort described in \sref{sec:QL}, %
integration by parts is typically performed under the assumption that all fields vanish in the limit of spatial co-ordinates going to infinity, and become arbitrarily rapidly oscillating in the limit of time co-ordinates going to infinity (or vice versa, depending on the signature of the metric). This condition is, however, stronger than necessary, and it suffices that there exists some boundary outside the area under study on which the boundary term evaluates to zero (or, in practice, sufficiently close to zero as to permit the boundary term to be ignored). Integration by parts may then be performed on the Lagrangian with elimination of boundary terms, with the resulting alternative form of the Lagrangian being valid {within this boundary, and also outside it if a suitable conformal transformation exists.}

The thermal background of \sref{sec:QL} comprises a set of %
fields which do not %
individually vanish as $r\rightarrow\infty$.
However, for a boundary characterised
by a length scale $r\gg\mc{L}_0$, the lack of long-range correlations in the pseudovacuum implies that for an appropriate choice of boundary, the average of most expressions on the pseudovacuum fields may be made arbitrarily small as $r$ %
grows arbitrarily large, permitting integration by parts of the spatial terms in the usual fashion. %
While there exist some expressions on the pseudovacuum which do not vanish on average, e.g.~$\la\vp^\mu(x)\vp_\mu(x)\ra_{\mc{L}}$ in \Eref{eq:E0}, these expressions are all necessarily of even length or energy dimension while the boundary term is of odd dimension, and thus no nonvanishing pseudovacuum boundary term can exist.

\subsection{The quasiparticle field\label{sec:quasi}}

The low-energy quasiparticle fields of the model constructed on $\Cwt$ are now (finally) introduced, and take the form of small perturbations to the pseudovacuum of \sref{sec:QL}. For simplicity these perturbations are first examined for the vector field $\vp_\mu$ in isolation, treating $\bar\psi_{\dot\alpha}$, $\psi^\alpha$, $\h$, and $\h^*$ as zero in both the perturbations and the pseudovacuum (\srefr{sec:fgbg}{sec:denseregime}). The spinor and complex scalar fields are then reintroduced, and it is possible at %
last to concisely write a Lagrangian for these excitations involving the full spectrum of particle interactions (\sref{sec:fgspinorscalar}), valid in the low-energy limit. Mass terms are briefly discussed in \sref{sec:masses}.
Owing to the existence of the pseudovacuum, this model has a preferred frame (the isotropy frame) in which the ``low-energy limit'' is specified. \Sref{sec:pushlimits} explores the use of high-energy excitations to probe this violation of Lorentz invariance, and also identifies %
the energy scale at which the quasiparticle model of the low-energy limit %
breaks down.

\subsubsection{Foreground and background vector boson fields\label{sec:fgbg}}

Consider a field
$\varphi_\mu(x)$ which may be decomposed into a sum of two terms,
\begin{equation}
\varphi_\mu(x)=\bgfield{\varphi_\mu(x)}+\fgfield{\varphi_\mu(x)},\label{eq:varphifgbg}
\end{equation}
where $\bgfield{\varphi_\mu(x)}$ satisfies the definition of the pseudovacuum, and is termed the ``background field'', and $\fgfield{\varphi_\mu(x)}$ is a low-energy perturbation around the pseudovacuum state, termed the ``foreground field''. %
Require that the peak energy of the foreground perturbation satisfies $\mc{E}\ll\mc{E}_0$ in the isotropy frame of the background field.

Given this definition, it may be understood that the foreground field perturbation represents the presence (on average, over length scales large compared with $\mc{L}_0$) of an over- or under-excitation of low-energy field modes relative to those extrapolated from the high-energy modes and Gaussian distribution $\bm{f}(x-y)$.

Now consider the evaluation of a correlator
\begin{equation}
\la\vp^\mu(x)\vp_\mu(y)\ra_{\mc{L}}\label{eq:2corr}
\end{equation}
which is purely spatial in the isotropy frame. Let $\mc{L}_c$ denote the distance $|x-y|$, and require $\mc{L}_c\gg\mc{L}_0$. %
Any realistic measurement of a correlator has a finite resolution $\mc{L}_p$ corresponding to the energy of the process used as a probe, $\mc{E}_p=\mc{L}_p^{-1}$ (which itself is also subject to uncertainty, though that is neglected here). For $\mc{E}_p\ll\mc{E}_0$, source and sink are averaged over a region of scale $\mc{L}_p\gg\mc{L}_0$. The probe scale is also assumed to satisfy $\mc{L}_p\ll\mc{L}_c$.

Averaging over length scale $\mc{L}_p\gg\mc{L}_0$ ensures that contributions 
\begin{equation}
\begin{split}
&\la\bgfield{\vp^\mu(x)\vp_\mu(y)}\ra_{\mc{L}_p},~~\\&\la\bgfield{\vp^\mu(x)}\fgfield{\vp_\mu(y)}\ra_{\mc{L}_p},\textrm{ and }\\&\la\fgfield{\vp^\mu(x)}\bgfield{\vp_\mu(y)}\ra_{\mc{L}_p}
\end{split}\label{eq:vanishingcorrterms}
\end{equation}
necessarily vanish, and thus for $\mc{L}=\mc{L}_p$ and $\mc{L}_c\gg\mc{L}_p\gg\mc{L}_0$, correlator~\eref{eq:2corr} reduces to the foreground correlator
\begin{equation}
\la\fgfield{\vp^\mu(x)\vp_\mu(y)}\ra_{\mc{L}_p}.\label{eq:2corrfg}
\end{equation}

In \sref{sec:pushlimits} it is argued that foreground fields are largely %
insensitive to the broken Lorentz symmetry of the pseudovacuum provided the energies of all foreground fields involved are small compared with an energy scale $\frac{1}{2}\mc{E}_\Omega:=\frac{1}{2}(N_0-\frac{1}{2})\mc{E}_0$ when evaluated in the isotropy frame. %
This result for correlators which are spacelike in the isotropy frame therefore extends also to correlators which are spacelike in some other frame provided this requirement is met, and the argument may be repeated for timelike correlators.

This result illustrates a broader trend, namely that many expressions involving both foreground and background fields vanish in the low-energy regime due to the limited range of correlations involving the background fields (with the exceptions to this trend becoming mass terms, as described in \sref{sec:masses}). This, in turn, motivates the construction of an effective Lagrangian for the foreground fields.

\subsubsection{Sparse perturbation regime\label{sec:sparseregime}}

To construct a Lagrangian for the foreground fields, recognise that quantities involving foreground fields are always evaluated on average over length scales $\mc{L}\gg\mc{L}_0$ in which regimes most background field terms vanish %
(up to the exceptions in \sref{sec:masses} which yield mass terms%
). In this context it is largely reasonable to assume that the background fields obey approximate distribution~\eref{eq:window}. [Situations where this approximation is not appropriate will be indicated; relaxation of \Eref{eq:window} to a more realistic (e.g.~Gaussian) distribution is discussed in \sref{sec:denseregime}.] Therefore, first consider a scenario in which $\RM$ is divided into hypercubes of side length $\mc{L}_0$ in the isotropy frame of the pseudovacuum, and at most one field in each volume ${\mc{L}_0}^4$ exhibits a foreground perturbation (or equivalently, that exactly one field in each volume exhibits a foreground perturbation, but this perturbation may be zero). In \sref{sec:operators} %
this regime will be identified with the existence of a single quantised excitation of the foreground fields. %

Observe that a perturbation of a single fundamental field
\begin{align}
&\varphi_q(x)\rightarrow\varphi_q(x)+\Delta\varphi_q(x) \\ %
&\|\Delta\varphi_q(x)\|\ll\|\varphi_q(x)\|~\forall~x
\end{align}
may be rewritten as
\begin{align}
\varphi_q(x)\rightarrow\varphi_q(x)\left[1+\frac{\Delta\vp_q(x)}{\vp_q(x)}\right],
\end{align}
modifying $\vp_\mu$ according to
\begin{align}
\vp_\mu(x)&\rightarrow\vp_\mu(x)+\Delta_q\vp_\mu(x)\\
\Delta_q\vp_\mu(x)&:=\frac{\Delta\vp_q(x)}{\vp_q(x)}\vp_\mu(x) + \vp(x)\,\partial_\mu\!\left[\frac{\Delta\vp_q(x)}{\vp_q(x)}\right]\label{eq:defDeltaq}
\end{align}
where the subscript $q$ on $\Delta_q$ indicates that this perturbation to $\vp_\mu$ arises from a perturbation to fundamental field $\vp_q$ on $\Cwt$.

Introduce perturbations of two fundamental fields $\vp_{q_1}$ and $\vp_{q_2}$ whose centres on $\RM$ are separated by a distance (or time) $\mc{L}\gg\mc{L}_0$, and let $x_{q_1}$ and $x_{q_2}$ denote the co-ordinates of these centres. %
Let the perturbations be correlated such that 
\begin{equation}
\la\vp^\mu(x_{q_1})\vp_\mu(x_{q_2})\ra_{\mc{L}_p},\label{eq:preDeltacorr}
\end{equation}
which would otherwise vanish, now yields the value of
\begin{equation}
\la\Delta_{q_1}\vp^\mu(x_{q_1})\Delta_{q_2}\vp_\mu(x_{q_2})\ra_{\mc{L}_p}.\label{eq:Deltacorr}
\end{equation}
Proceed by introducing further perturbations across $\RM$ with a density such that for an arbitrarily-chosen hypercube in the grid %
there is precisely one perturbed field whose centre lies within this cube, and the mean pairwise correlators between these perturbed fields %
\eref{eq:preDeltacorr} %
correspond to the correlators between the desired foreground field at the same co-ordinates as the perturbed fields' centres. Thus the perturbations $\{\Delta_q\vp_\mu\}$ encode a foreground field $\fgfield{\vp_\mu}$, and the requirement that $\fgfield{\vp_\mu}$ has an energy scale small compared with $\mc{E}_0$ ensures that correlators such as \Eref{eq:Deltacorr} are in general nonvanishing  over distances and times large compared with $\mc{L}_0$.

Again consider a pair of these fields, $\vp_{q_1}$ and $\vp_{q_2}$ with centres at $x_{q_1}$ and $x_{q_2}$, and let $\{\vp_q\}$ denote the set of all other perturbed fields with centres %
not at $x_{q_1}$ or $x_{q_2}$. Note that on average across a sufficiently large number of choices for $\vp_{q_1}$ and $\vp_{q_2}$, the distance to the closest other perturbed field centre has been chosen to be at least $\mc{L}_0$. Introduce a third field $\vp_{q_3}\in\{\vp_q\}$ with centre at $x_{q_3}$, initially neglecting its perturbation $\Delta\vp_{q_3}$, and recognise that %
the signs of $\la\vp^\mu(x_{q_1})\vp_\mu(x_{q_3})\ra_{\mc{L}_p}$ and $\la\vp(x_{q_1})\vp(x_{q_3})\ra_{\mc{L}_p}$ are random. The sign and gradient of $\Delta\vp_{q_3}(x_{q_3})$ in \Eref{eq:defDeltaq} are then chosen %
to compensate for these random signs, %
making the value of
$\la\Delta_{q_1}\vp^\mu(x_{q_1})\Delta_{q_3}\vp_\mu(x_{q_3})\ra_{\mc{L}_p}$ consistent with the desired foreground field, as described above. %
The signs and gradients of the perturbations $\Delta\vp_{q_i}$ are thus correlated with those of the field centres $\vp_{q_i}(x_{q_i})$ and their gradients $\partial_\mu\vp_{q_i}(x_{q_i})$. 
However, the sign of the central value of a field $\vp_{q_i}$ is uncorrelated with %
its sign in the far field, %
and consequently at $x_{q_2}$
the signs of $\{\Delta\vp_{q_i}(x_{q_2})/\vp_{q_i}(x_{q_2})\,|\,i>2\}$ %
are random, and when summed, make no net contribution to the correlator $\la\vp^\mu(x_{q_1})\vp_\mu(x_{q_2})\ra_{\mc{L}_p}$. %

Likewise, the second term of \Eref{eq:defDeltaq} depends on $\partial_\mu[\Delta\vp_{q_3}(x_{q_2})/\vp_{q_3}(x_{q_2})]$ so is similarly of arbitrary sign, %
and the perturbations $\{\Delta\vp_{q_i}\,|\,{i>2}\}$ therefore make a vanishing net contribution to correlator~\eref{eq:preDeltacorr}. Note that the vanishing of these contributions is dependent upon the mean density of perturbations being no more than one per hypervolume ${\mc{L}_0}^4$.

This on-average vanishing of foreground correlators involving far field contributions, in conjunction with the on-average vanishing of correlators where a summed index appears on both a foreground and a background field, suffices to establish the vanishing of cross-terms \erefr{eq:addconstr1}{eq:addconstr2s} involving one or more foreground fields when averaged across a probe scale $\mc{L}_p\gg\mc{L}_0$. This vanishing of cross-terms, in turn, suffices to validate Lagrangian~\eref{eq:Lvector} for free vector bosons in the sparse perturbation regime as well as in the pseudovacuum regime.

Recalling that terms having the forms of \Eref{eq:vanishingcorrterms} disappear when averaged over length scales large compared with $\mc{L}_p$, it follows that in the sparse perturbation regime, over length scales large compared with $\mc{L}_p$, the free vector boson Lagrangian~\eref{eq:Lvector} decomposes into %
independent foreground and background terms, %
\begin{equation}
\begin{split}
\mscr{L}_{\vp}&=\mscr{L}_{{\vp_\fg}}+\mscr{L}_{{\vp_\bg}}\\
\mscr{L}_{{\vp_\fg}}&=\fgfield{\varphi_{\mu}\triangle^{\mu\nu}\varphi_{\nu}}\label{eq:Lvectorfgbg}\\ 
\mscr{L}_{{\vp_\bg}}&=\bgfield{\varphi_{\mu}\triangle^{\mu\nu}\varphi_{\nu}}, 
\end{split}
\end{equation}
where the foreground terms are small perturbations of the product field, $\fgfield{\vp_\mu}\equiv\Delta\vp_\mu$, and are composed from the perturbations of appropriately selected individual fields $\Delta\vp_q$.
The decoupling of foreground and background fields in \Eref{eq:Lvectorfgbg} justifies the identification of the thermal background state as a pseudovacuum, across which foreground perturbations propagate as if it were free space.

\subsubsection{Dense perturbation regime\label{sec:denseregime}}

Now consider a situation where more than one fundamental field may be perturbed per grid volume ${\mc{L}_0}^4$. In this situation it is helpful to reframe the proposed vector Lagrangian~\eref{eq:Lvector} as
\begin{equation}
\mscr{L}_{\vp} = {\partial_\mu\vp}(g^{\mu\nu}\partial^\rho\partial_\rho-\partial^\mu\partial^\nu)\partial_\nu\vp,
\end{equation}
and again to recall that this Lagrangian only gives a true description of the behaviour of the system in a regime in which cross-terms vanish, causing it to reduce to the vector boson component of $\mscr{L}_\Sigma$~\eref{eq:LSum}.

First, recognise that previously, on expansion of $\vp$ according to \Eref{eq:defvp}, sparsity implied that the only terms with non-vanishing foreground contributions were those in which all derivatives acted on the single perturbed field with centre in the target hypervolume.
Now that there are two or more perturbed fields with centres in the same hypervolume, cross-terms involving multiple foreground fields may exist, and due to the long-range correlations of the foreground perturbations, these cross-terms will not necessarily vanish on averaging over $\mc{L}_p$, and thus Lagrangian~\eref{eq:Lvector} does \emph{not} hold in the dense perturbation regime.

In these cross-terms, derivatives acting on separate perturbed fundamental fields are both eligible for promotion to linearly independent vector fields, e.g.~in the expansion of $\partial_\mu\partial_\nu\vp$,
\begin{equation}
\partial_\mu\vp_q\partial_\nu\vp_{q'}\rightarrow\vp_{q\mu}\vp_{q'\nu}.\label{eq:multipsparse}
\end{equation}
However, where both derivatives act on a single fundamental field this still yields e.g.{}
\begin{equation}
\partial_\mu\partial_\nu\vp_{q}\rightarrow\partial_\mu\vp_{q\nu}
\end{equation}
as before.\footnote{This behaviour foreshadows the identification, in \psref{sec:operators}, %
of \protect{$n$} perturbations per volume \protect{${\mc{L}_0}^4$} with the presence of \protect{$n$} excitation quanta. The requirement that fields \protect{$\vp_{q\mu}$} and \protect{$\vp_{q'\nu}$} in \pEref{eq:multipsparse} arise from separate superposed sparse perturbations then corresponds to recognising that you can't annihilate the same quantum twice.}
This non-vanishing of cross-terms is problematic, as it implies that in the dense perturbation regime, the correct Lagrangian for foreground vector fields on $\RM$ (with all spinor and complex scalar fields elided) can no longer be obtained simply by replacing $\vp_{q\mu}$ with $\fgfield{\vp_\mu}$ in \Eref{eq:Lq1} and deleting all spinor and scalar boson terms to obtain the foreground component of \Eref{eq:Lvectorfgbg}. 

To address this, the replacement
\begin{equation}
\partial_\mu\rightarrow\partial_\mu-\rmi f\vp_\mu
\end{equation}
conveniently cancels out all of the arising cross-terms, %
and yields the Lagrangian
\begin{align}
\mscr{L}_{\vp_\fg}&=\fgfield{\vp_\mu\tilde{{\triangle}}^{\mu\nu}\vp_\nu}\label{eq:Lvectorfggauge}\\
\tilde{{\triangle}}^{\mu\nu}&=g^{\mu\nu}\bm{\partial}^\rho\bm{\partial}_\rho-\bm{\partial}^\mu\bm{\partial}^\nu\\
\bm{\partial}_\mu&=\partial_\mu-\rmi f\vp_\mu.\label{eq:newpartialmu}
\end{align}
This Lagrangian is more closely suitable. Since it cancels out all arising cross-terms, including those involving multiple perturbed fields in the same hypervolume, it comes substantially closer to exhibiting the required reduction to the vector boson terms of Lagrangian~$\mscr{L}_\Sigma$~\eref{eq:LSum} (which, in turn, is obtained from antisymmetry of co-ordinate indices on $\Cwt$). However, 
if viewed as a classical construction then Lagrangian~\eref{eq:Lvectorfggauge} now contains some new superfluous terms when compared with $\mscr{L}_\Sigma$. %
These additional terms are those in which consecutive derivative operators acting on the same instance of $\vp$ yield repeated first derivatives of the same fundamental field, e.g.{}
\begin{equation}
\vp^{q\mu}\partial^\nu\vp_{q\nu}\textrm{ arising within }\tilde{{\triangle}}^{\mu\nu}\vp_\nu. 
\end{equation}
If the purpose of constructing a quasiparticle model on $\Cwt$ is to emulate a classical Lagrangian~\eref{eq:Lvectorfggauge}, then this model can at best emulate classical Lagrangian~\eref{eq:Lvectorfggauge} \emph{up to} those terms, which are present in $\mscr{L}_{\vp_\fg}$ on $\RM$ but absent in $\mscr{L}_\Sigma$ on $\Cwt$. Conversely, the presence of these terms in \Eref{eq:Lvectorfggauge} limits the ability of $\mscr{L}_{\vp_\fg}$ to describe the behaviour of the quasiparticles arising in fields on $\Cwt$.

However, in \sref{sec:semiclassical} it is seen that the model on $\Cwt$ is in fact better understood emulating a semiclassical approximation to a quantum field theory (QFT), and in %
\sref{sec:operators} multiple perturbations within a single ${\mc{L}_0}^4$ hypervolume are identified with multiple excitation quanta of the field $\vp_\mu$. In QFT a single quantum cannot be annihilated multiple times, and consequently the omission of terms repeating the same vector field is %
in fact \emph{necessary}. %
[The only permitted duplication is across inbound and outbound fields %
$\vp_\mu$ and $\vp_\nu$ of $\vp_\mu\tilde{\Delta}^{\mu\nu}\vp_\nu$, in keeping with the form of \Eref{eq:Lq1}.] %
In conjunction with other results presented in \sref{sec:semiclassical}, the omission of these prohibited terms %
permits identification of the vector boson sector of the model on $\Cwt$ as an analogue to the low-energy regime of a \emph{quantum} field theory on $\RM$ which has Lagrangian~\eref{eq:Lvectorfggauge}.
Or conversely, it is likewise necessary if the QFT counterpart to Lagrangian~\eref{eq:Lvectorfggauge} is to provide an effective description of the vector boson component of the foreground perturbations of the model on $\Cwt$ (assuming all spinor and complex scalar fields vanish).%
\footnote{Note that this restriction does not exclude four-point functions or loop diagrams in the QFT. Repetition of a single field \protect{$\vp_q$} implies repetition of a field \protect{$\vp_\mu$} with identical momentum, analogous to two lines at a Feynman diagram vertex carrying the same label, even when these lines do not form a loop. In QFT each line at a vertex carries a different label, with momenta independent up to a delta function on the total momentum at the vertex, and any delta functions arising from closed loops. These excitations with independent momenta therefore map to gradients of different fields, \protect{$\vp_q$}, \protect{$\vp_{q'}$}, etc., rather than being repetitions of the same field.}

Thus, when \Eref{eq:Lvectorfggauge} is understood as a QFT Lagrangian, this provides the correct field couplings for consistency with~\Eref{eq:LSum} and with the origin of this latter Lagrangian in the anticommutation of basis vectors on $\Cwt$.

As a further addendum, note that for realistic (e.g.~Gaussian) background field correlators, as opposed to the sharply truncated approximation of \Eref{eq:window}, %
the terms given in \Eref{eq:vanishingcorrterms} truly vanish only in the limit $\mc{L}_p\rightarrow\infty$. 
Therefore, given a realistic dataset comprising $\varphi_\mu$ over a finite submanifold $S\subset\RM$ [with the assumption that decomposition~\eref{eq:varphifgbg} exists], study of $\vp_\mu$ does not in general permit decomposition~\eref{eq:varphifgbg} to be exactly recovered. %
However, it may be approached arbitrarily closely for sufficiently large $S$, with the residual ambiguity which cannot be resolved on $S$ giving rise to an equivalence class of decompositions whose foreground field behaviours on $S$ differ only on the scale of the unresolved ambiguity, converging to a single member if $S$ is made large in a way which causes the ambiguity to vanish. %
For $S$ sufficiently large in all dimensions when compared with $\mc{L}_0$, the uncertainty arising from the tails of the Gaussians---which are characterised by $\mc{L}_0$---will in general be small for any foreground process on this submanifold.

\subsubsection{Foreground spinors and complex scalars\label{sec:fgspinorscalar}}

It is, of course, unrealistic to completely ignore the spinor and complex scalar fields as in \srefr{sec:fgbg}{sec:denseregime}. However, having obtained Lagrangian~\eref{eq:Lvectorfggauge} it is now possible to re-establish these fields in a much more cohesive fashion.

First, recognise that the definitions of the product fields
\begin{align}
&\varphi_\mu(x) := -\rmi\partial_\mu\varphi(x)\tag{\ref{eq:defvarphimu}}\\
&\psi^\alpha(x):=\partial^\alpha\vp(x)\quad\qquad\bar\psi_{\dot\alpha}(x):=\bar\partial_{\dot\alpha}\vp(x) \tag{\ref{eq:defspinors}}\\
&\h(x) := \partial_\rmU\vp(x)~\,\quad\qquad\h^*(x):=\bar{\partial}_\rmU\vp(x)\tag{\ref{eq:defHproduct}}
\end{align}
imply the substitutions
\begin{align}
\partial_\mu&\rightarrow \rmi f\vp_\mu(x)\label{eq:opsubvp}\\
\partial^\alpha&\rightarrow f\psi^\alpha(x)\qquad\bar\partial_{\dot\alpha}\rightarrow f\bar\psi_{\dot\alpha}\label{eq:opsubpsi}\\
\partial_\rmU&\rightarrow f\h(x)\qquad~\,\bar\partial_\rmU\rightarrow f\h^*(x)\label{eq:opsubH}
\end{align}
which are valid when the derivative operators act directly on the product field $\vp(x)$, or when they are linearly independent of any operators which lie between them and $\vp(x)$. Thus, for example, linear independence of the holomorphic and antiholomorphic sectors permits substitution
\begin{equation}
\begin{split}
\partial_\mu\vp(x) = \bar\partial\bsmm\partial\vp(x) \longrightarrow f\bar\partial\vp(x)\bsmm\partial\vp(x)=f\bar\psi\bsmm\psi
\end{split}\label{eq:vphipreonsub}
\end{equation}
in addition to \Eref{eq:defvarphimu}. One interpretation of this interchangeability is that the vector bosons $\vp_\mu$ are composite bosons assembled from pairs of spinor preons, and that notation $\vp_\mu$ is no more than a shorthand for a preon and an antipreon at the same co-ordinate.

Similarly, any instance of the complex scalar boson field $\h$ may be understood as a pair of chiral spinor derivative operators acting on the scalar field $\vp$. Since the $\h$ and $\h^*$ bosons always appear in conjugate pairs, integration by parts may always bring both components of these bosons to places where they too may be substituted by spinor fields, for example
\begin{align}
\begin{split}
\h^*(x)\square\h(x)&=\h^*(x)\square\partial\partial\vp(x)%
\\
&\longrightarrow
\rmi\partial^\alpha\h^*(x)\square\psi_\alpha\\&\longrightarrow\rmi f\psi^\alpha\h^*\square\psi_\alpha
\end{split}\\
\h(x)\h^*(x)\h(x)\h^*(x)&\longrightarrow f^{-1}\partial\partial\h^*(x)\h^*(x)\h(x)\nn
\\&\longrightarrow-\rmi f^{-1}\partial_\alpha\h^*(x)\partial^\alpha\h^*(x)\h(x) %
\\&\longrightarrow \rmi f\psi^\alpha(x)\h^*(x)\psi_\alpha(x)\h^*(x)\h(x).\nn
\end{align}

\begin{figure}
\includegraphics[width=\linewidth]{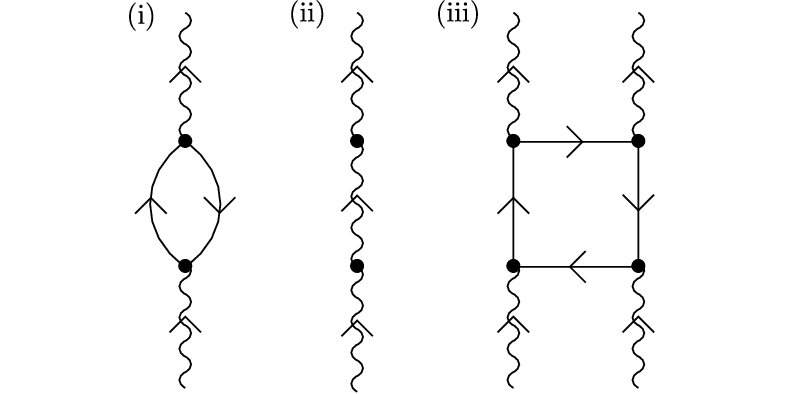}
\caption{Examples of elimination of redundant preon processes. Diagram~(i) is redundant, as both preons share the same source and sink. The equivalency \protect{$\vp_\mu(x)=\bar\psi(x)\bsmm\psi(x)$} then implies that this figure describes the exact same process as diagram~(ii). Diagram~(iii) is not redundant as there is no way to pairwise replace the preonic constituents with equivalent scalar or vector bosons.\label{fig:doublecount}}
\end{figure}

Now let perturbations of the fundamental fields $\{\vp_q\}$ on $\Cwt$ also give rise to foreground spinor and complex scalar boson fields, again being low-energy perturbations of the pseudovacuum state having correlation lengths large compared with $\mc{L}_0$. Reintroducing the complex scalar boson into the derivative operator and recognising that both the vector boson and scalar boson fields are no more than shorthands for their preon equivalents, it is convenient to write the derivative
\begin{align}
\bm{D}_\mu:=\partial_\mu&-\rmi f(\vp_\mu-\rmi f\bar\psi\bsmm\psi)\label{eq:Dbold}
\\&-\frac{\rmi f}{2}[\Upsilon_\mu(\h+\rmi f\psi\psi)+\bar\Upsilon_\mu(\h^*+\rmi f\bar\psi\bar\psi)],\nn
\end{align}
with the understanding that the preon forms are to be used only in contexts where the bosonic forms are inapplicable, in order to avoid double counting.
Anticipating that $\bm{D}_\mu$ will appear in a Lagrangian $\mscr{L}_\fg$~(\ref{eq:Lfgintexp},\ref{eq:Lexp}) which is used in construction of a generating functional $\Z$~\eref{eq:cplxZ}, this would eliminate the Feynman diagram of \fref{fig:doublecount}(i) as a duplicate of \fref{fig:doublecount}(ii), but permit \fref{fig:doublecount}(iii) as a process with no vector or scalar analogue.
As a diagram rule, this condition for avoidance of double counting may be stated as \emph{no two preon lines may have in common both terminating vertices.} Given their equivalence to preon pair exchanges, it might seem somewhat redundant to have introduced the vector and complex scalar fields at all---but in %
later chapters it is seen that for some values of $\N$ in $\Cwn$ these pairs may propagate as bound doublets on account of mass-generating interactions with the pseudovacuum.

Recognising that the action of derivative operator $\partial_\mu$ on $\bar\psi\bsmm\psi$ expands as
\begin{equation}
(\partial_\mu\bar\psi)\bsm\psi + \bar\psi \bsm (\partial_\mu\psi)\label{eq:derivpsis}
\end{equation}
and thus vanishes under integration by parts,
the introduction of $\bm{D}_\mu$
permits the Lagrangian to be written in a particularly concise form,
\begin{align}
F_{\mu\nu}&=\bm{D}_\mu\bm{D}_\nu 1-\bm{D}_\nu\bm{D}_\mu 1\label{eq:I:L1}\\
\mscr{L}&=-\frac{1}{4}F^{\mu\nu}F_{\mu\nu}+\rmi(\bm{D}^\mu\bar\psi)\,\bmDslash(\bm{D}_\mu\psi)\label{eq:I:L}
\end{align}
up to the following caveats:
\begin{itemize}
\item Expansion of the first term of $\mscr{L}$ sets $K^{(3)}=K^{(1)}$. The value of $K^{(2)}$ is not fixed directly, but its magnitude is set by transforming one copy of $\vp_\mu$ in $\frac{1}{2}\vp^\mu\square\vp_\mu$ into spinors, and the other into the derivative operator $\dslash$. There are two choices of how this is to be done, cancelling the factor of $\frac{1}{2}$.
Expansion~\eref{eq:derivpsis} then yields two terms, one governing the dynamics of $\dslash(\partial_\mu\psi_\alpha)$ and $\partial_\mu\bar\psi_{\dot\alpha}$, and the other governing the dynamics of $\dslash(\partial_\mu\bar\psi_{\dot\alpha})$ and $\partial_\mu\psi_{\alpha}$. Since the corresponding equations of motion are equivalent up to hermitian conjugation,
one of these terms is redundant. Since the pair sum to zero~\eref{eq:derivpsis}, both in fact vanish, and one is then explicitly reintroduced up to a phase determined by convention.
\item Recognising that $\bar\psi\bsm\psi$ may be substituted for $\vp_\mu$, and that $\h\h^*$ may likewise be substituted for $\vp^\mu \vp_\mu$ and then in turn replaced by $\bar\psi\bar\psi\psi\psi$, a further superselection rule is required to prevent double counting of some spinor interaction terms. %
In the context of determining the Lagrangian,
the replacements
\begin{align}
\vp^\mu\vp_\mu &\longrightarrow -f^2\bar\psi\bsm\psi\bar\psi\bsmm\psi=2f^2\bar\psi \bar\psi \psi \psi\\
\vp^\mu\vp_\mu &\longrightarrow 2\h\h^* \longrightarrow 2f^2\bar\psi \bar\psi \psi \psi 
\end{align}
represent two different ways to compute one single re-expression of the same term $\vp^\mu\vp_\mu$, rather than being distinct physical processes. Accidental duplication is avoided by requiring that within a single term of the Lagrangian, all spinors associated with upsilon markers must be of the same parity. (Spinors not associated with upsilon markers may be of diverse parity, as before.)
\end{itemize}
Restricting to the foreground fields only, the expanded form of Lagrangian~\eref{eq:I:L} is
\begin{align}
\begin{split}
{\mscr{L}}_\fg=&\p{+}\frac{1}{2}\fgfield{\varphi_{\mu}\tilde{\underline{\triangle}}^{\mu\nu}\varphi_{\nu}}+\fgfield{\h^*\tilde{\underline{\square}}\h}\\
&+\rmi \fgfield{(\bm{D}^\mu\bar\psi)\,\bmDslash(\bm{D}_{\mu}\psi)}
\end{split}\label{eq:Lfgintexp}
\\
\tilde{\underline{\triangle}}^{\mu\nu}&:=\eta^{\mu\nu}\tilde{\underline{\square}}-\bm{D}^\mu \bm{D}^\nu\label{eq:I:ttr}\\
\tilde{\underline{\square}}&:=\bm{D}^\mu \bm{D}_{\mu}\label{eq:I:tsq}
\end{align}
with equations of motion
\begin{align}
\tilde{\underline\triangle}^{\mu\nu}\fgfield{\vp_\nu}=0\label{eq:fgEOM1massless}\\
\bmDslash^{\dot\alpha\alpha}(\bm{D}_{\mu}\fgfield{\psi_{\alpha}})&=0\\
\bmDslash_{\alpha\dot\alpha}(\bm{D}_{\mu}\fgfield{\bar\psi^{\dot\alpha}})&=0\\
\tilde{\underline\square}\fgfield{\h}=0\qquad\qquad&\tilde{\underline\square}\fgfield{\h^*}=0.\label{eq:fgEOM5massless}
\end{align}

Of note, elimination of cross-terms through substitution~\eref{eq:Dbold} implies that
$\bm{D}_\mu\h$ and $\bm{D}_\mu\h^*$
independently vanish, as did $\partial_\mu\h_q$ and $\partial_\mu\h_q^*$~\erefr{eq:RMconstr4}{eq:RMconstr5}. However, this %
in turn implies that unlike the individual fields $\{\h_q,\h_q^*\}$, the product fields $\h$ and $\h^*$ do not have individually vanishing derivatives $\partial_\mu\h$ and $\partial_\mu\h^*$, and so \emph{do} exhibit an $x$ co-ordinate dependency. Indeed, these fields may be of arbitrary order in $x^\mu$ in much the same way as $\vp_\mu$. The same is likewise true of the product spinor fields: Although the component spinor fields $\{\psi_q,\bar\psi_q\}$ were at most linear, the product spinor fields may also be of arbitrary order in $x^\mu$.

It is also worth revisiting substitutions~\erefr{eq:opsubvp}{eq:opsubH} for the foreground fields. These substitutions were originally introduced for the total fields, but are likewise valid for the foreground fields in the low-energy limit. Recognising that in the low-energy limit the value of $f$ is dominated by the pseudovacuum, %
\begin{equation}
f\approx\la\bgfield{\vp^{-1}}\ra_\mc{L},\label{eq:deff2}
\end{equation}
the error associated with using this approximate value of $f$ may be computed by substituting~\erefr{eq:opsubvp}{eq:opsubH} into the corresponding definitions~\erefr{eq:defvarphimu}{eq:defHproduct} and approximating $f$ using \Eref{eq:deff2}. In each case the discrepancy is a factor of
\begin{equation}
1-\frac{\fgfield{\vp(x)}}{\la\bgfield{\vp}\ra_\mc{L}}\label{eq:opsuberror}
\end{equation}
which may be neglected if the foreground perturbations to $\vp(x)$ are %
sufficiently small, or are periodic or fluctuating over a timescale which is likewise sufficiently small.

\subsubsection{Foreground particle masses\label{sec:masses}}

The foreground field Lagrangian~\eref{eq:Lfgintexp} incorporates all interactions between foreground quasiparticles which arise in the low-energy effective description of the $\Cwt$ model on $\RM$.
However, it overlooks one further class of interactions. %
In addition to \Eref{eq:Lfgintexp}, the full Lagrangian~\eref{eq:I:L} %
contains terms proportional to expressions such as
\begin{equation}
\fgfield{\vp^\mu\vp_\mu}\bgfield{\h\h^*}\quad\textrm{and}\quad\fgfield{\vp^\mu\vp_\mu}\bgfield{\vp^\nu\vp_\nu}
\end{equation}
which involve both background and foreground fields, and which do \emph{not} vanish when averaged over length scales large compared with $\mc{L}_0$. %
Similar terms may also be constructed involving the spinor component of the pseudovacuum, or for other foreground species (though both processes are a little more complex when spinors are involved).
In the low-energy limit, the background field components in these expressions may be replaced by their mean field values [e.g.~\eref{eq:Exy}] to give effective mass terms. This mass arises from interactions between the foreground particles and the pseudovacuum, and exhibits an explicit dependency on the characteristic energy scale of the pseudovacuum, $\mc{E}_0$. This effect is somewhat analogous to the Higgs coupling of the Standard Model, where particles interact with the nonvanishing Higgs expectation value in the vacuum state (though no symmetry breaking is present in the current model, with the pseudovacuum being introduced by fiat).
Incorporating this effect, Lagrangian~\eref{eq:Lfgintexp} may be expanded to yield
\begin{align}
\begin{split}
{\mscr{L}}_\fg=&~\frac{1}{2}\fgfield{\varphi_{\mu}(\tilde{\underline{\triangle}}^{\mu\nu}\!\!\!-\eta^{\mu\nu}m^2_\vp)\varphi_{\nu}}+\fgfield{\h^*(\tilde{\underline{\square}}-m^2_\h)\h}\\
&+\rmi \fgfield{(\tilde{\bm{D}}^\mu\bar\psi)\,\tilde{\bmDslash}(\tilde{\bm{D}}_{\mu}\psi)}
\end{split}\label{eq:Lexp}\\
\tilde{\bm{D}}_\mu=&~\bm{D}_\mu+\frac{\rmi}{2} \tilde{\Upsilon}_{\mu}m_\psi\qquad
\tilde\Upsilon_\mu\tilde\Upsilon_\nu=\eta_{\mu\nu}\label{eq:bmDtilde}
\end{align}
with equations of motion
\begin{align}
\tilde{\underline\triangle}^{\mu\nu}\fgfield{\vp_\nu}&=m^2_\vp\fgfield{\vp^\mu}\label{eq:I:fgEOM1}\\
\bmDslash\,\bmDslash\,\fgfield{\psi_{\alpha}}&=m^2_\psi \fgfield{\psi_\alpha}\label{eq:I:fgEOM2}\\
\bmDslash\,\bmDslash\,\fgfield{\bar\psi^{\dot\alpha}}&=m^2_\psi\fgfield{\bar\psi^{\dot\alpha}}\label{eq:I:fgEOM3}\\
\tilde{\underline\square}\fgfield{\h}=m^2_\h\fgfield{\h}\quad&\quad\tilde{\underline\square}\fgfield{\h^*}=m^2_\h\fgfield{\h^*}.\label{eq:I:fgEOM5}
\end{align}
Derivation of \Eref{eq:bmDtilde} and calculation of the effective masses of the foreground species is non-trivial, so a more comprehensive treatment of this effect is deferred to \crefr{ch:fermion}{ch:detail}.

\subsubsection{Cauchy surfaces and initial conditions\label{sec:cauchy}}

Having identified the quasiparticles of the model, namely small foreground perturbations of the pseudovacuum state, and constructed their effective Lagrangian, it is now possible to identify a set of initial conditions on a Cauchy surface. %

First select an appropriate spacelike submanifold of $\RM$, for example $\RT$, to act as a Cauchy surface. Specification of the fields
\begin{equation}
\fgfield{\vp_\mu}\quad\fgfield{\psi^\alpha}\quad\fgfield{\bar\psi_{\dot\alpha}}\quad\fgfield{\h}\quad\fgfield{\h^*}\label{eq:fgfieldset}
\end{equation}
and their derivatives
\begin{equation}
\partial_\nu\fgfield{\vp_\mu}\quad\partial_\nu\fgfield{\psi^\alpha}\quad\partial_\nu\fgfield{\bar\psi_{\dot\alpha}}\quad\partial_\nu\fgfield{\h}\quad\partial_\nu\fgfield{\h^*}\label{eq:fgfieldderivs}
\end{equation}
now suffices to permit full reconstruction of the foreground fields over all of $\RM$, %
up to errors which are small [$\OO{\mc{E}/\mc{E}_0}$] when averaged over length scales greater than $\mc{L}_0$.
This reconstruction is limited to the foreground vector boson, spinor, and complex scalar boson fields only---the background fields obviously cannot be reconstructed from this data, and it also does not suffice to reconstruct $\fgfield{\vp}$ or any of the component fields $\{\vp_q\}$.
Nevertheless, %
provided these specified foreground fields, %
in the low-energy regime, are the objects of interest on $\RM$, %
then for this purpose the parameters~\erefr{eq:fgfieldset}{eq:fgfieldderivs} and Lagrangian~\eref{eq:Lexp} suffice. By construction, if a set of fundamental fields and their gradients on $\Cwt$ are pulled back to $\RT$ and foreground fields on $\RT$ are calculated from these initial conditions (or better yet, if these values are averaged over a time interval $\mc{T}$ which is short compared to the highest energy scale of the foreground fields but nevertheless long compared with $c^{-1}\mc{L}_0$), then the extrapolation to all of $\RM$ from these initial conditions yields foreground fields which are consistent (up to errors small over regions large compared with $\mc{L}_0$) with the direct pullbacks of the fundamental fields on $\Cwt$.

Note that the vector boson propagator term in $\mscr{L}_\fg$, %
which may be rewritten
\begin{equation}
\frac{1}{2}\fgfield{(\partial_\mu\varphi)\underline\triangle^{\mu\nu}(\partial_\nu\varphi)},\label{eq:LK1}
\end{equation}
resembles both the propagator of a free vector field $\fgfield{\varphi_{\mu}}$ and---up to an integration by parts---also a candidate propagator for the unitless product field $\fgfield{\varphi}$ on $\RM$. 
However, on extending the construction presented here from $\Cwt$ to $\Cw{2\N}$ (in \cref{ch:colour})%
, the scalar product field ceases to have a uniquely-defined derivative on a Minkowski submanifold for $\N>1$. In contrast, $\fgfield{\varphi_{\mu}}$ is replaced by a family of $\N^2$ vector fields (with $\fgfield{\varphi_{\mu}}$ being the single member of this family for $\N=1$), and each vector field in the family has a well-defined derivative on $\RM$. Interpreting term~\eref{eq:LK1} as the Lagrangian for vector field $\fgfield{\varphi_{\mu}}$ therefore corresponds to the $\N=1$ limit of a more general construction.

\subsection{Extension to higher energies\label{sec:pushlimits}}

Up to this point, foreground excitations have been assumed to have energies $\mc{E}\ll\mc{E}_0$. However, this constraint is in fact over-stringent. While the energy scale $\mc{E}_0$ has an important role to play in determining the bounds of the quasiparticle regime, it turns out that well-behaved foreground excitations with energies larger than $\mc{E}_0$ may also be sustained.

\subsubsection{Nature of foreground excitations\label{sec:naturefg}}

Over timescales large compared to $\mc{L}_0$, the presence of a foreground excitation in the $\varphi$ field must in principle be taken only as implying that some correlators are non-vanishing over appropriate length and timescales. If the quasiparticles were free particles, individual quanta could in principle be identified as foreground or background excitations. However, in the interacting quasiparticle model which arises from free fields on $\Cwt$, even if a foreground quasiparticle with appropriate wavefunction is explicitly introduced over a background of the pseudovacuum at some initial time $t$, subsequent scattering interactions between this foreground quasiparticle and quasiparticles in the pseudovacuum may result in any one---or more than one---of the involved quasiparticles ending up correlating with the original foreground field at some later time $t'$. The resulting correlations may be understood either as reflecting the transmission of the foreground excitation as a collective excitation of the quasiparticles present, or as reflecting that at time $t'$ some number of the quasiparticles may possibly be identified with the foreground quasiparticle, each with some specific probability amplitude. Either way, the propagation of a foreground excitation is observed to be a collective phenomenon.

This distinction is for the time being largely academic, and %
in this and subsequent chapters, %
individual quasiparticles will %
mostly be identified as foreground or background without explicitly articulating this detail. However, it is of importance when %
exploring interactions between high-energy foreground particles and the pseudovacuum in
\sref{sec:probelorentz} below, and when calculating the effect of the background fields on boson coupling strengths in \sref{sec:interactions}. %

\subsubsection{Probing Lorentz symmetry breaking\label{sec:probelorentz}}

To probe the Lorentz invariance of the pseudovacuum, consider a massive test quasiparticle which for definiteness is assumed to be a foreground perturbation of field $\vp_\mu$, with the energy scale of the perturbation being $\mc{E}$ in some rest frame $F$. A description of the quasiparticle's propagation in the presence of the pseudovacuum necessarily involves both the energy scale of the particle, $\mc{E}$, and that of the pseudovacuum, $\mc{E}_0$.

Since frame $F$ need not coincide with the isotropy frame, %
the mean magnitude of the energy component of the pseudovacuum will appear to the particle as $\gamma\mc{E}_0$ for some boost $\gamma$, with a mean momentum component $\beta\gamma\mc{E}_0$ homogeneously oriented in some direction (for sake of argument, $+x$). In the rest frame of the test particle, there is a pseudovacuum ``wind'', or equivalently, in the rest frame of the pseudovacuum, the particle experiences drag from propagating through the static pseudovacuum state. In both descriptions, momentum is transferred from the foreground particle to the pseudovacuum fields. However, with this momentum transfer come correlations with the original foreground field and its source, and these correlations will persist over length and time scales large compared with $\mc{L}_0$. Thus the perturbations to the pseudovacuum induced by this drag effect are themselves foreground in character, and this represents a process already discussed in \sref{sec:naturefg} where a ``foreground excitation'' is a collective property of multiple underlying scalar fields. This excitation may contain on- and off-shell excitations of multiple particle species, and in light of \sref{sec:4momflucs}, may be taken to represent the full dressed propagator of the foreground excitation.

\subsubsection{Extending the foreground energy regime\label{sec:extendfgE}}

In \sref{sec:quasi}, the adoption of an energy bound $\mc{E}_0$ for foreground particle excitations is motivated by the definition of $\mc{E}_0$ as the characteristic energy scale of the pseudovacuum. However, in practice foreground fields may be constructed with energies substantially exceeding $\mc{E}_0$ with very little consequence.

First, recognise that there is a limit on the spatial resolution of background particle wavefunctions. Although the field $\vp(x)$ is the product of an infinite number of unitless scalar fields $\vp_q$ and thus is of infinite order in $x$, over scales large compared with $\mc{L}_0$ these fields and their gradients are uncorrelated. The foreground field excitations are gradients of $\vp(x)$, and consequently receive multiple cancelling contributions. In essence, the wavefunction of the background field at a point $P$ is determined by the $N_0$ correlated fields with centres within a hypervolume of size ${\mc{L}_0}^4$ containing $P$. It is therefore reasonable to approximate $\vp(x)$ as being constructed from local patches of dimension $\mc{L}_0$ and being of order $\OO{x^{2N_0}}$, implying on average a maximum of $2N_0-1$ inflection points per distance $\mc{L}_0$ and thus bounding the energy of background field excitations from above by 
\begin{equation}
\mc{E}_\Omega:=\left(N_0-\frac{1}{2}\right)\mc{E}_0\qquad\mc{L}_\Omega=\mc{E}_\Omega^{-1}.\label{eq:I:EOmega}
\end{equation}
(This bound may fluctuate slightly if the centres of the unitless scalar fields on $\Cwt$ are not evenly distributed.)

It should be noted that this constraint does not prohibit the existence of foreground excitations at energies large compared with $\mc{E}_0$ or $\mc{E}_\Omega$---foreground excitations exhibit correlations outside the local region, and thus are able to recruit contributions from unitless scalar fields outside the local ${\mc{L}_0}^4$ hypervolume. In principle any number of fields may contribute to a foreground excitation within a given hypervolume, yielding a foreground field which is an arbitrary power of $x$. 
However, this must be reconciled with the observation in \sref{sec:operators} that the average number of unitless scalar fields actually contributing to a single-particle excitation at any given point is one, and thus becomes a commentary on a nonuniform distribution of the scalar fields' centres. 

In \sref{sec:sparseregime} the restriction to a single unitless scalar field was achieved by perturbing only one field per hypervolume ${\mc{L}_0}^4$, an approach which suffices for excitations with energies small compared with $\mc{E}_0$ whose wavefunction may be approximated by a quadratic in $x$ over the relevant hypervolume. However, for energies $\mc{E}>\mc{E}_0$ it is in general necessary to perturb more than one scalar field per hypervolume, and for $\mc{E}>\mc{E}_\Omega$ it is necessary to perturb scalar fields over a wider area in order to yield the requisite $x$ dependency within the target region. In each instance these perturbations must be chosen
such that, like the background fields, they cancel outside of the target region.
They should also predominantly mutually cancel within the target region, 
such that even though more than one unitless scalar field %
is perturbed, at any given point $P$ within the region the foreground field receives
uncancelled contributions from on average only one scalar field at a time. For $\mc{E}\geq\mc{E}_0$ or $\mc{E}\geq\mc{E}_\Omega$, which scalar field is participating changes over an average length- or time-scale $\mc{L}\leq\mc{L}_0$ or $\mc{L}\leq\mc{L}_\Omega$ respectively. Note that it is not necessary that the relevant scalar field be the one with its centre closest to $P$, or indeed even within $\mc{L}_0$ of $P$ given the long-range nature of foreground field correlations.

Note that it is possible to apply this construction even at energy scales small compared with $\mc{E}_0$, though it is in general unnecessary to do so.

\subsubsection{Meaning of \prm{\mc{E}_\Omega}\label{sec:meaningofEOmega}}

Regarding the existence of a bound of $\OO{\mc{E}_\Omega}$ on the instantaneous (as opposed to the mean-square) energy of the background fields, note that in \sref{sec:operators} the foreground fields are seen to model the fields of a quantum field theory,
and in \sref{sec:4momflucs} the local 4-momentum of the pseudovacuum is identified as being available to model 
\begin{itemize}
\item quantum fluctuations in the 4-momentum of these particles, and 
\item quantum fluctuations in the vacuum itself.
\end{itemize}
The existence of an upper bound of order $\ILO{\mc{E}_\Omega}$ therefore represents the existence of a physical UV cutoff $\Lambda$ for the magnitude of these fluctuations, such that the model on $\Cwt$ is unable to accurately model quantum processes involving energy fluctuations of order $\mc{E}_\Omega$ or higher, whereas these would usually be permitted %
over scales small compared with $\mc{L}_\Omega$ by %
the Schr\"odinger time/energy uncertainty relation. However, the finite nature of the UV cutoff does ensure that the $\Cwt$ model is not troubled by unrenormalisable infinities.

To determine an expression for the cutoff energy in terms of $\mc{E}_\Omega$, consider that $\mc{E}_\Omega$ is a reparameterisation of a timescale $c^{-1}\mc{L}_\Omega$, and that in the pseudovacuum rest frame, the local product field is dominated by a consistent set of unitless scalar fields over this timescale and thus $c^{-1}\mc{L}_\Omega$ is the mean time over which pseudovacuum fields are correlated with consistent sign. [This is distinct from scale $c^{-1}\mc{L}_0$, over which the mean values of the correlators are consistent but the participating fields, and thus local values and signs, may vary subject to this overall constraint. The correlators are, however, approximated as being of consistent sign under the window approximation~\eref{eq:window}.]
By comparison, for a foreground field of definite energy $\mc{E}$ the scale over which consistency of sign of correlators is achieved corresponds to a half-period. An energy scale analogous to that of the pseudovacuum would associate such an excitation with an energy of $2\mc{E}$, and thus it is necessary to introduce factors of~2 or~$\frac{1}{2}$ when comparing foreground and pseudovacuum energy scales. A pseudovacuum of energy $\mc{E}_\Omega$ may thus transiently emulate foreground excitations of energies up to a UV cutoff of $\frac{1}{2}\mc{E}_\Omega$.

\subsubsection{Probing Lorentz symmetry breaking---revisited\label{sec:ProbeOmegaScale}}

Having identified $\frac{1}{2}\mc{E}_\Omega$ with a finite ultraviolet cutoff, the nature of the Lorentz symmetry breaking in the $\Cwt$ model becomes apparent: In rest frames which do not coincide with the isotropy frame of the pseudovacuum, the value of the UV energy cutoff is augmented, and the UV momentum cutoff becomes anisotropic. Effects of these properties are likely to be subtle, but to become increasingly pronounced as energy scales approach $\ILO{\mc{E}_\Omega}$.

Of particular note, virtual processes requiring the borrowing of energies larger than $\frac{1}{2}\mc{E}_\Omega$ have limited support in the $\Cwtn$ model series. For $\mc{E}_\Omega$ large, this is unlikely to be a major concern. However, in models where there exist particles with rest energies larger than $\frac{1}{2}\mc{E}_\Omega$ (see e.g.~\sref{sec:heavyWZH}), %
virtual processes involving these particles will be heavily suppressed. Whether this has an impact on the utility of the $\Cwtn$ model will depend on whether there are processes of interest at smaller energy scales, for which the contributions of these particles are anyway likely to be negligible, or whether all physics of interest in the model occurs at energy scale $\frac{1}{2}\mc{E}_\Omega$ or above, in which context the correspondence between the $\Cwtn$ model and the QFT will break down.

Having thus established the role of $\mc{E}_\Omega$ as the energy scale around which the behaviour of the foreground quasiparticles begins to significantly deviate from that of the quantum field theory they represent, it is useful to review what role $\mc{E}_0$ now plays. Although $\mc{E}_0$ no longer bounds either the energies of the foreground fields or the instantaneous energies of the background fields (see \sref{sec:4momflucs}), it remains a key descriptor of the background fields. An essential property of the background fields is that when averaged over a sufficiently large region or duration, their contribution to many correlators disappears. The scale for this to take place continues to be $\mc{L}_0={\mc{E}_0}^{-1}$.

When studying the behaviour of high-energy foreground excitations having energies large compared with $\ILO{\mc{E}_0}$, with associated length scales %
small compared with $\ILO{\mc{L}_0}$, it is tempting to assume that measurements involving these particles will %
be able to detect the pseudovacuum. However, a realistic description of measuring apparatus for high-energy processes typically involves detectors with considerable spatial extent, and measurements collected over an extended period of time or a large number of repetitions (cumulatively equivalent to an extended duration). Thus the correlation length $\mc{L}_c$ in \sref{sec:fgbg} satisfies $\mc{L}_c\gg\mc{L}$, and while particles with energies large compared with $\ILO{\mc{E}_0}$ could in principle be used to probe the pseudovacuum,\footnote{More precisely, large compared with \prm{\frac{1}{2}\mc{E}_0} as per \psref{sec:meaningofEOmega}. The factor of \prm{\frac{1}{2}} will frequently be omitted when employing the comparators ``much greater than'' or ``much less than'', \prm{\gg} or \prm{\ll}.} this is unlikely to happen without deliberate intent.

\subsubsection{Goldstone's theorem\label{sec:goldstone}}

It is noted that the pseudovacuum has a preferred isotropy frame and therefore breaks the $\SLTC$ symmetry of the emergent Lagrangian~\eref{eq:Lexp}. It is also noted that the $\Cwt$ model has an emergent complex spin-0 boson, $\fgfield{\h}$, which is intrinsically massless as per \Eref{eq:fgEOM5massless} (in keeping with the expectations of Goldstone theorem), and then acquires mass through a Higgs-like mechanism as a direct consequence of the broken symmetry of the pseudovacuum (\sref{sec:masses}). The $\Cwt$ model therefore respects Goldstone's theorem up to the same exemption which permits the Higgs boson of the Standard Model to acquire mass through coupling to a non-zero vacuum expectation value. In subsequent chapters analogous behaviour is seen in all members of the $\Cwtn$ series, and this mass mechanism is explored in much greater detail in \crefs{ch:boson}{ch:detail}. %

\subsection{A simplified construction for the pseudovacuum\label{sec:entropic}}

All fields in this model are decomposed into foreground and background components, with the understanding that the foreground components, characterised by an energy scale $\mc{E}$ and length scale $\mc{L}$, decouple from the background components (save for the interactions giving rise to mass terms) when averaged over regions or durations characterised by $\mc{L}\gg\mc{L}_0$, noting that for this to hold, it is sufficient but not necessary that $\mc{E}\ll\mc{E}_0$. In practice no such limit is ever fully realised for $\mc{L}>0$ and $\mc{L}_0$ finite, so there exists some residual coupling between foreground and background fields which grows weaker for larger $\mc{L}$ (and, for $\mc{E}\ll\mc{E}_0$, smaller $\mc{E}$).

Suppose that there exist initial conditions with decomposition of the fields into foreground and background components, but the foreground components do not satisfy $\mc{E}\ll\mc{E}_0$. Coupling of foreground and background fields will initially be strong, but the entropy of the foreground fields will increase over time.
In effect, as entropy of the fields increases, high-entropy foreground modes will increasingly approximate the behaviour of background modes. For most initial configurations they may subsequently be rewritten as an additional contribution to the pseudovacuum and a (smaller) perturbation around the pseudovacuum state, so the energy remaining in the foreground modes decreases over time.
Consequently there exists a time direction in which the energy of the foreground fields leaks into the background fields.

This does not always guarantee the existence of a regime with arbitrarily small ratio $\mc{E}/\mc{E}_0$ after a sufficiently long period of time---for example, if considering the behaviour of fields over time on a disc $D\subset\RT\subset\RM$ then high-energy excitations may propagate into disc $D$ from outside its boundaries. Nevertheless, if disc $D$ is sufficiently large then an increase in entropy within the inner regions of $D$ may be noticed before such excitations reach the centre.

Given these considerations, it may be reasonable to set up a manifold $\RT$ with random initial conditions and evolve this forward in time, identifying local regions of higher entropy in which $\mc{E}\ll\mc{E}_0$ is more effectively satisfied. In these regions, the decoupling of foreground and background fields will be more complete. This simplifies the construction of the model, in that instead of specifying the existence of a high-energy pseudovacuum and low-energy perturbations, it is enough to specify an arbitrary field configuration on $\Cwt$, select an arbitrary submanifold $\RT\subset\RM$, and evolve forward in time until a region of $\RM$ (and hence of $\Cwt$) can be identified in which the foreground/background decomposition holds to the desired degree and the effective description given in this chapter may then be applied.

\subsection{From classical to emulated semiclassical\label{sec:semiclassical}}

\subsubsection{Requirements}

A semiclassical theory is one which exhibits some properties of a classical theory and some more commonly associated with a quantum theory; typically, %
enough properties of the quantum theory %
to permit %
calculation of scattering amplitudes within some domain of validity. It may therefore be thought of as a form of analogue model.
The current Section demonstrates %
that the model on $\Cwt$ %
incorporates all of the necessary features for the foreground fields to
implement a numerically useful 
semiclassical theory in the low-energy regime.

To reproduce scattering amplitudes computed using the method of Feynman diagrams it suffices that the $\Cwt$ model must exhibit
\begin{itemize}
\item identified classical counterparts to the observable quantum field operator(s) and wavefunction(s), 
with appropriate interaction terms,
\item bosonic or fermionic excitation statistics as appropriate to each exhibited field species, and
\item energy/momentum fluctuations governed by the Heisenberg uncertainty principle.
\end{itemize}
While the calculation of
Green functions through variation of %
the generating functional
\begin{equation}
\Z=e^{\rmi S}\qquad S=\!\int\!\rmd^4x\,\mscr{L}%
\label{eq:cplxZ}
\end{equation}
is also often considered a hallmark
of semiclassical field theories, %
once the above features are demonstrated
this construction follows directly %
from formulation of an action principle to describe the evolution of the foreground fields
as per Dirac~\cite{dirac1933} and Schwinger~\cite{schwinger1951,schwinger1953,schwinger1960,schwinger1966}%
. %

Demonstration of these properties for the decoupled foreground fields observed in the low-energy ($\mc{E}\ll\mc{E}_0$) regime establishes these foreground fields as yielding a semiclassical approximation to the equivalent QFT, %
and thus shows
that classical scalar fields on $\Cwt$ may %
act as an analogue model for a simple QFT with vector, complex scalar, and spinor fields. %
Extension to $\mc{E}_0<\mc{E}\ll\mc{E}_\Omega$ then follows as per the discussion of \sref{sec:pushlimits}.

\subsubsection{Operators\label{sec:operators}}

\paragraph[Background fields]{Background fields:\label{sec:bgfieldsequiv}}
First consider just the vector boson field and the pseudovacuum state, and recall that the classical field $\varphi_\mu(x)$ is constructed by acting the operator $-\rmi\partial_\mu$ on field $\varphi(x)$ \eref{eq:defvarphimu}. This field, in turn, is composed from the product of an infinite number of fundamental fields~$\varphi_q(x)$~\eref{eq:defvp}, and thus when computing a potentially nonvanishing unitful correlator on the background $\vp_\mu$ fields, e.g.~$\la\bgfield{\vp^\mu(x)\vp_\mu(y)}\ra$~\eref{eq:Exy}, this is in general a sum over an infinite number of terms. As discussed in \sref{sec:QL} this expression is dominated by contributions from fields with centres close to $x$ and $y$, with %
longer-range contributions mostly cancelling to leave a finite result. 

When $x$ is close to $y$, approximation~\eref{eq:window} yields some insight: under this approximation there are on average exactly $N_0$ fundamental scalar fields $\vp_q$ which each contribute equally to correlator~\eref{eq:Exy}. Compare this %
with the %
more realistic model where $\bm{f}(x-y)$ is a Gaussian distribution, %
with
an infinite number of fields contributing. Each field is then weighted by $\bm{f}(x-y)$, %
and in the limit $y\rightarrow x$ the sum over these weights %
tends on average towards $N_0$.

Recall that any realistic measurement of a correlator over distance $\mc{L}_c$ has a finite resolution $\mc{L}_p$ assumed to satisfy $\mc{L}_p\ll\mc{L}_c$.
In the limit of sufficiently low probe energy it is convenient to adopt approximation~\eref{eq:window} and represent short-range background correlators as arising from $N_0$ independent contributions.

In this approximation,
recognise that the operator $-\rmi\partial_\mu$ used in construction of the classical field $\varphi_\mu$
has an action on $\vp(x)$ of %
\begin{align}
\begin{split}
\vp_\mu(x)&=-\rmi\partial_\mu\vp(x)=-\frac{\rmi\partial_\mu\vp(x)}{\vp(x)}\,\vp(x) \\
&=-\sum_{q=1}^\infty\sum_{q'=1}^\infty\frac{\rmi\bar\partial\vp_q(x)\bsmm\partial\vp_{q'}(x)}{\vp_q(x)\vp_{q'}(x)}\,\vp(x),
\end{split}\label{eq:Nterms}
\end{align}
where no special treatment is required for $q=q'$ due to linear independence of the derivative operators $\partial$ and $\bar\partial$.

In the absence of foreground fields, 
and applying approximation~\eref{eq:window}, the contribution of a given instance of $\vp_\mu$ to a non-vanishing instance of correlator~\eref{eq:Exy} is accounted for entirely by $N_0$ of the fundamental scalar fields, such that \Eref{eq:Exy} is invariant under the substitution
\begin{equation}
\vp_\mu(x)\longrightarrow-\sum_{q=1}^{N_0}\sum_{q'=1}^{N_0}\frac{\rmi\bar\partial\vp_q(x)\bsmm\partial\vp_{q'}(x)}{\vp_q(x)\vp_{q'}(x)}\,\vp(x)\label{eq:vpsub0}
\end{equation}
so long as the $N_0$ participating fields are chosen appropriately.
This substitution corresponds to recognising that at short range, $\bm{f}(x-y)$ is~1 for $N_0$ of these fields and~0 for the rest.
Following the approach used in \sref{sec:QL}, this is a near field expression which will vanish on averaging over a sufficiently large area; this corresponds with satisfaction of \Eref{eq:<>0}. However, the magnitude of \Eref{eq:vpsub0} is nonvanishing by \Eref{eq:E0}.
Under substitution~\eref{eq:vpsub0}, correlator~\eref{eq:Exy} becomes
\begin{equation}
\begin{split}
\Bigg<&\vp(x)\sum_{q=1}^{N_0}\sum_{q'=1}^{N_0}\left[\frac{\rmi\bar\partial\vp_q(x)\bsm\partial\vp_{q'}(x)}{\vp_q(x)\vp_{q'}(x)}\right]\\
&\times\vp(y)\sum_{r=1}^{N_0}\sum_{r'=1}^{N_0}\left[\frac{\rmi\bar\partial\vp_r(y)\bsmm\partial\vp_{r'}(y)}{\vp_r(y)\vp_{r'}(y)}\right]
\Bigg>_{\!\!\mc{L}_p}
\end{split}
\label{eq:vpsubsep}
\end{equation}
and for $N_0$ sufficiently large this reduces to
\begin{equation}
\begin{split}
\Bigg<\vp(x)\sum_{q=1}^{N_0}\sum_{q'=1}^{N_0}\bigg[&\frac{\rmi\bar\partial\vp_q(x)\bsm\partial\vp_{q'}(x)}{\vp_q(x)\vp_{q'}(x)}\frac{\rmi\bar\partial\vp_{q'}(y)\bsmm\partial\vp_{q}(y)}{\vp_q(y)\vp_{q'}(y)}\bigg]
\vp(y)\Bigg>_{\!\!\mc{L}_p}
\end{split}
\end{equation}
via \Erefr{eq:<>0}{eq:E0}.
Replacing each of the ${N_0}^2$ terms in this sum %
with the average contribution yields
\begin{align}
\begin{split}
\delta&^4_{\mc{L}_0}(x-y)\,{\mc{E}_0}^2\\
&=-{N_0}^2\Bigg<\vp(x)\bigg[\frac{\rmi\bar\partial\vp_q(x)\bsm\partial\vp_{q'}(x)}{\vp_q(x)\vp_{q'}(x)}%
\frac{\rmi\bar\partial\vp_{q'}(y)\bsmm\partial\vp_{q}(y)}{\vp_q(y)\vp_{q'}(y)}\bigg]
\vp(y)\Bigg>_{\!\!\mc{L}_p,q}\\
&=:\delta^4_{\mc{L}_0}(x-y)\,{N_0}^2{\omega_0}^2,
\end{split}\label{eq:N0terms}
\end{align}
where subscript ${\mc{L}}_p,q$ indicates that the mean is both over the probe region and over the different participating fundamental fields $q$, and this equation defines $\omega_0$.

Although the derivatives of individual background fields $-\rmi\partial_\mu\vp_q$ %
are not normalised, a short-range correlator of form~\eref{eq:Exy} receives on average ${N_0}^2$ contributions of average value ${\omega_0}^2$. Collectively, this result matches that obtained on evaluating a quantum field theory correlator $\la\hat\vp^\mu(x)\hat\vp_\mu(y)\ra$ in the presence of $N_0$ independent excitations of quantum field $\hat\vp_\mu$ of average energy $\omega_0$. %

In what follows, 
the analogy between independent fields $-\rmi\partial_\mu\vp_q$ and independent excitations of a quantised field $\hat\vp_\mu$ is extended to the normalised foreground fields, %
for which it holds field by field as well as collectively. %
Furthermore, at the lower energy scales associated with foreground fields, only a collective description of the background fields is required. %
Thus %
when studying the behaviour of the foreground fields, the background fields' contributions to boson correlators %
may be modelled by (or may be thought of as modelling) %
a sea of normalised excitations having mean energy $\omega_0$ and %
correlators %
that %
adequately approximate the behaviour of distribution $\bm{f}(x-y)$ in \Eref{eq:Exy}. %

Although \Eref{eq:N0terms} only demonstrates this analogy under approximation~\eref{eq:window},
it extends directly to more general distributions through
the introduction of varying correlation weights $\bm{f}(x-y)$%
, and thus the analogy between fields and quanta may be made regardless of whether approximation~\eref{eq:window} is applied or not. %
Thus $\la\vp^\mu(x)\vp_\mu(y)\ra_{\mc{L}_p}$ invites identification as a semiclassical approximation to an expectation value $\la\hat\vp^\mu(x)\hat\vp_\mu(y)\ra$, with the latter receiving contributions from an average of $N_0$ %
quanta in the limit $y\rightarrow x$ regardless of distribution. This average is to be understood probabilistically if $\bm{f}(x-y)$ varies smoothly, for example if it is a Gaussian. %

This analogy may be made more illuminating by 
identifying $\vp(x)$ as the (non-normalised) analogue of the vacuum state~$|0\ra$, i.e.~at $x$, %
\begin{equation}
\begin{split}
\la0|0\ra &\equiv %
\mc{N}\la\varphi(x)\varphi(x)\ra_{\mc{L}_p}
\end{split}\label{eq:defqpv}
\end{equation}
where $\mc{N}$ is a normalisation factor.
Note that 
separability follows from maximisation of pseudovacuum entropy,
\begin{equation}
\la\varphi(x)\varphi(y)\ra_{\mc{L}_p} = \la\varphi(x)\ra_{\mc{L}_p}\,\la\varphi(y)\ra_{\mc{L}_p}, %
\end{equation}
and that homogeneity implies co-ordinate independence of $\la\vp(x)\ra_{\mc{L}_p}$, supporting definition of the convenient constant 
\begin{equation}
f={\la\vp(x)\ra_{\mc{L}}}^{-1}\tag{\ref{eq:deff}}
\end{equation}
for any $\mc{L}\gg\mc{L}_0$. Next, identify
\begin{equation}
\hat\vp_\mu(x)\equiv-{\rmi}{f^{-1}}\partial_{\mu|x}\label{eq:vpop}
\end{equation}
as being the operator analogue of $\vp_\mu(x)$ in regime $\mc{L}_p\gg\mc{L}_0$. In \Eref{eq:vpop}, 
$_{|x}$ indicates that the derivative is applied to the wavefunction analogue at $x$, i.e.{} %
\begin{equation}
\hat\vp_\mu(x)|0\ra\equiv -\rmi f^{-1}\partial_\mu\vp(x). \label{eq:vpop2}
\end{equation}
An operator may also act to the left, and consistency requires that when it does so, this corresponds to the hermitian conjugate of this expression,
\begin{equation}
\la 0|\hat\vp_\mu(x)\equiv \rmi f^{-1}\partial_\mu\vp(x). \label{eq:vpop3}
\end{equation}
Taking care to explicitly normalise with respect to the vacuum in order to avoid factors of $\mc{N}$ from \Eref{eq:defqpv}, it follows immediately that
\begin{equation}
\begin{split}
\la\hat\vp^\mu(x)\hat\vp_\mu(y)\ra %
&=\frac{\la 0|\hat\vp^\mu(x) \hat\vp_\mu(y)|0\ra}{\la 0|0\ra}\\
&=\la\vp^\mu(x)\vp_\mu(y)\ra_{\mc{L}_p}.
\end{split}
\end{equation}
Note that the related low-energy substitution
\begin{align}
\partial_{\mu|x}&\longrightarrow \rmi f\varphi_\mu(x)\label{eq:lesub1}
\end{align}
is %
valid only where the derivative operator acts on $\vp(x)$ %
within $\la\,\cdot\,\ra_{\mc{L}_p}$, or can be brought to do so using integration by parts.
Thus, for example,
\begin{align}
\begin{split}
\la\partial_\mu\vp(y)\ldots\partial_\nu\partial_\rho\vp(x)\ra_{\mc{L}_p}
&\rightarrow \rmi f\la\partial_\mu\vp(y)\ldots\partial_\nu\vp_\rho(x){\vp(x)}%
\ra_{\mc{L}_p}\\
&\!\!\!\!\!\stackrel{\mc{L}_p\gg\mc{L}_0}{\approx}\!\!\!\!- \la\partial_\mu\vp(y)\ldots\partial_\nu\vp_\rho(x)\ra_{\mc{L}_p}\\&\not\rightarrow -\rmi f\la\partial_\mu\vp(y)\ldots\vp_\nu(x)\,\vp_\rho(x)\ra_{\mc{L}_p}.
\end{split}
\label{eq:inversesub}
\end{align}
Instead, the effect in the penultimate line of the derivative operator $\partial_\nu$ %
is to compute the space-time derivatives of the components of $\vp_\rho(x)$. %
These requirements are exactly the same as for the corresponding
classical operator identity~\eref{eq:opsubvp}. 

In keeping with the interchangeability of particle species implicit in their common construction from spinor derivatives and manifest in \Erefs{eq:Dbold}{eq:I:L1}, similar arguments may be pursued for scalar bosons and spinor pairs, with appropriate caution around spinor statistics, yielding operator analogues
\begin{align}
\hat{\h}(x)\equiv f^{-1}\partial_{\rmU|x}\qquad\hat{\h}^*(x)\equiv f^{-1}\bar{\partial}_{\rmU|x}\\
\hat\psi_\alpha(x)\equiv f^{-1}\partial_{\alpha|x}\qquad\hat{\bar{\psi}}_{\dot{\alpha}}(x)\equiv f^{-1}\bar{\partial}_{\dot{\alpha}|x}
\end{align} 
where the right hand sides are understood to inhabit $\Cwt$ and act at the point $\bthetas$ associated to a point $x$ by mapping $\G$ (\sref{sec:defpullback})%
.

\paragraph[Foreground fields]{Foreground fields:\label{sec:fgwavf}}

Now reintroduce foreground perturbations of the $\varphi_\mu(x)$ field as per \sref{sec:quasi}. 
As previously noted, a near-arbitrary foreground field of energy $\mc{E}\ll\mc{E}_0$ may be constructed by introducing customised perturbations of one fundamental field per hypervolume ${\mc{L}_0}^4$ with the density of centres of the perturbed fields being macroscopically homogeneous (when viewed at length scales $\mc{L}\gg\mc{L}_0$). 
This density is %
consistent with the decision in \sref{sec:operators} %
to identify the number of fundamental fields making a non-cancelled contribution to a correlator with the number of quanta %
present in the equivalent quantum field theory. 
Additional quanta may be represented by perturbing fundamental fields with more than one centre per hypervolume ${\mc{L}_0}^4$, and the nonvanishing derivatives of the perturbation determine the species being represented (vector boson, scalar boson, or spinor).\footnote{Note that under approximation~\peref{eq:window} a specific grid of hypercubes must be chosen, with exactly one perturbed field per foreground quantum having a centre in each hypercube of the grid. For Gaussian correlators this requirement need only hold on average, but comes at the expense of increasing uncertainty over how many quanta are in a given region as the probe scale approaches \protect{$\mc{L}_0$}.}

At first glance the limitation on density of perturbations appears to introduce a Nyquist-like ambiguity over the energy of a foreground excitation \cite{whittaker1915,nyquist1928,shannon1949}, but this is resolved as in \sref{sec:extendfgE} by permitting a foreground field to involve more than one fundamental field per hypervolume provided all the fields involved in the excitation undergo a net cancellation such that on average only one yields a detectable contribution to the foreground field at any given point. On account of these cancellations between the different perturbations, the domains in which individual fundamental fields contribute to the foreground field may be smaller than ${\mc{L}_0}^4$ apiece.

It is %
worth noting that even with distribution of excitations across multiple fundamental fields, in the analogue model the average number of foreground excitations which may occupy a volume ${\mc{L}_0}^4$ is bounded above by $N_0$, which must therefore be sufficiently large.

\paragraph[Multi-point correlators of foreground fields]{Multi-point correlators of foreground fields:}
Regarding the evaluation of operator expressions, it is also illuminating to consider an equal-time four-point correlator having the form
\begin{equation}
\la\hat\vp^\mu(w)\hat\vp_\mu(x)\hat\vp^\nu(y)\hat\vp_\nu(z)\ra\label{eq:4expec}
\end{equation}
where $w$, $x$, $y$, and $z$ are all distinct.
The equal-time commutation relationships of the bosonic field operators,
\begin{equation}
\left[\hat\vp_\mu(\mbf{x},t),\hat\vp_\nu(\mbf{y},t)\right]=0\quad\forall\quad\mu,\nu,
\end{equation}
imply that ordering of the operators within correlator~\eref{eq:4expec} is irrelevant, and hence all operators may in principle be substituted as if acting directly on the vacuum state to either the left or the right. However, their action at four different co-ordinates necessitates the simultaneous application of \Erefs{eq:vpop}{eq:lesub1} to directly replace $\hat\vp_\mu(x)\hat\vp^\nu(y)$ in \Eref{eq:4expec} with $\vp_\mu(x)\vp^\nu(y)$. The end result is to identify
\begin{equation}
\begin{split}
\la\hat\vp^\mu(w)\hat\vp_\mu(x)\hat\vp^\nu(y)\hat\vp_\nu(z)\ra&=\frac{\la 0|\hat\vp^\mu(w)\hat\vp_\mu(x)\hat\vp^\nu(y)\hat\vp_\nu(z)|0\ra}{\la 0|0\ra}\\
&\equiv \la\vp^\mu(w)\vp_\mu(x)\vp^\nu(y)\vp_\nu(z)\ra_{\mc{L}_p}.
\end{split}
\end{equation}

\subsubsection{Wavefunctions\label{sec:wavefunctions}}

Following \rcite{ryder1996} the usual bosonic ladder operators may be defined in terms of the field operator $\hat\vp_\mu(x)$ and the plane wave solutions to the equations of motion $f_k(x)$, 
\begin{align}
f_k(x)&:=\frac{e^{-\rmi kx}}{\left[(2\pi)^3\,2k_0\right]^\frac{1}{2}}\\
A(t)\!\stackrel{\leftrightarrow}{\partial}_0\!B(t)&:=A(t)\partial_0B(t)-\partial_0A(t)B(t)\\
\hat a_\mu(k)&:=\int\!\!{\rmd^3x}\left[(2\pi)^3\,2k_0\right]^\frac{1}{2} f^*_k(x)\,\rmi\!\stackrel{\leftrightarrow}{\partial}_0\!\hat\vp_\mu(x)\\
\hat a^\dagger_\mu(k)&:=\int\!\!{\rmd^3x}\left[(2\pi)^3\,2k_0\right]^\frac{1}{2} \hat\vp_\mu(x)\,\rmi\!\stackrel{\leftrightarrow}{\partial}_0\!f_k(x),
\end{align}
such that %
\begin{equation}
\hat\vp_\mu(x)=\int\!\!\frac{\rmd^3k}{(2\pi)^3\, 2k_0}\left[\hat a_\mu(k)e^{-\rmi kx}+\hat a^\dagger_\mu(k)e^{\rmi kx}\right].
\end{equation}
When a creation operator acts to the right on the vacuum state, this permits the identifications
\begin{align}
\hat a^\dagger_\mu(k)&\equiv \int\!\!\rmd^3x\,e^{-\rmi kx}\,f^{-1}\! \stackrel{\leftrightarrow}{\partial}_0\! \partial_{\mu|x}\\
\hat a^\dagger_\mu(k)|0\ra&\rightarrow \int\!\!\rmd^3x\,e^{-\rmi kx}\,f^{-1}\!\stackrel{\leftrightarrow}{\partial}_0\! \vp_{\mu}(x)\label{eq:defadagger}
\end{align}
which may be verified by substituting \Eref{eq:defadagger} into the inverse Fourier transform (with annihilator terms eliminated)
\begin{align}
\hat\vp(x)|0\ra&=\!\int\!\!\frac{\rmd^3k}{(2\pi)^3\,2k_0}\,\hat a^\dagger_\mu(k)e^{\rmi kx}|0\ra,\label{eq:toreduce}
\end{align}
and recognising that if the resulting expression is nonvanishing in the context of an expectation value $\la 0|\ldots\hat\vp(x)|0\ra$ then this permits the replacement $\!\stackrel{\leftrightarrow}{\partial}_0\rightarrow 2\partial_0$,
thus confirming that \Eref{eq:toreduce} then reduces to $-\rmi f^{-1}\partial_\mu\vp(x)$ as per \Eref{eq:vpop2}.
Identifying
\begin{equation}
|p_\mu\ra = \hat a^\dagger_\mu(p)|0\ra,\label{eq:defpmu}
\end{equation}
the single-particle wavefunction for an excitation of momentum $p_\mu$ is then given by the usual expression (again with explicit normalisation with respect to the vacuum state)
\begin{equation}
\psi(x) = \frac{\la 0|\hat\vp^\mu(x)|p_\mu\ra}{\la 0|0\ra}.
\end{equation}
Applying \Ereft{eq:defqpv}, \eref{eq:vpop3}, \eref{eq:defadagger}, and \eref{eq:defpmu} yields
\begin{equation}
\psi(x) = e^{-\rmi px}
\end{equation}
as expected. (Wavefunction normalised as per \rcite{ryder1996}.) The tools of \sref{sec:operators} permit similar arguments to be pursued for spinor and scalar fields.

\subsubsection{Particle statistics}

Recognising that the analogue to the quantum field operator is constructed from the classical derivative operator, it follows that the operator $\hat\vp_\mu$ automatically inherits the commutation properties of the classical derivative operator $\partial_\mu$. Similarly, the spinor field operators $\hat{\psi}%
$ and hermitian conjugate %
inherit the anticommuting behaviour of the classical operator $\partial^\alpha$ and conjugate $\bar{\partial}_{\dot\alpha}$, while the scalar bosons inherit commuting behaviour from $\partial_\rmU$ and $\bar\partial_\rmU$. Thus the emergent foreground fields of analogue models constructed using the $\Cwt$ construction shown here (and, it will be seen, its $\Cw{2\N}$ generalisations) obey the spin-statistics theorem, consistent with the quantum fields they represent.

\subsubsection{Energy/momentum fluctuations\label{sec:4momflucs}}

The next semiclassical feature exhibited by the fields $\varphi_\mu$ is that of fluctuations in energy and momentum. To appreciate these fluctuations, recognise that the effective decoupling of foreground and background field components---e.g.~(for free vector bosons)
\begin{equation}
\begin{split}
\mscr{L}_{\vp,m}=&~\bgfield{\varphi_{\mu}(\triangle^{\mu\nu}-\eta^{\mu\nu}m^2_\vp)\varphi_{\nu}}\\
&~+\fgfield{\varphi_{\mu}(\triangle^{\mu\nu}-\eta^{\mu\nu}m^2_\vp)\varphi_{\nu}}
\end{split}
\end{equation}
---holds only on average over a region large compared with $\mc{L}_0$. Furthermore, even the total product field Lagrangian~\eref{eq:I:L}
only holds approximately when averaged over such a region. 
However, %
the fields surrounding any given point $x$ may then be decomposed into a portion for which the equations of motion of the product fields~\erefr{eq:I:fgEOM1}{eq:I:fgEOM5}
do hold both at $x$ and on average,
and a portion for which they do not hold at $x$, and vanish on average. 
If this latter portion is viewed as a perturbation, it may involve fundamental fields $\varphi_q$ which carry foreground excitations as per \sref{sec:operators}, %
and also fundamental fields which do not otherwise carry foreground excitations.

A perturbation in a fundamental field $\varphi_q$ which already contributes to a foreground excitation at $x$ may be understood as modifying the local value of that non-pseudovacuum excitation, with consequences for its energy and momentum. %
Perturbations in fields which do not carry foreground excitations %
are akin to the construction of transient excitations with borrowed energy, corresponding to virtual particles. 
By \Eref%
{eq:E0}
the energy and momentum scales associated with these perturbations range from at least 
$\ILO{-\mc{E}_0}$ to $\ILO{\mc{E}_0}$. 
The finite density of field centres across $\RM$
means that $\varphi_\mu(x)$ cannot support step functions or their fourier transforms, so %
the transition between positive and negative energies cannot be abrupt; consequently, fluctuations must %
take on all values %
of energy or momentum in the range %
$[-\frac{1}{2}\mc{E}_0,\frac{1}{2}\mc{E}_0]$ where the factor of $\frac{1}{2}$ is as per \sref{sec:meaningofEOmega}. Further, by the extension discussed in \sref{sec:extendfgE} these values may potentially range as widely as $[-\frac{1}{2}\mc{E}_\Omega,\frac{1}{2}\mc{E}_\Omega]$, and the maximally entropic nature of the pseudovacuum implies that all such excitation energies/momenta will indeed appear, with distributions appropriate to a thermal state insofar as this can be supported consistent with the upper bound on energy discussed in \srefs{sec:extendfgE}{sec:meaningofEOmega}.
These fluctuations have lifespans in the range $(2c)^{-1}[-\mc{L}_\Omega,\mc{L}_\Omega]$ in keeping with the Heisenberg uncertainty principle. 
These fluctuations %
belong to the background fields, but %
the involvement of a range of energies means that %
approximation~\eref{eq:window} may not %
be incautiously applied.

Note that this argument applies identically to spinor and scalar fields. Also note that when the foreground and background fields are taken to include the fluctuations and virtual excitations described above, total 4-momentum is then conserved pointwise on $\RM$ as well as on average. This ensures conservation of 4-momentum at the vertices of Feynman diagrams in interactions on any scale. Over any length scale large compared with $\mc{L}_0$, foreground particles may freely borrow energy and momentum up to scale $\frac{1}{2}\mc{E}_\Omega$
due to field fluctuations,
but this vertex-by-vertex conservation of 4-momentum indicates that such a borrowing comes at the expected cost of creating the corresponding (often virtual) foreground antiparticle with average lifespan naturally limited by the Heisenberg time/energy uncertainty relationship. %

\subsubsection{Normalisation of the generating functional\label{sec:normWrtBgFields}}

When %
evaluating correlators describing the behaviour of foreground fields in the presence of a background thermal pseudovacuum, if only the behaviour of the foreground fields is of interest then it is convenient to normalise with respect to a generating functional which incorporates not only any pseudovacuum loop diagrams (as would normally be the case), but also all interactions involving only particles from the background field, and 
all different ways in which the background sources/sinks can be connected. When using such a normalisation, the background fields are implicitly assumed to remain on average unchanged over the length and time scales associated with the process being studied, as is %
appropriate %
in the low-energy limit. 

A minor caveat applies in two-vertex mass interactions of foreground particles (see \crefr{ch:fermion}{ch:detail})%
, where the background may remain unchanged on average only with respect to the process as a whole, and not vertex-by-vertex, provided %
the existence of these nonvanishing modifications of the background field is restricted to length and time scales less than or equal to $\ILO{\mc{L}_0}$ in the isotropy frame of the pseudovacuum. %

There is one other behaviour worth mentioning for interaction multiplets separated by length and time scales less than or equal to $\ILO{\mc{L}_0}$ in the isotropy frame. Until now, it has been tacitly assumed that both of the derivative operators $\bm{D}_\mu$ participating in the construction of a $\h\h^*$ pair, or of a complex scalar boson and a chiral fermion pair ($\h\bar\psi\bar\psi$ or $\h^*\psi\psi$), must arise from the same term of Lagrangian $\mscr{L}_\fg$. However, coupling to the vacuum fields gives rise to uncertainty in position over length scales of $\ILO{\mc{L}_0}$ or less, allowing two vertices separated by distance and time small compared to this scale in the isotropy frame to be effectively treated as collocated. The symbols $\Upsilon^\mu$ and $\bar\Upsilon^\mu$ may then come from separate vertices, though the superselection constraint prohibiting $\Upsilon^\mu\psi\psi$ and $\bar\Upsilon^\mu\bar\psi\bar\psi$ appearing in the same term (\sref{sec:fgspinorscalar}) now applies collectively to the product of all effectively collocated vertices.

\subsubsection{Completing the Lagrangian}

Finally, in order to perform useful calculations, it is necessary to supplement Lagrangian~\eref{eq:Lexp} with source and sink terms for the foreground fields (and perhaps also background fields if not using the normalisation of \sref{sec:normWrtBgFields}). This is performed \emph{ad hoc} in the usual fashion, recognising that these terms are a mathematical convenience encapsulating the introduction and removal of particles from the vicinity of an interaction.

The Lagrangian of a full quantum field theory also necessarily attracts renormalisation terms; the present semiclassical model does not require these additional terms, on account of the existence of a real UV cutoff corresponding (up to a possible boost) to half the peak energy $\mc{E}_\Omega$ of the background fields. Physically relevant results may of course only be %
obtained in situations where %
expressions involving this upper bound %
vanish as the bound is taken to infinity, and are negligible at the actual value of $\frac{1}{2}\mc{E}_\Omega$ %
adopted in the analogue system. %
Fortunately this corresponds %
to the domain of interest established in \sref{sec:pushlimits}, in which the model admits a quasiparticle description of long-range correlations corresponding to the foreground fields. %
Hypothetically, if a realisation of the $\Cwt$ model were achieved in a lab,
then the value of $\mc{E}_\Omega$ would naturally be finite %
and the application of the model would be constrained to the computation of physically observable processes independent of $\mc{E}_\Omega$ such as low-energy scattering amplitudes ($\mc{E}\ll\mc{E}_\Omega$).

\subsubsection{Collapse of the wavefunction}

This model contains no intrinsic mechanism for the collapse of the wavefunction, only being capable of computing the evolution of the unobserved wavefunction over time. However, the same %
could arguably be said of quantum theory as a whole, in which %
the collapse of the wavefunction is likewise generally inserted as an \emph{ad hoc} postulate (with some notable exceptions, such as decoherence- and gravitation-based %
models of wavefunction collapse \cite{zurek1991,schlosshauer2005,penrose1996,bassi2017}).

Similarly, in an analogue model the collapse of the wavefunction may likewise be introduced as an \emph{ad hoc} postulate, requiring manual intervention to set the model to the collapsed state once an observation is carried out.

It is notable %
that the $\Cwt$ model is endowed with a plenitude of background fields at all points in space and time, with potential to act as a reservoir for purposes of decoherence-mediated effective wavefunction collapse.

\section{Physical applications and implications\label{sec:physapp}}

The present chapter has introduced the basics of the $\Cwtn$ model series with particular emphasis on the simplest member of that series, $\Cwt$. Full details of construction of the higher-$\N$ members of this series are deferred to \cref{ch:colour}, %
yet from the properties of $\Cwt$ it is nevertheless possible to infer some implications and applications of the $\Cwtn$ model series. 
The author's primary purpose in constructing the $\Cwt$ series is to support a particular analogue to the Standard Model in curved space--time which is described in \crefr{ch:SM}{ch:gravity}. However, some other applications may also be briefly speculated on here.

\subsection{Applications in model building}

First, consider the relationship between the $\Cwt$ model and conventional supersymmetric classical (``superclassical'') models as discussed by e.g.~DeWitt in \rcite{dewitt1986}.
A DeWitt-type supermanifold always has a real space--time as its body, and thus admits both commuting and anticommuting co-ordinates, e.g.~$\mbb{R}^{4|4}$. 
In \sref{sec:repCwt} it was stated that $\Cwn$ also admits an embedding in $\mbb{R}^{4|4}$, and that this embedding inhabits the restriction of the Minkowski subspace to a single point, e.g.~$\mbb{R}^{4|4}|_{x=(0,0,0,0)}$. Indeed, given the embedding of vectors on $\Cwn$ within a single copy of the supernumbers, the $\Cwn$ model may be completely contained within a submanifold of $\mbb{R}^{4|4}$ such as $\mbb{C}^{0|1}$ or $\mbb{R}^{0|2}$, having only anticommuting co-ordinates. %

It is frequently presumed that purely anticommuting manifolds cannot support normalised excitations due to the exhaustion of the Taylor series discussed in \sref{sec:quantumdust}. In the present model the construction of normalisable excitations was made possible through 
the %
use %
of techniques from condensed matter physics, 
with normalised \emph{quasi}particle excitations being collectively supported by an arbitrarily large number of non-normalised fields.

Second, recognise that the superspace (or Grassmann space) inhabited by a $\Cwn$ model has no commuting degrees of freedom and no time axis, and it is therefore necessary to identify a subspace of this manifold which is {isomorphic} to $\RM$ on which the quasiparticle excitations are to be studied.

Thus this chapter introduces a technique which permits an entirely new class of manifolds to host emergent quasiparticle excitations and space--times.

\subsection{Applications in supersymmetry}

Perhaps one of the most intriguing aspects of this model is that a rich universe of interacting (quasi)particles can be constructed in the ancillary, anticommuting sector of superspace. While it is unremarkable that a space isomorphic to $\RM$ can be found over every point in the $\RM$ submanifold of any superspace, %
given their restricted field content it is perhaps more surprising that these bundles of spaces should be capable of hosting normalisable excitations with nontrivial dynamics. %
These emergent behaviours appear to be restricted to the high-particle-number regime, and it is possible that the study of these emergent excitations might therefore be able to give insights into particle behaviours in this %
regime as a result.

\subsection{Applications in quantum gravity}

Another point of interest is the emergence of normalisable particle fields on Minkowski space--time from an anticommuting manifold hosting no normalisable fields and having no time axis. Given the tension between quantum field theory and relativistic gravitation, entire fields of research such as spin networks/loop quantum gravity~\cite{rovelli2004} are predicated on the principle that an emergent space--time might circumvent problems with renormalisation of the gravitational interaction. 
In loop quantum gravity, space--time emerges from the connections of nodes, whereas in the present model, normalisable excitations on Minkowski space--time emerge from unnormalisable classical fields on a timeless and anisotropic manifold. Although these space--times are constructed in different ways, the $\Cwtn$ series also provides a situation in which a pseudo-Riemannian space--time emerges from a less problematic environment, namely one on which the Taylor series of all fundamental fields terminate. (\Cref{ch:gravity} %
extends this construction to also include the natural emergence of curved space--times.) %
This work may therefore provide a new direction of research in quantum gravity.

\subsection{Applications in foundations of physics}

Why does the interaction Lagrangian (or indeed, any Lagrangian at all) even exist? Appeal is customarily made to N\oe{}ther's theorem and to local symmetries, these being a natural consequence of conservation laws in a reality whose structure takes the form of bundles over space--time.

The construction of the $\Cwt$ model presented in this chapter provides an alternative answer to this question, at least for the $\Cwtn$ series of manifolds. In these models there is no Lagrangian at all on the underlying manifold, and all fields are completely free to take any values whatsoever, subject only to the constraint that the manifold co-ordinates anticommute and the fields must be functions of these co-ordinates. Surprisingly, from these near-arbitrary fields and minimal constraints an effective Lagrangian emerges, complete with local symmetry. (In the $\Cwt$ model this symmetry is $\GL{1}{R}$, mediated by the vector boson $\vp_\mu$. Higher values of $\N$ in $\Cwtn$ yield larger symmetry groups---see \crefr{ch:colour}{ch:SM}.) %

\subsection{Applications in particle physics\label{sec:appPphys}}

The construction of the $\Cwt$ model from excitation gradients on a manifold with anticommuting degrees of freedom has the effect of implying a preon substructure for all fields of dimension greater than $L^{-1/2}$. The intrinsic symmetries of the anticommuting manifold then guarantee description of the emergent spinor, vector, and scalar fields under a single, explicit unifying symmetry group, corresponding to the local symmetry of the manifold. 

In \cref{ch:colour}, extending this construction to $\Cw{6}$
is seen to generate a local $\SU{3}$ symmetry, with the resulting composite bosons acting as the gluons of the emergent interaction. This model may perhaps therefore provide a novel environment in which to study confinement, with a natural upper limit on energy scale corresponding to the energy at which the composite bosons themselves must be described in decomposed form.

\section{Conclusion\label{sec:lookforward}}

The construction demonstrated in this chapter serves four main purposes.
First, %
it demonstrates that classical field theories on manifolds with anticommuting co-ordinates are capable of exhibiting interesting (and normalisable) quasiparticle excitations.
Second, by mapping to a simple quantum field theory on $\RM$ it provides a familiar %
description of an unfamiliar system, %
namely perturbations %
of collections of fields on anticommuting manifold $\Cwt$. 
Third, %
it introduces %
concepts and constructions which will be key to the %
$\Cw{2\N}$ series of analogue models, all of which have behaviours mapping to effective quasiparticles described by semiclassical field theories on $\RM$.
Higher-$\N$ systems map to more complex physical systems, with $\N=3$ anticipated to represent a Yang-Mills gauge theory, and $\N={9}$ %
incorporating %
all the interactions of the Standard Model through %
a 
construction 
described in \crefr{ch:colour}{ch:gravity}. %

Finally,
for any model it is necessary to motivate the construction of the Lagrangian. 
For the model on $\Cwt$, and indeed for all models in the $\Cw{2\N}$ series, there is a particularly close connection between dynamics and geometry, and the $\Cwt$ model demonstrates how the equations of motion and hence the Lagrangian may arise entirely from constraints imposed by the anticommutation rules of the manifold co-ordinates, and the construction of the effective degrees of freedom (the quasiparticle fields).

Following chapters extend the present model to manifolds $\Cw{2\N}|_{\N>1}$, including the construction of a more familiar fermion sector consisting of species with dimension $L^{-3/2}$. Of particular interest is $\Cw{18}$, which contains an analogue of the Standard Model and is capable of predicting the masses of the tau particle and the weak sector bosons. Some other areas for potential application of the $\Cwtn$ model series are discussed in \sref{sec:physapp}, above.

\notchap{
\section*{Acknowledgements}
This research was supported in part by the Perimeter Institute for Theoretical Physics.
Research at the Perimeter Institute is supported by the Government of Canada through Industry Canada and by the Province of Ontario through the Ministry of Research and Innovation.
The author thanks the Ontario Ministry of Research and Innovation Early Researcher Awards (ER09-06-073) for financial support.
This project was supported in part through the Macquarie University Research Fellowship scheme.
This research was supported in part by the ARC Centre of Excellence in Engineered Quantum Systems (EQuS), Project No.~CE110001013.
}

\bibliographystyle{apsrev4-2}
\bibliography{Paper.bib}

\end{document}